\documentclass[12pt]{article}
\usepackage[margin=2 cm]{geometry}
\usepackage{comment}
\usepackage[skip=0.5ex, belowskip=1ex,
                labelformat=brace,
                singlelinecheck=off]{subcaption}
                
\usepackage{caption}
\usepackage{amsmath}
\usepackage{graphicx}
\usepackage{dsfont}
\usepackage[font={small,it}]{caption}
\usepackage{mwe}
\usepackage[nottoc]{tocbibind}
\usepackage{amsmath,amssymb,extarrows,mathtools,graphicx,setspace}
\usepackage{cite}
\usepackage{stackengine}
\usepackage{tikz}\usetikzlibrary{calc}
\usepackage{braket}
\usepackage{epsfig}
\usepackage[section]{placeins}
\usepackage{slashed}
\usepackage{color}
\usepackage{caption}
\usepackage{amsmath}
\usepackage{hyperref}
\makeatother
\newmuskip\pFqmuskip

\newcommand{\be}{\begin{equation}}
\newcommand{\bea}{\begin{eqnarray}}
\newcommand{\eea}{\end{eqnarray}}
\newcommand{\ba}{\begin{array}}
\newcommand{\ea}{\end{array}}
\newcommand{\ee}{\end{equation}}
\newcommand{\bes}{\begin{equation*}}
\newcommand{\beas}{\begin{eqnarray*}}
\newcommand{\eeas}{\end{eqnarray*}}
\newcommand{\bas}{\begin{array*}}
\newcommand{\eas}{\end{array*}}
\newcommand{\ees}{\end{equation*}}

\newcommand*\pFq[6][8]{%
  \begingroup 
  \pFqmuskip=#1mu\relax
  \mathcode`\,=\string"8000
  \begingroup\lccode`\~=`\,
  \lowercase{\endgroup\let~}\pFqcomma
  {}_{#2}F_{#3}{\left[\genfrac..{0pt}{}{#4}{#5};#6\right]}%
  \endgroup
}
\newcommand{\pFqcomma}{\mskip\pFqmuskip}

\numberwithin{equation}{section}

\begin{document}

\onehalfspacing
\vfill
\begin{titlepage}
\vspace{10mm}

\begin{center}

\vspace*{10mm}
\vspace*{1mm}
{\Large  \textbf{Photonic Exceptional Points in Holography and QCD}} 
 \vspace*{1cm}

{$\text{Mahdis Ghodrati}$}

\vspace*{8mm}
{ \textsl{ $ $ International Centre for Theoretical Physics Asia-Pacific,
University of Chinese Academy of Sciences, 100190 Beijing, China}}

 \vspace*{0.45cm}

\textsl{e-mails: {\href{mahdisghodrati@ucas.ac.cn}{mahdisghodrati@ucas.ac.cn}}}
 \vspace*{2mm}

\vspace*{1.7cm}

\end{center}

\begin{abstract}
In this work, based on an analogy with holographic confining geometries and using complexified fields, we build a holographic toy model of third order photonic exceptional points (EPs) of ternary coupled microrings with gain and loss, which makes an open, non-Hermitian quantum system. In our model, we discuss the Ferrell-Glover-Tinkham sum rule for various combinations of gain and loss systems, and numerically find the behavior of spectra which matches  with the experiments. We also discuss the inhomogeneous case of a holographic lattice for three-site photonic EPs. Additionally, we numerically find the behavior of phase rigidity and the Petermann factor around EPs versus various parameters of the model. We also discuss the connections between recent developments in complexified, time-dependent entanglement entropy and EPs, and then, we connect EPs and the $\theta$-vacuum of QCD through topological structures, partition functions, and winding numbers, and find a second-order EP in a perturbed $\theta$-vacuum model. Finally, we examined a controlled non-Hermitian deformation of $\theta$-vacuum toy model, by using the Lindblad formalism and Liouvillian eigenvalues.

 \end{abstract}

\end{titlepage}

\tableofcontents

\section{Introduction}

There are interesting connections between quantum computing and quantum chromodynamics (QCD) concepts that could be implemented in studying various problems in physics such as deriving the phase diagrams of QCD, confinement, strong CP problem, calculating hadron properties, testing string theory in the lab, etc.  For instance, the lattice gauge theories can be used to study the properties of QCD on quantum computers \cite{Ghodrati:2015rta} or on the other hand, formulation of the problem for quantum simulations can lead to new perspectives about the quantum structure of gauge theories.
Similar other connections have been studied in \cite{Ghodrati:2014spa,Ghodrati:2016tdy,Ghodrati:2017roz, Ghodrati:2018hss, Ghodrati:2019hnn,Zhou:2019jlh,Ghodrati:2019bzz,Ghodrati:2020vzm,Ghodrati:2021ozc,Ghodrati:2022kuk,Ghodrati:2022hbb,Ghodrati:2023uef, Ghodrati:2020mtx,Ghodrati:2016ggy,Ghodrati:2016vvf,Ghodrati:2022lnd}.

Using ideas from photonic quantum computing or optics, one could also gain novel insights to simulate various QCD structures such as meson decays. In this work, we aim to pursue such an unconventional approach. We aim to connect the non-Hermitian Hamiltonians and their corresponding exceptional points to Anti deSitter (AdS)/Condensed Matter (CMT) and AdS/QCD-like models.

\begin{figure}[ht!]   
\begin{center}
\includegraphics[width=0.82\textwidth]{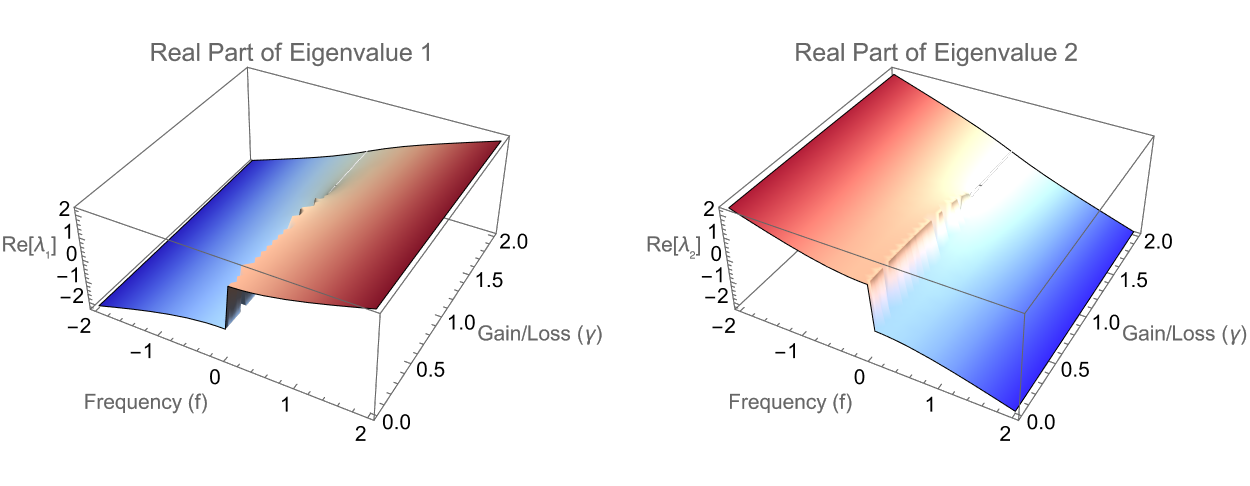}  \ \ \ 
\includegraphics[width=0.82\textwidth]{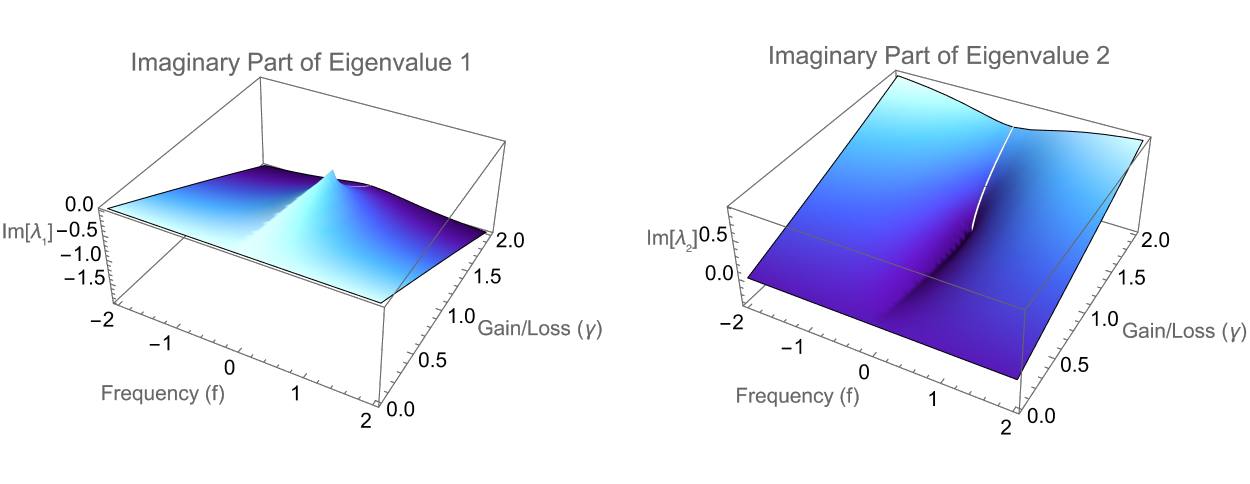} 
\caption{The relationship between the real and imaginary parts of the eigenvalues versus frequency $f$ and the gain/loss parameter $\gamma$ is shown.}
\label{fig:EPplots}
\end{center}
\end{figure}

Exceptional points (EPs) \cite{Heiss_2012}, as shown in figure \ref{fig:EPplots}, are branch-point singularities in the parameter space of a system where two or more eigenvalues and their corresponding eigenvectors merge and become degenerate.
They are the boundaries between regions of unbroken $\mathcal{PT}$-symmetry, where a Hamiltonian $H$ has all real eigenvalues, and regions of broken $\mathcal{PT}$-symmetry
where $H$ has at least one pair of complex conjugate eigenvalues.

EPs appear in many different physical problems, and they correspond to exceptional topologies \cite{Bergholtz:2019deh}. These points are also associated with symmetry breaking for $\mathcal{P}\mathcal{T}$-symmetric Hamiltonians. Such symmetry-breaking effects have been observed in numerous experiments in QCD, optics, atomic and molecular physics, as well as in quantum phase transitions and quantum chaos.

In several confining holographic models, on the other hand, the transition from the free phase to the confined phase has been modeled by a UV cut-off dubbed the “end-wall” in the geometry, which significantly affects the system as one gets close to it.

Here, we make an analogy between exceptional points and exceptional lines with the end-wall of holographic confining geometries, which can simulate the confinement/deconfinement phase transitions \cite{Fatemiabhari:2024aua}. We note that, as both EPs and the confining end-wall cause similar phase transitions, induce chaos, and cause chiral symmetry breaking, each one could modeled using the other one. Also, it has been suggested that at EPs, the metric ceases to exist \cite{Geyer:2008sug}, similar to the effects of the end-wall in AdS/QCD models.

There are several other reasons which indicate such connections between AdS/QCD models and EPs. First, is the behavior of the quantum measure “fidelity.” In quantum information, one could use fidelity to find the exceptional points and their orders, as they only require ground-state calculations. The definition of fidelity is
\begin{gather}
F= | \langle \psi_0 (\lambda) | \psi_0(\lambda+\epsilon) \rangle |^2,
\end{gather}
where $\epsilon$ is small, and it can define the distance between the quantum states and detect quantum phase transitions in the parameter space. 

After series expansion,
\begin{gather}
F= 1- \chi_F \epsilon^2 + \mathcal{O}(\epsilon^3),
\end{gather}
the fidelity susceptibility $\chi_F$ can also be defined, which diverges toward positive infinity as the parameters get close to the quantum phase transition point, i.e., $ \lim\limits_{\lambda \to \lambda_{\text{QCP}} } \mathbb{R}e \chi_F = \infty $, while for the case of EPs, the real part of the fidelity susceptibility diverges toward negative infinity, i.e., $ \lim\limits_{\lambda \to \lambda_{\text{EP}} } \mathbb{R}e \chi_F = -\infty $. So its absolute value for both scenarios would diverge. Therefore, the holographic dual of exceptional points should be based on such models of phase transitions.

Another reason is that the non-Hermitian Dirac operators in QCD can be modeled by (higher-ordered) EPs. At finite quark chemical potential, the QCD Dirac operator, $D(\mu) = \gamma^\mu (\partial_\mu + ig A_\mu) + \mu \gamma^0$ becomes non-Hermitian, as its eigenvalues become complex and can exhibit spectral degeneracies. These degeneracies can coalesce and so the chiral phase transition in QCD could be approached via a spectrum collapse similar to an EP.

The similar behavior of chaotic systems near exceptional points and around the confining walls is another evidence for this analogy. In \cite{Bianchi:2021sug}, the chaos in the scatterings of open strings has been discussed, where they noticed that the bound-state creation in the process would leave its signatures on the plots of the amplitude $A_4$ versus the Mandelstam variable $s$. They noted that the peaks in the plots are more pronounced at small $s$, but they fade away at large $s$, which they interpreted as the effects of asymptotic freedom, where the gauge coupling becomes weaker, and therefore the binding energy of the bound states becomes smaller. This behavior can also be modeled by the non-Hermitian skin effect and its bound states inside a loop \cite{Gong:2022krq,Gong:2022oce}. Note that both in \cite{Bianchi:2021sug} and \cite{Gong:2022krq}, a similar power-law decay has been found, which only matches with the non-Hermitian behavior, as it would be invisible for Hermitian baths.

Based on the above ideas and based on bottom-up holographic QCD, in this work, we build a toy model which can simulate third order EPs. First, we explain the optical setup of anti-$\mathcal{PT}$ symmetry consisting of ternary coupled optical resonators and microrings. Based on this system, we then build a holographic toy model to simulate EPs. We discuss various results coming from this model such as lasing spectra, eigenvalue trajectories, $\mathcal{PT}$ phase diagram, Ferrell-Glover-Tinkham (FGT) sum rule, transfer fraction to coherent peak and holographic Petermann factor and phase rigidities.

Then, we discuss the connections between exceptional points and timelike entanglement entropy and the behavior of Kirkwood-Dirac-type distribution in the presence of EPs.

Finally, we focus on QCD itself and first define a winding number for EPs based on the winding number in QCD and then we look for EPs in $\theta$-vacuum QCD. We actually could not find any EP in the pure case, however after adding a small angular coupling, we found a second-order EP there. We finish with a brief conclusion.

So we should emphasize that in sections \ref{sec:Holoconf} and \ref{sec:Hol2}, we build a bottom-up phenomenological, holographic-like toy model and present an approximate dictionary, while in section \ref{sec:Kaons} we construct an analogy based on comparing two matrices for two different open quantum systems, where there is no holography argument in sections \ref{sec:Kaons} and \ref{sec:EPtime}. Then, in section \ref{sec:numEP}, we look for EPs in a perturbed or open $\theta$-vacuum QCD model directly (without any use of holography), first by a numerical search \ref{sec:numEP} and then analytically in \ref{sec:controlled}.

\section{Photonic setup of anti-$\mathcal{PT}$ symmetry and higher-order exceptional point}\label{Observation of antiPT}

\begin{figure}[ht!]   
\begin{center}
\includegraphics[width=0.3\textwidth]{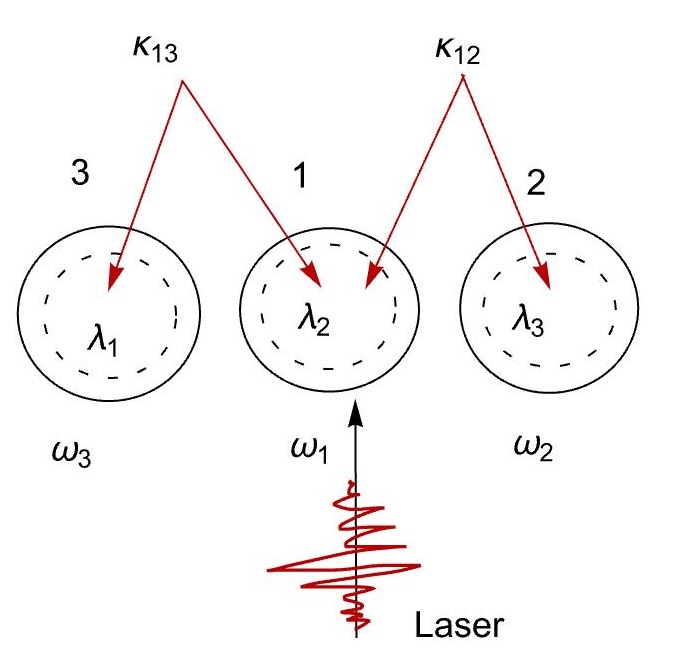} \ \ \ \ \ \ \  \ \ \ \ \ \ 
\includegraphics[width=0.31\textwidth]{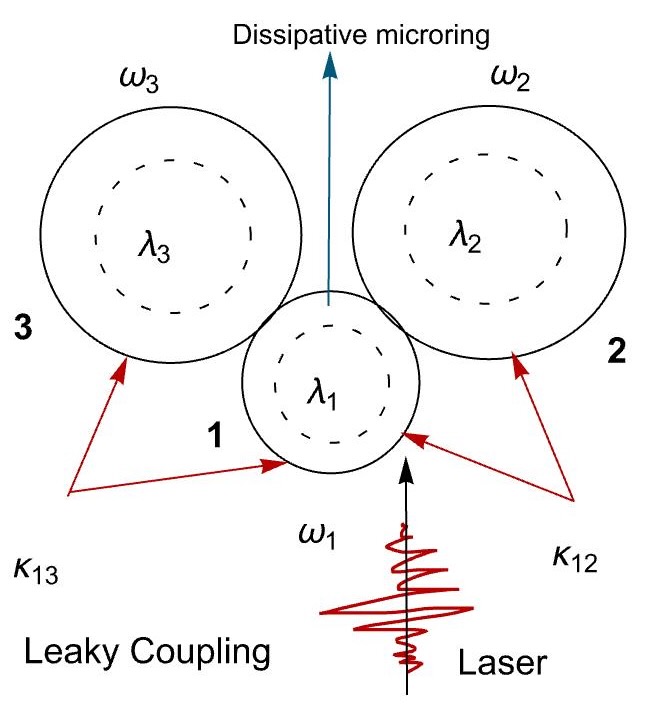} 
\caption{A ternary system, consisting of three coupled microrings is shown. On the left, a $\mathcal{PT}$-symmetric and on the right, an anti-$\mathcal{PT}$-symmetric system with dissipative coupling are depicted. This setup with single-mode operation could lead to various orders of exceptional points.}
\label{fig:EPmicrorings}
\end{center}
\end{figure}

In \cite{citekeyJahangiri,Miri2019ExceptionalPI}, for single-mode operations and for ternary systems, the observation method of anti-parity-time (APT) symmetry and higher-order exceptional point has been investigated. Their experimental setup for different configurations of ternary systems synthesizing a third-order exceptional point is shown in figure \ref{fig:EPmicrorings}.

In the vicinity of a third-order exceptional point, and when the $\mathcal{PT}$ symmetry is being broken, as has been argued in \cite{citekeyJahangiri}, one can experimentally achieve a stable single-mode laser, which could then be used for different applications.

For three coupled resonators, the energy exchanges between these optical resonators can be written as
\begin{gather}
\frac{i d a_1}{dt}=-\omega_1 a_1 - i \lambda_1 a_1 + \kappa a_2, \nonumber\\
\frac{i d a_2}{dt}=-\omega_2 a_2 - i \lambda_2 a_2 + \kappa a_1, \nonumber\\
\frac{i d a_3}{dt}=-\omega_3 a_3 - i \lambda_3 a_3 + \kappa a_1,
\end{gather}
where $a_n$ are the energy amplitudes in the cavities, $\omega_n$ is the resonance frequency, $\lambda_n$ is the net gain/loss in microcavities, and $\kappa$ is the coupling factor between micro-resonators.

For the triple-coupled microring, the non-Hermitian $3\times 3$ Hamiltonian is
\begin{gather} \label{eq: quarkH}
H_{PT} + H_{\epsilon_{\lambda} }  = \begin{pmatrix}
\omega_1+i \lambda_1 + i \epsilon_{\lambda}  & \kappa_{12} & \kappa_{13} \\
\kappa_{21} &\omega_2+ i \lambda_2+ i \epsilon_{\lambda}  & \kappa_{23}=0 \\
\kappa_{31}  & \kappa_{32}=0 & \omega_3 + i \lambda_3
\end{pmatrix},
\end{gather}

\begin{gather} \label{eq: quarkH}
H_{APT} + H_{\epsilon_{\lambda} }  = \begin{pmatrix}
\omega_1+i \lambda_1 + i \epsilon_{\lambda}  & i \kappa_{12} & i \kappa_{13} \\
i \kappa_{21} &\omega_2+ i \lambda_2+ i \epsilon_{\lambda}  & \kappa_{23}=0 \\
i \kappa_{31}  & \kappa_{32}=0 & -\omega_3 + i \lambda_3
\end{pmatrix},
\end{gather}
where $w_{1-3}$ are the main resonance frequencies in three microrings, $\lambda_{1-3}$ are the net loss or gain in each microring, and $\kappa_{12}= \kappa_{21}=\kappa, \kappa_{13}=\kappa_{31}$ are the coupling coefficients between the microrings. Also, $\epsilon_\lambda$ is the gain–loss difference caused by the perturbation.

The lasing spectra of symmetric and asymmetric gain-gain-gain and gain-loss-loss microrings, evenly pumped or partially pumped triple microrings, and unbroken or broken parity-time symmetric coupled microring lasers are shown in \cite{citekeyJahangiri}. Now our goal is to build a holographic toy model to simulate their results and find a similar spectra, which we show in the next section.

\section{Holographic confining toy model of coupled ternary resonators and EPs}\label{sec:Holoconf}

In holographic QCD models, flavor branes are introduced to account for the presence of quarks. Similarly, each microresonator in the optical system can be viewed as a flavor brane, with its own dynamics influenced by the gain and loss perturbations. The coupling coefficients between the microresonators determine the interactions between the resonator modes. In a bottom-up, phenomenological holographic toy model, these coupling coefficients are related to bulk interactions between the flavor branes, affecting the meson spectrum and dynamics. Exceptional points in the optical system correspond to singularities in the eigenvalue spectrum, marking phase transitions between different symmetry phases. In a holographic dual perspective, these exceptional points can be interpreted as critical points or the end-wall in the bulk geometry, associated with phase transitions in the boundary side.

To construct a bottom-up, phenomenological holographic toy model, we can start with a 5-dimensional $\text{AdS}$ spacetime with a metric that incorporates the effects of the flavor branes. The action governing the dynamics of the system includes terms for the bulk gauge fields, the flavor branes, and the interactions between them. By solving the equations of motion derived from this action, we can obtain the qualitative meson spectral and response features observed in non-Hermitian photonic systems with exceptional points, and then we can study the phase transitions corresponding to the presence of exceptional points in the optical system.

The $(d+1)$-dimensional AdS metric can be written as
\begin{gather}
ds^2= \frac{R^2}{z^2} (\eta_{\mu \nu} dx^\mu dx^\nu + dz^2), \ \ \ \ \ \ \ \mu, \nu =0, 1, ... d.
\end{gather}
The wave equation for a scalar particle of mass $M$ in the bulk of AdS is
\begin{gather}
\psi_k (z) = z^{\frac{d}{2}} J_{\Delta - \frac{d}{2} } (mz),
\end{gather}
where
\begin{gather}
\Delta = \frac{d}{2} + \sqrt{\frac{d^2}{4} + M^2 R^2},
\end{gather}
and $J_\alpha$ is the Bessel function. Here the spectrum of mass is $m^2 = - k^2$. Note that in the free AdS geometry, the system is gapless and the spectrum is continuous, while in QCD the system is gapped and the spectrum is discrete, which is similar to the case near the EPs.

As for the boundary condition, one can assume that either the wave function vanishes on the wall, i.e., $\psi_k(z_0) =0$, or it becomes peaked at the wall, which lets us determine the spectrum by zeroes of the appropriate Bessel function \cite{Bianchi:2021sug}. The choice of the boundary condition affects the structure of the amplitude only at low energies.
For $M^2=0$ and $d=4$, the wave function is
\begin{gather}
\psi_k(z) = z^2 J_2(mz).
\end{gather}
However, in the hard-wall model, the excitation spectrum is not in the form of $m_n^2 \sim n$, as one would expect. To achieve such a spectrum, one can use the ``soft-wall'' model instead. In the soft-wall model, the effective potential for a dilaton $D=\lambda^2 z^2$ is in the form of \cite{Bianchi:2021sug}
\begin{gather}
V_{\text{soft}} (z) = V_{\text{AdS}}(z) + \delta V_{\text{soft}}(z)= \frac{4\mu^2 L^2 + \kappa(\kappa+2)}{4z^2} + \lambda^4 z^2 + (\kappa-1)^2 \lambda^2.
\end{gather}
 
Instead of using a hard wall model, the AdS horizon then could be cut softly with an exponential function $e^{-D}$. Then, the eigenvalues and eigenfunctions would be \cite{Bianchi:2021sug}
\begin{gather}
\psi_{n, \ell} (z) = z^{\ell+ \frac{d}{2} } L_{n, \ell} (\lambda^2 z^2), \
m^2_{n, \ell}= 2 \lambda^2 (2n + \ell + 2).
\end{gather}
Here, $L_{n, \ell}(x)$ are the Laguerre polynomials. One should note that at EPs the norm of these wave functions vanishes and there is a square-root branch point in the interaction strength.
When at the EP, the norm of a wave function vanishes, the two resonances coalesce, and the spectroscopic factors increase significantly. This effect has been used to model nuclei near the drip line \cite{Michel_2010}, which specifically is an open system. Here we use such soft-wall models to construct our analogy.

Another interesting top-down holographic model which could be used in future works, is Witten's model with the metric
\begin{gather}
ds^2 = \left( \frac{U}{L} \right)^{\frac{3}{2} } \left( \eta_{\mu \nu} dx^\mu dx^\nu + f(U) dx_4^2\right) + \left( \frac{L}{U} \right)^{\frac{3}{2} } \left( \frac{dU^2}{f(U)} + U^2 d\Omega_4^2\right),
\end{gather}
where $f(U) = 1- \frac{U_\Lambda^3}{U^3}$, and $U_\Lambda$ is the position of the wall, and $L^3 = \pi g_s N_c \ell^3$. The coordinate $x_4$ has circular radius $R = \frac{3 U_\Lambda^{1/2} }{2 \pi L^{3/2} }$, which sets the mass scale $M_{gb} = 1/R$. The spectrum, i.e., the wave functions of the first three scalar modes of this model, are shown in figure 1 of \cite{Bianchi:2021sug}, where all modes peak at $r=r_0$.

For this model, in \cite{Bianchi:2021sug}, it has been shown that the spectra are not of a Regge form but mostly in the form of a KK-like spectrum as $m^2 \approx aN^2 + bN +c$, again similar to the $N$ levels coalescing of an N-dimensional matrix, where for complex symmetric matrices, $(N^2 + N -2)/2$ parameters are needed to enforce the coalescence of $N$ levels \cite{Heiss_2008}. So this point again suggests a connection between EPs and the holographic end-walls, as both show similar functional behaviors in $N$. This wall at $U_\Lambda$ also causes various phase transitions, drastic changes in the entanglement entropy, mutual information, and other quantum information quantities, and induces chaos. 

Then, to include the flavor branes, we introduce a warp factor in a soft-wall model as
\begin{gather}
ds^2 = \frac{L^2}{z^2} e^{- \phi(z)} (\eta_{\mu \nu} dx^\mu dx^\nu - dz^2),
\end{gather}
where $\phi(z)$ is a dilaton-like background simulating the effect of the microresonators’ gain/loss, which is analogous to the flavor branes’ backreaction. For a “soft-wall” model, a common choice is $\phi(z) = \kappa^2 z^2$, where $\kappa$ sets a scale that can be tuned to reflect coupling strengths or exceptional points.
The bulk action for a gauge field $A^a_M$, which simulate mode dynamics of each resonator, and scalar $X$, simulating the symmetry-breaking effects, or the $\mathcal{PT}$/anti-$\mathcal{PT}$ transition, would be $S= S_{\text{bulk}}+S_{\text{flavor}}$. 

Here the bulk action is
\begin{gather}
S_{\text{bulk}} = - \frac{1}{4g_5^2} \int d^5 x \sqrt{-g} \ \text{Tr} \ \lbrack F_{MN} F^{MN} \rbrack + \int d^5 x \ \sqrt{-g} \ \text{Tr} \ \left \lbrack | D_M X|^2 - m_5^2 |X|^2 \right \rbrack,
\end{gather}
where
\begin{gather}
F_{MN} = \partial_M A_N - \partial_N A_M - i \lbrack A_M, A_N \rbrack, \nonumber\\
D_M X = \partial_M X - i A_M X + i X A_M,
\end{gather}
and $m_5^2$ is the $5D$ mass squared of the scalar field, which is related to the conformal dimension $\Delta$ of the dual operator as $m_5^2 L^2 = \Delta (\Delta-4)$. Note that here the complex bulk parameters are only computational devices.

Note that here, the complex bulk couplings and effective masses are used as calculational devices to encode dissipative or gain-loss effects of the boundary theory. We are not claiming the existence of a fundamental non-Hermitian bulk gravitational theory. Rather, physical non-Hermiticity should be understood as entering through boundary conditions, external sources, or effective open-system descriptions, in line with Schwinger-Keldysh or influence-functional approaches.

Then, for the three flavor branes, which represent the ternary system, the action would be
\begin{gather}
S_{\text{flavor}} = -T_3 \sum_{i=1}^3 \int d^4 x \ dz \ \sqrt{- \text{det} \left(g_{MN} + 2\pi \alpha' F_{MN}^{(i)} \right)},
\end{gather}
where $T_3$ is the brane tension, and $F_{MN}^{(i)}$ is the field strength on the $i$-th brane, each corresponding to a resonator. The coupling between branes, which represents the anti-$\mathcal{PT}$ symmetry and exceptional point interactions, can be included via off-diagonal terms in $F_{MN}$ or interaction terms in $X$.

Then, the gauge field equation is
\begin{gather}
\frac{1}{\sqrt{-g}} D_M ( \sqrt{-g} F^{MN}) - i \lbrack X, D^N X^\dagger \rbrack = 0,
\end{gather}
which describes mode interactions between resonators in the bulk.
The off-diagonal terms in $X$ or $F^{MN}$ capture exceptional point couplings.

Then, the scalar field equation is
\begin{gather}
\frac{1}{\sqrt{-g}} D_M ( \sqrt{-g} D^M X) - m_5^2 X = 0,
\end{gather}
which controls the symmetry-breaking pattern, mapping to $\mathcal{PT}$/anti-$\mathcal{PT}$ transitions.

So, the 3-mode microresonators are related to the 3 flavor branes or gauge fields $A_M^{(i)}$, the gain/loss is related to the background dilaton $\phi(z)$, and the coupling/anti-$\mathcal{PT}$ symmetry is related to the off-diagonal components in $F_{MN}$ or $X$. The exceptional points are related to the critical points in the bulk potential $X(z)$ or interaction term. The single-mode lasing would be related to the normalized lowest-energy mode in the bulk spectrum.

Then, the mode amplitudes along $z$, $a_i(z)$, would be
\begin{gather}
\partial_z \left (\frac{e^{- \phi(z)} }{z} \partial_z a_i(z) \right ) + \frac{e^{- \phi(z)} }{z} \sum_{j=1}^3 \mathcal{M}_{ij} (z) a_j(z) = 0,
\end{gather}
where $\mathcal{M} (z)$ is the coupling matrix, representing gain/loss and anti-$\mathcal{PT}$ symmetry. For the anti-$\mathcal{PT}$ symmetry with ternary resonators, it would be
\begin{gather}
\mathcal{M} (z) =  \begin{pmatrix}
i\gamma & \kappa_{12} & \kappa_{13}\\
\kappa_{12} & 0 & \kappa_{23} \\
\kappa_{13} & \kappa_{23} & -i\gamma
\end{pmatrix},
\end{gather}
where the explicit coupled equations are
\begin{gather}
\partial_z \left ( \frac{e^{- \phi(z)} }{z} \partial_z a_1 (z) \right )+ \frac{e^{- \phi(z)} }{z} (i \gamma a_1 + \kappa_{12} a_2 + \kappa_{13} a_3 ) = 0, \nonumber\\
\partial_z \left ( \frac{e^{- \phi(z)} }{z} \partial_z a_2 (z) \right )+ \frac{e^{- \phi(z)} }{z} (\kappa_{12} a_1 + 0 \ . \  a_2 + \kappa_{23} a_3 ) = 0, \nonumber\\
\partial_z \left ( \frac{e^{- \phi(z)} }{z} \partial_z a_3 (z) \right )+ \frac{e^{- \phi(z)} }{z} (i \gamma a_1 + \kappa_{13} a_1 + \kappa_{23} a_2 - i \gamma a_3 ) = 0. 
\end{gather}

One should note that these bulk equations are along the holographic direction $z$, and the off-diagonal couplings $\kappa_{ij}$ generate interactions between the branes. The imaginary entries $i \gamma$ encode the gain/loss, which realizes the anti-$\mathcal{PT}$ symmetry in the bulk. In the UV boundary, $z \to 0$, the Dirichlet boundary condition leads to $a_i(0) = a_i^{(0)}$, which corresponds to the driving/resonator amplitude input. In the IR boundary, $z=z_{IR}$, the normalizable solution $\partial_z a_i (z_{IR})=0$ selects the single-mode lasing.

Then, for a higher-order exceptional point, EP3, one can write
\begin{gather}
\text{det} ( \mathcal{M} - \lambda I ) =0,  \ \ \text{with triple root at \ } \lambda= \lambda_{\text{EP}},
\end{gather}
which gives the constraints on $\gamma$ and $\kappa_{ij}$ for which the three bulk modes coalesce in both eigenvalues and eigenvectors. For example, for the symmetric couplings $\kappa_{12} = \kappa_{23} = \kappa_{13} = \kappa$, an EP3 occurs at $\gamma = \sqrt{2} \kappa$.

\begin{figure}[ht!]   
\begin{center}
\includegraphics[width=0.4\textwidth]{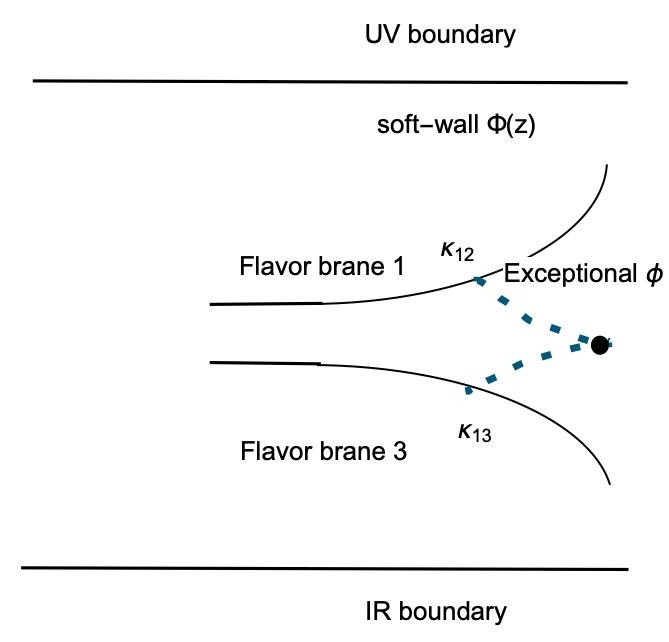}  \ \ \ 
\caption{Modeling EPs using flavor branes in the bulk, showing the couplings and where the exceptional point occurs along $z$.}
\label{fig:EPHoloModel}
\end{center}
\end{figure}

Therefore, starting from a $5d$ AdS metric with soft-wall, in the from of $ds^2 = \frac{L^2}{z^2} e^{- \kappa^2 z^2} ( \eta_{\mu \nu}dx^\mu dx^\nu - dz^2)$, we can find numerically three coupled bulk equations for flavor brane gauge fields $a_1(z), a_2(z), z_3(z)$, and the coupling matrix $\mathcal{M}(z)$ with gain/loss for anti-$\mathcal{PT}$ symmetry. We also can find the EP3 condition related to the triple degeneracy in $\mathcal{M}$ and also we can find the boundary conditions which select the physical modes. It is important though to mention that the complex bulk fields here are just computational devices we use to simulate the spectral properties and they are not actually physical solutions.

The holographic diagram showing the three branes, the couplings, and where the exceptional point occurs along $z$ is shown in Figure \ref{fig:EPHoloModel}.

Note that in the UV boundary ($z \to 0$), the fields map to sources or VEVs of the dual operators, such as resonator amplitudes, and in the IR boundary ($z=z_{IR}$) one can impose normalizability or Dirichlet/Neumann condition, which corresponds to mode selection, such as single-mode operation.
The dictionary for our specific setup of three-coupled microresonators is shown in table \ref{tab:tableMap}.

\begin{table}[h!]
  \begin{center}
    \caption{Mapping to Optical System}
    \label{tab:tableMap}
    \begin{tabular}{l|l} 
      \textbf{Optical System} & \textbf{Holographic Dual} \\
      \hline
      Microresonatores (3 modes) & 3 Flavor branes/ gauge fields $A_M^{(i)}$\\
      Gain/Loss & Background dilaton $\varphi(z)$ \\
      Coupling/anti-PT symmetry & Off-diagonal components in $F_{MN}$ or $X$ \\
      Exceptional points & Critical points in bulk potential $X(z)$ or interaction term \\
      Single-mode lasing & Normalizable lowest-energy mode in bulk spectrum
    \end{tabular}
  \end{center}
\end{table}

\subsection{The holographic lasing spectra of various pumping patterns}
Now, based on our model of three coupled flavor-brane gauge modes $a_i(z)$, we can extend the work of \cite{Miri2019ExceptionalPI} and find analytical results for several symmetric cases which are gain-gain-gain, gain-loss-loss, and gain-gain-loss patterns, with partially pumped or evenly pumped structures, $\mathcal{PT}$-symmetric gain-loss arrangements (unbroken to broken), and then we can identify EP2/EP3.

Note that these results are based on the three coupled flavor-brane gauge modes $a_i(z)$ obeying a Sturm-Liouville problem in the soft-wall AdS background, with a $3 \times 3$ non-Hermitian coupling matrix $\mathcal{M}$ encoding gain/loss couplings.

If we assume the time dependence in the system is $e^{- i \omega t}$, then we get the $5d$ profiles as
\begin{gather}
A_\mu^{(i)} (x^\mu, z) = \epsilon_\mu a_i(z) e^{-i \omega t}.
\end{gather}

If the $4D$ dispersion is encoded in $\omega$ and the holographic transverse profile satisfies a standard soft-wall Sturm-Liouville equation, a separation of variables leads to the scalar ODE as
\begin{gather}
\partial_z \left ( \frac{e^{- \phi (z)} }{z} \partial_z \psi (z) \right ) + \frac{e^{- \phi(z)} }{z} \lambda \psi(z) = 0,
\end{gather}
where $\phi(z) = \kappa^2 z^2$ is the soft-wall dilaton and $\psi(z)$ is the radial profile. The algebraic part is
\begin{gather}
\mathcal{M} \ \mathbf{v} = \lambda \mathbf{v},
\end{gather}
where $\mathbf{v} = (v_1, v_2, v_3)^T$ is the eigenvector across the three branes in our setup and $\mathcal{M}$ is the $3 \times 3$ non-Hermitian coupling matrix that encodes gain/loss and inter-brane couplings. Thus, the full normal modes separate into radial profiles $\psi_n (z)$ with eigenvalues $\lambda_n$ given by the eigenvalues of $\mathcal{M}$. Physically, one can then map $\lambda_n$ to a complex frequency/growth rate of the $4D$ mode, where modes with $\text{Im} \lambda_n > 0$ correspond to amplifying (lasing) channels in the holographic mapping.

So, first we choose an appropriate $\mathcal{M}$ for a specific pumping pattern, and then we compute its eigenvalues $\lambda_n$ and eigenvectors $\mathbf{v}_n$. Next, for each $\lambda_n$, we solve the SL equation with boundary conditions to determine whether a normalizable radial mode exists that selects the physical discrete set, and then we can interpret $\text{Re} \lambda_n$ and $\text{Im} \lambda_n$ as the mode frequency and growth rate.

For a physically motivated ansatz in the simplest bottom-up approach, we have
\begin{gather}
\mathcal {M} = \begin{pmatrix}
i \gamma_1 & \kappa_{12} & \kappa_{13} \\
\kappa_{12} & i \gamma_2 & \kappa_{23}\\
 \kappa_{13} & \kappa_{23} & i \gamma_3 
\end{pmatrix},  \ \ \ \ \ \ \ \gamma_i \in \mathbb{R}, \ \  \kappa_{ij} \in \mathbb{R},
\end{gather}
where $i \gamma_i$ on the diagonal parts encodes gain $(+i \gamma)$ or loss $(-i \gamma)$ on brane $i$, and the off-diagonal $\kappa_{ij}$ which are real coupling strengths between branes $i$ and $j$.

\begin{figure}[ht!]   
\begin{center}
\includegraphics[width=0.464\textwidth]{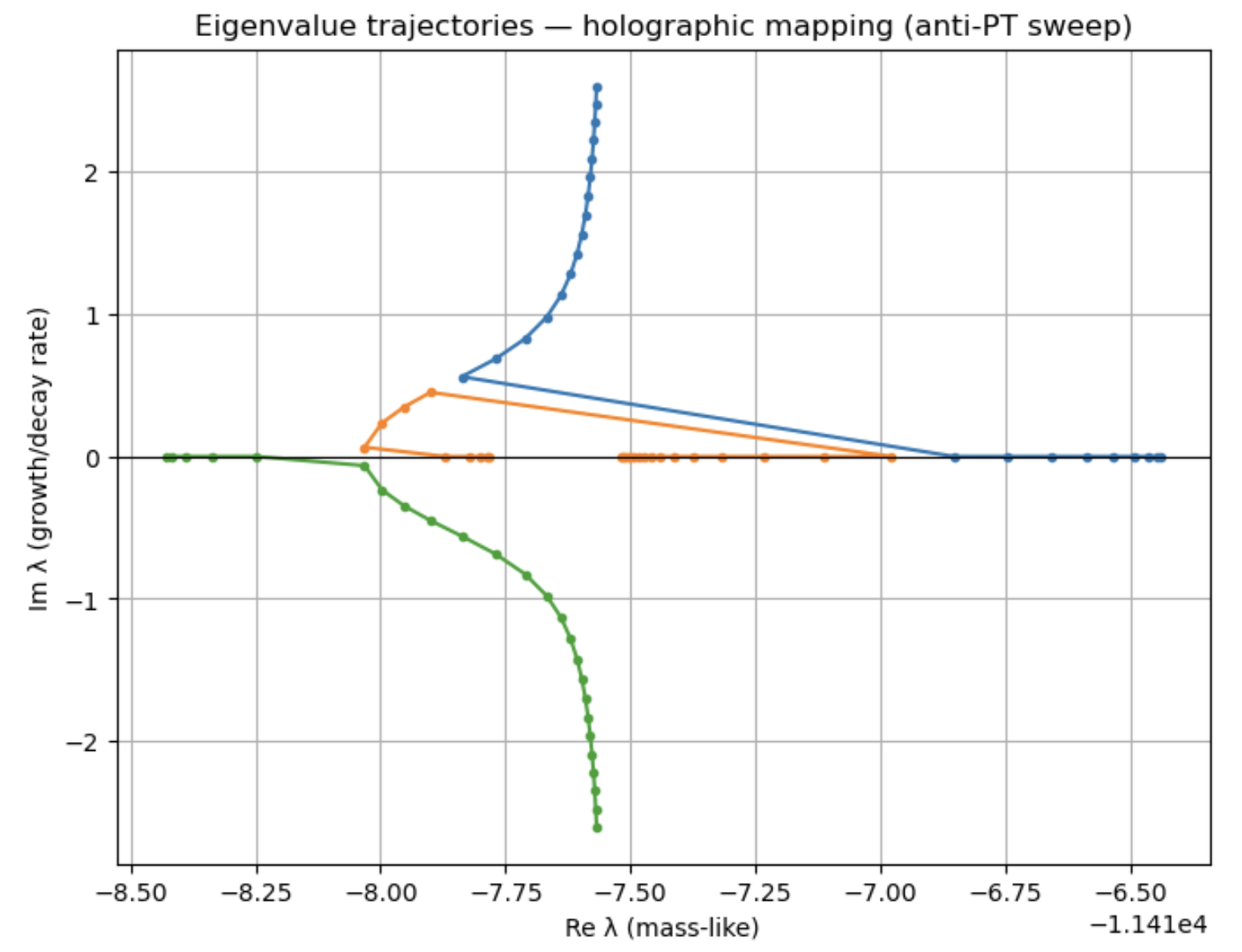} 
\includegraphics[width=0.496\textwidth]{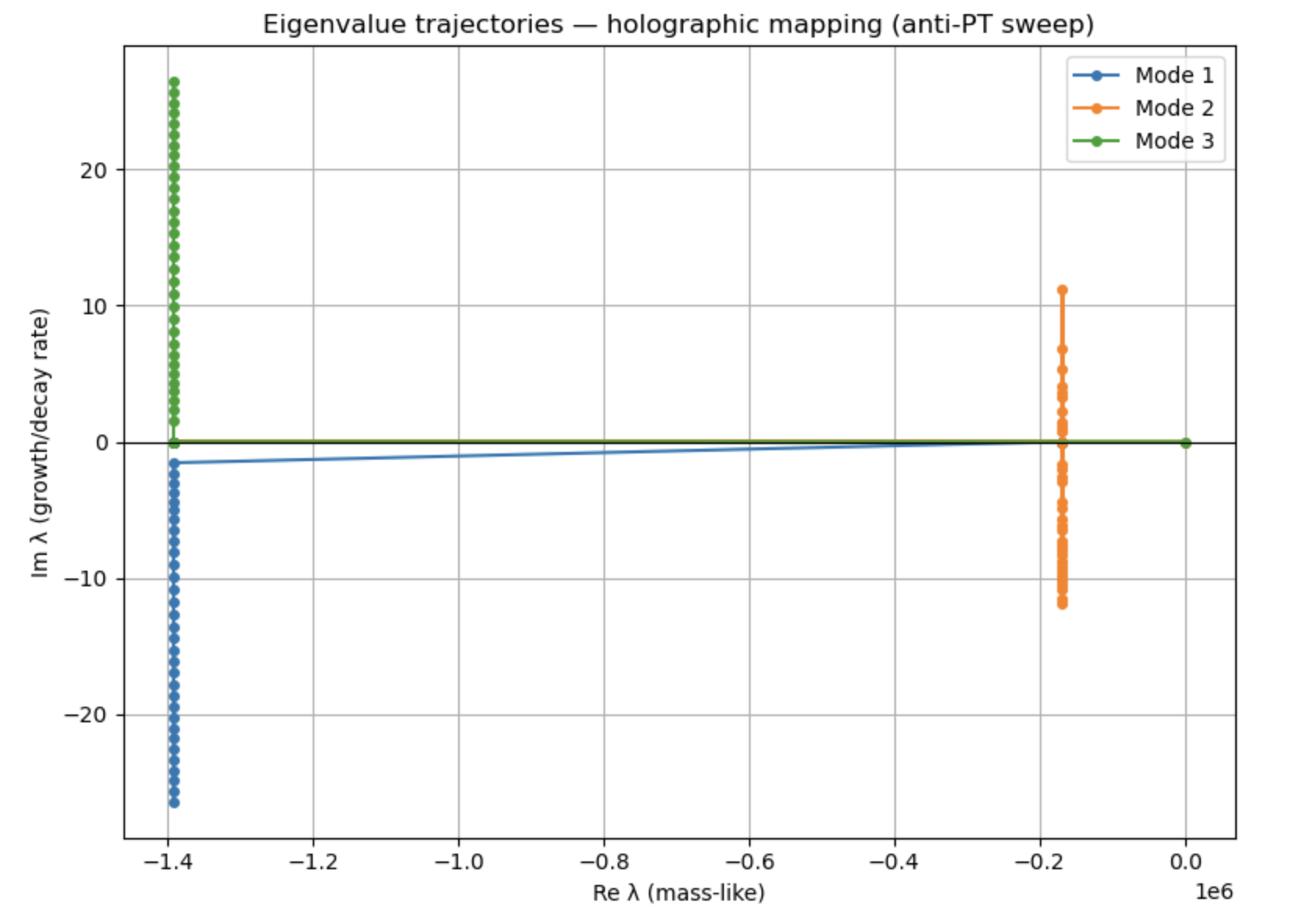} 
\caption{Two runs for finding eigenvalue trajectories in the complex plane as $\gamma$ is varied for the ternary coupled microcavity model (anti-$\mathcal{PT}$ sweep). In the left $N_z=20$ (grid points along holographic $z$) and in the right $N_z=200$.}
\label{fig:eigenvalues1}
\end{center}
\end{figure}

\begin{figure}[ht!]   
\begin{center}
\includegraphics[width=0.45\textwidth]{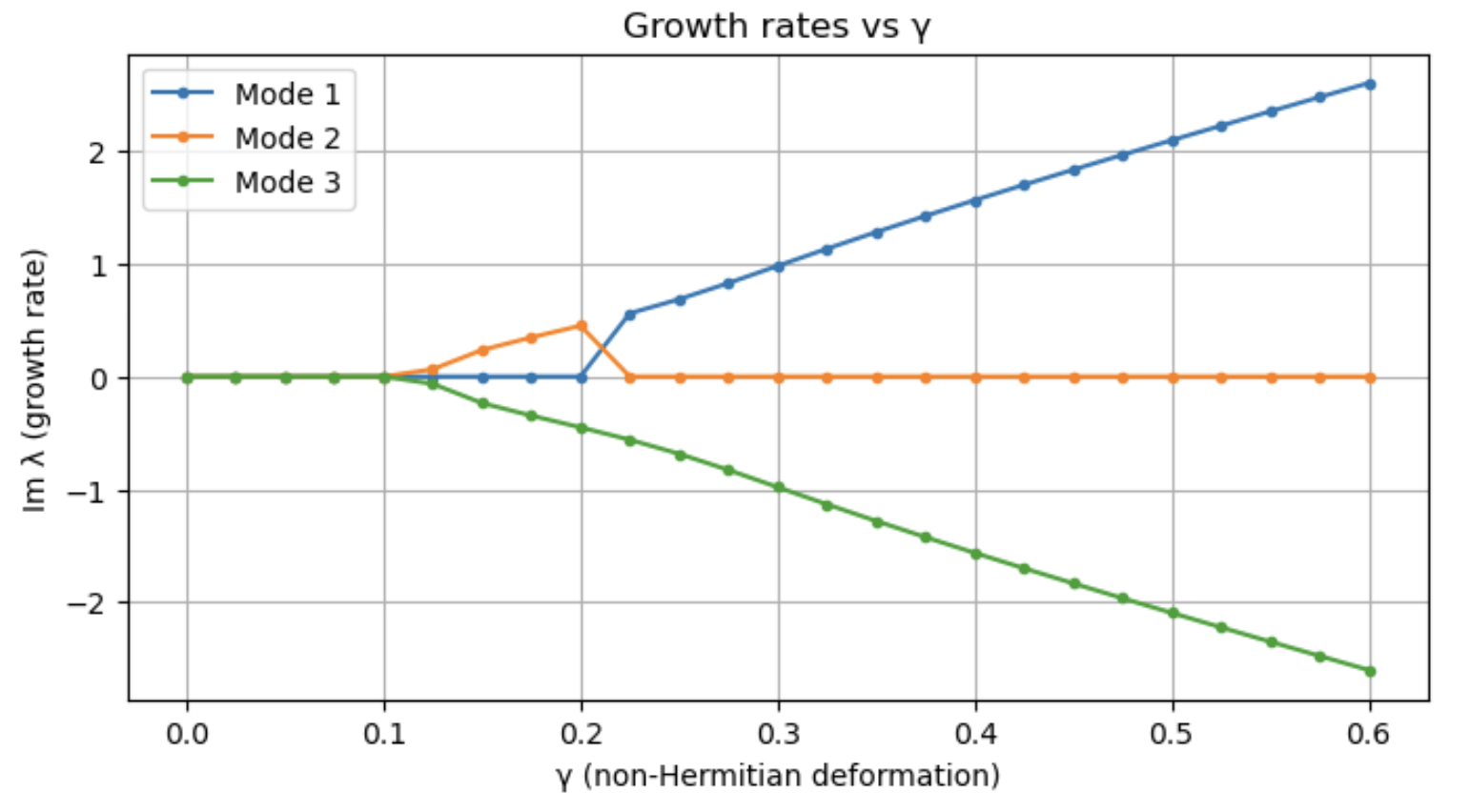} 
\includegraphics[width=0.53\textwidth]{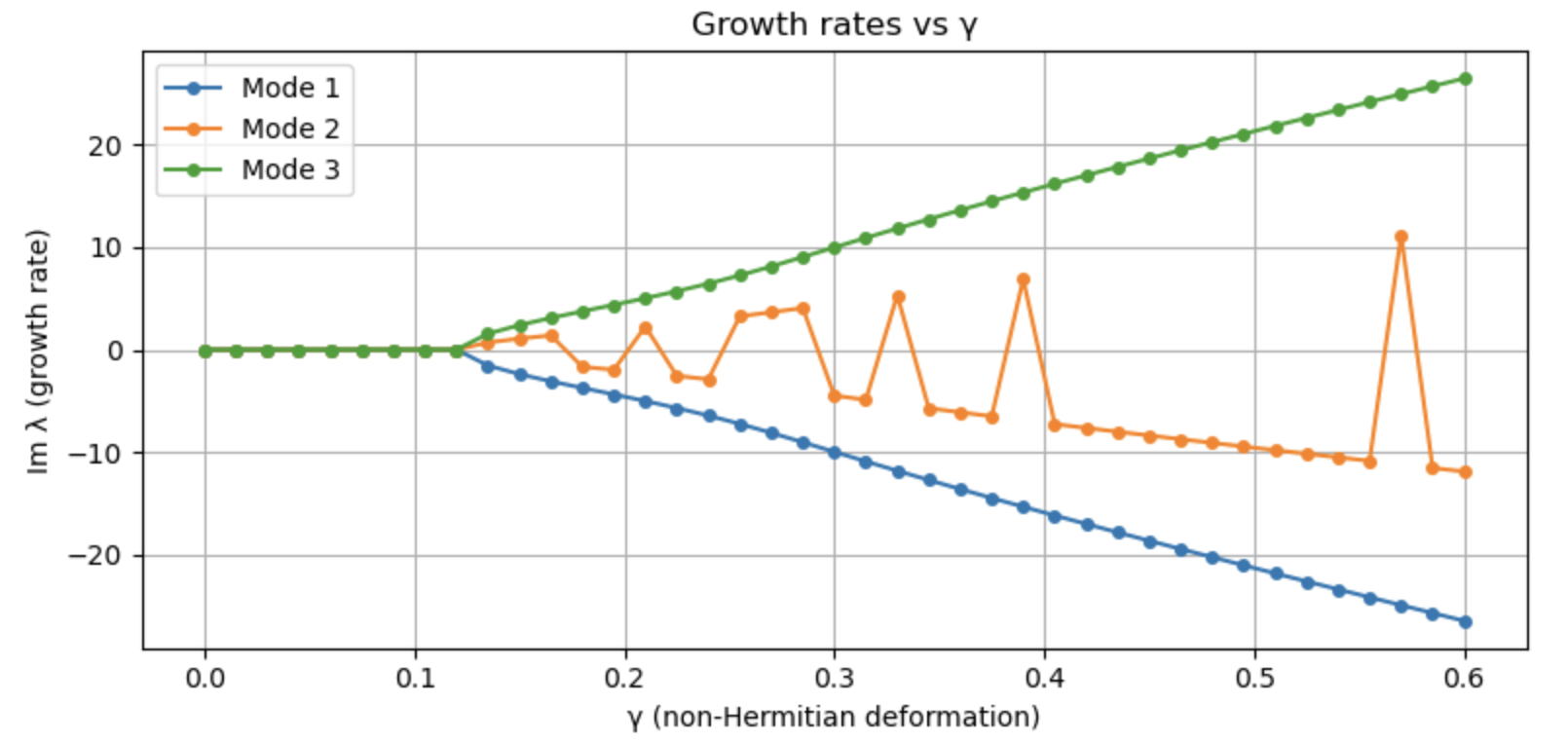} 
\caption{Two runs for finding imaginary parts versus $\gamma$ (growth rates of modes) for the ternary coupled microcavity model (anti-$\mathcal{PT}$ sweep). In the left $N_z=20$ (grid points along holographic z) and in the right $N_z=200$.}
\label{fig:eigenvalues2}
\end{center}
\end{figure}

\begin{figure}[ht!]   
\begin{center}
\includegraphics[width=0.457\textwidth]{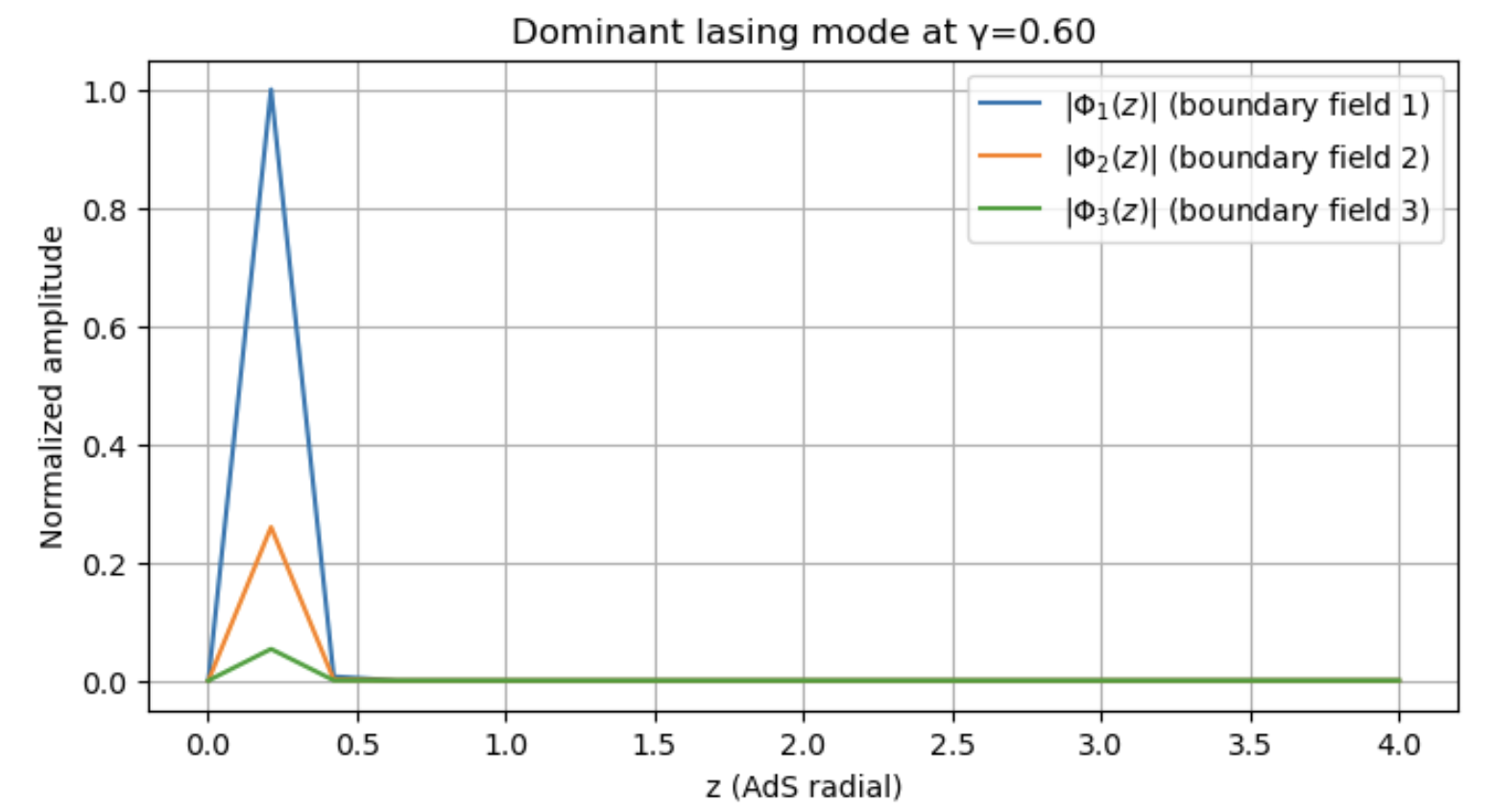}
\includegraphics[width=0.52\textwidth]{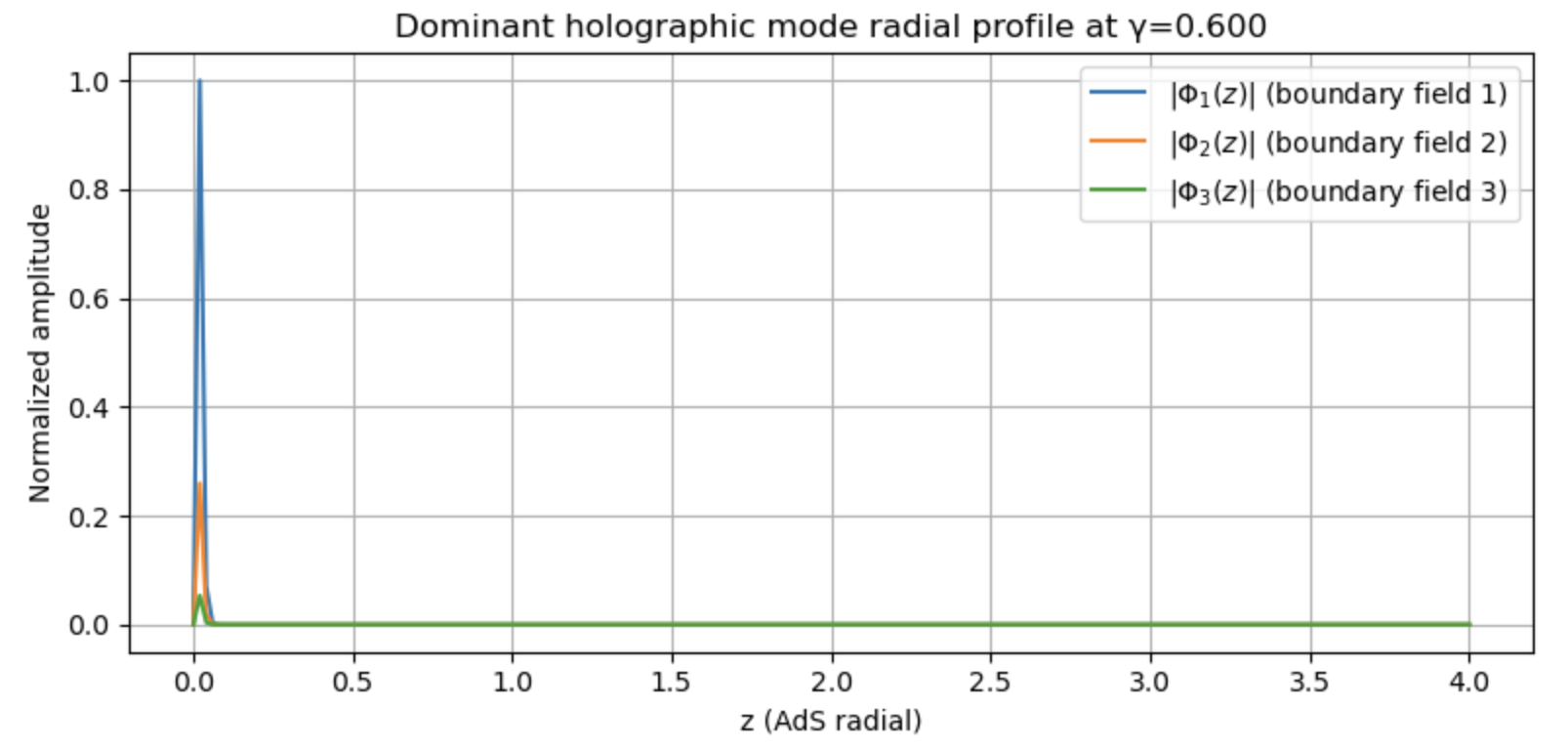}
\caption{Two runs for finding radial profiles of the dominant lasing eigenmode at the $\gamma$ where the growth is maximal for the ternary coupled microcavity model (anti-$\mathcal{PT}$ sweep). In the left $N_z=20$ (grid points along holographic z) and in the right $N_z=200$. One could see that similar to confining models, most of the modes are accumulated close to the wall, around $z=0$.}
\label{fig:eigenvalues3}
\end{center}
\end{figure}

For the case of uniform couplings $\kappa_{12}=\kappa_{23}=\kappa_{13}=\kappa$, and even pumping of gain-gain-gain, $\gamma_1=\gamma_2=\gamma_3= \gamma$, we get
\begin{gather}
\mathcal{M} = i \gamma \mathbb{I}_3 + \kappa (\mathbf{1}\mathbf{1}^T - \mathbb{I}_3), \ \ \ \ \ \mathbf{1}= (1,1,1)^T.
\end{gather}

In the above matrix, the off-diagonal $\kappa_{ij}$ are the real coupling strengths between branes $i$ and $j$, and $i \gamma_i$ on the diagonal parts encodes gain $(i \gamma)$ or loss $(-i\gamma)$ on brane $i$.

For even pumping (gain-gain-gain), we have $\gamma_1 = \gamma_2 = \gamma_3 = \gamma$. For the partial pumping/gain-loss-loss, we get $\gamma_1 = \gamma$, $\gamma_2 = -\gamma$, $\gamma_3 = -\gamma$, or $\gamma_1 = \gamma$, $\gamma_2 = \gamma$, $\gamma_3= 0$.  For the anti-$\mathcal{PT}$ ternary system, we get $\gamma_1= \gamma$, $\gamma_2 =0$ and $\gamma_3 = - \gamma$. For the $\mathcal{PT}$ pair plus a spectator case, we get a $\mathcal{PT}$ pair on modes 1 and 2, while the mode 3 is neutral or weakly coupled. So we have $\gamma_1 = \gamma$, $\gamma_2 = - \gamma$, $\gamma_3 = 0$.

The eigenvalue trajectories in the complex plane, the growth rates of the modes, and the radial profiles of the dominant lasing eigenmode are shown in figures \ref{fig:eigenvalues1}, \ref{fig:eigenvalues2}, and \ref{fig:eigenvalues3}.

\begin{figure}[ht!]   
\begin{center}
\includegraphics[width=0.52\textwidth]{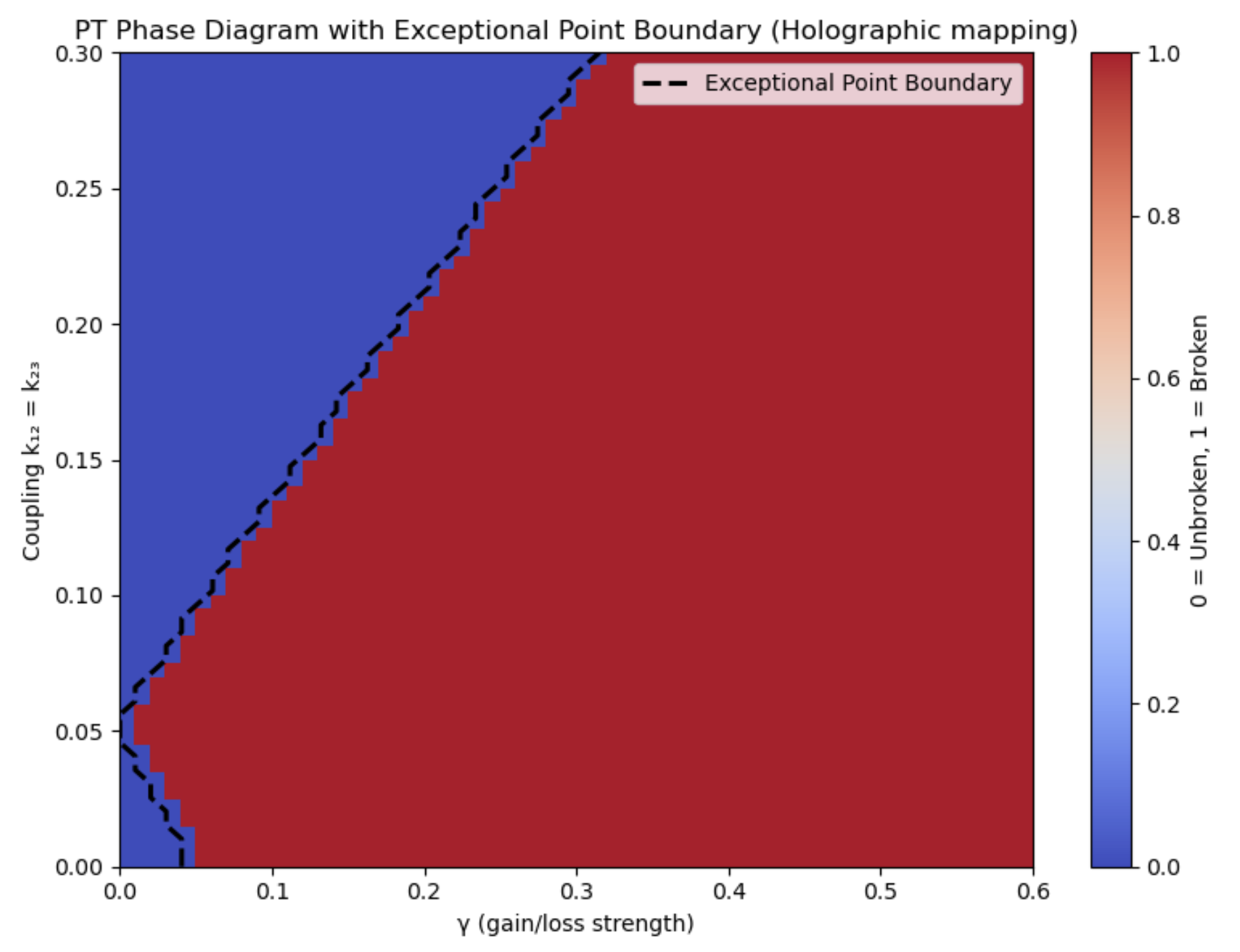}
\caption{Phase diagram distinguishing the unbroken $\mathcal{PT}$ phase, where all eigenvalues are real, to the broken $\mathcal{PT}$ phase, where at least one eigenvalue is complex.}
\label{fig:phasediagramKappa}
\end{center}
\end{figure}

In figure \ref{fig:phasediagramKappa}, a phase diagram heatmap is shown. In the blue region (0), there is an unbroken $\mathcal{PT}$ phase where all eigenvalues are real, corresponding to symmetric lasing. In the red region (1), the $\mathcal{PT}$ symmetry is broken, with complex-conjugate eigenvalue pairs indicating symmetry breaking. This illustrates where the exceptional points occur as transitions between the phases are consistent with the physics of ternary microrings of \cite{citekeyJahangiri}.

\subsection{Analytic eigenvalues for symmetric coupling}

For even pumping corresponding to gain-gain-gain, where $\gamma_1 = \gamma_2 = \gamma_3 = \gamma$, and for the non-degenerate symmetric mode, we get $(1,1,1): \lambda_+ = i \gamma + 2 \kappa$, while for the doubly degenerate orthogonal subspace with degeneracy 2, we have $\lambda_- = i \gamma - \kappa$.

So, for this case, all eigenvalues share the same imaginary part, $\gamma$, while increasing $\kappa$ lifts the real parts and splits the symmetric and antisymmetric sets. The lasing growth rate is determined by $\text{Im} \lambda = i \gamma$, where all modes are amplified equally. Thus, the mode selection must arise from the radial SL quantization.

For the fully symmetric anti-$\mathcal{PT}$ diagonal pattern $(i\gamma, 0 , - i \gamma)$ and uniform $\kappa$, the characteristic cubic must be solved. In this case, the eigenvalues are generally distinct complex numbers, and there is no triple root for nonzero $\kappa$ under fully symmetric conditions. Therefore, one finds an approximately symmetric mode and two complex modes whose imaginary parts determine which mode crosses the lasing threshold first.

From this model, we can also deduce that an EP3 is not generically reachable with strictly uniform $\kappa_{ij} = \kappa$ and the diagonal pattern $(i\gamma, 0, -i\gamma)$. Since an EP3 requires satisfying multiple algebraic constraints, it typically demands adjusting at least two independent parameters. Therefore, to realize an EP3 in a ternary non-Hermitian system, one must carefully tune asymmetries in the couplings and/or the diagonal terms. This result is consistent with the literature, which confirms that achieving EP3 requires more precise engineering than EP2.

\section{Soft-wall model of EPs and the Ferrell-Glover-Tinkham sum rule}\label{sec:Hol2}

One could write the above action in a slightly different way to simulate EPs. To imitate the photonic EPs by our holographical toy model more accurately, one should note that the boundary encodes the non-Hermitian features through “complex sources” for operators, while the bulk parameters capture the emergence of EPs via coalescence in the spectrum of quasinormal modes (QNMs) or via phase transitions in the free energy, similar to confinement/deconfinement as discussed above.

The gain and loss can be modeled in this setup by the imbalances in the inflow/outflow of a conserved charge (such as a $U(1)$ global symmetry on the boundary), which is dual to a bulk gauge field.
The $\mathcal{PT}$ symmetry can be imposed by ensuring that the sources and their conjugates are complex mirrors (e.g., $s$ and $s^*$), which makes the boundary action $\mathcal{PT}$-invariant. Here, we introduce a minimal model to capture the core behavior of EPs based on an AdS/QCD-inspired analogy. In this model, for a $4d$ boundary theory, we have a probe backreacting scalar in $\text{AdS}_5$, tunable to QCD-like scales via the bulk potential. The theory has a $U(1)$ global symmetry and a complex scalar operator $\mathcal{O}$ of dimension $\Delta = 3$, which, for instance, can mimic a mesonic operator in AdS/QCD.

The action here is 
\begin{gather}\label{eq:actionEP}
S= \int d^5 x \sqrt{-g}\left \lbrack R +12 -\frac{1}{4}F_{MN} F^{MN} - | D_M \phi|^2-V(| \phi|) \right \rbrack,
\end{gather}
where $D_M \phi =\nabla_M \phi - i A_M \phi$, $F_{MN} = \partial_M A_N - \partial_N A_M$, and the scalar potential is 
\begin{gather}
V(| \phi |) = m^2  | \phi |^2 + \frac{\lambda}{2}  | \phi |^4, \ \ \ \ \ \ \ m^2 =- \Delta ( \Delta-4)= -3.
\end{gather}

This action is the standard dilaton-like potential in holographic QCD models, but here $\phi$ introduces non-Hermiticity in the system. 

The equations of motion can be obtained by varying the action as
\begin{equation} \label{eq1}
\begin{split}
& \text{Scalar: } \ \nabla^M D_M \phi = \frac{\partial V}{\partial \phi^*}, \nonumber\\
& \text{Gauge: } \ \nabla^M F_{MN} + iq (\phi^* D_N \phi - \phi (D_N \phi)^*)=0, \nonumber\\
& \text{Einstein: } \  R_{MN} - \frac{1}{2} g_{MN} R + 6 g_{MN} = T_{MN},
\end{split}
\end{equation}
where the stress tensor $T_{MN}$ comes from the matter fields.

Then, at the AdS boundary ($z$ $\to$ 0, where $z$ is the radial coordinate, with the UV at the boundary and the IR deep in the bulk), we can impose Dirichlet boundary conditions for sources as $\phi(z, x^\mu) \sim s(x^\mu) z^\Delta + \langle \mathcal{O} \rangle z^{4-\Delta} + ... $, and similarly for $\phi^* \sim s^*(x^\mu) z^\Delta + \langle \mathcal{O}^\dagger \rangle z^{4-\Delta}$.

In order to have $\mathcal{PT}$ symmetry, we should take $s(x^\mu)$ real and keep it even under parity $(P: x^1 \to -x^1)$ and time-reversal $(T : i \to -i)$, making the boundary action $\int s\mathcal{O} + s^* \mathcal{O}^\dagger$ invariant. 

Non-Hermiticity enters via the ``imbalance parameter", $x= \frac{s-s^*}{s+s^*}$, which is real for $\mathcal{PT}$ symmetry. The gain/loss can be modeled as $|s| > 0$, which drives the charge inflow/outflow. The positive $\text{Im}(s)$ acts as gain (particle creation), while negative $\text{Im}(s)$ acts as loss or absorption.

For the three-resonator system, one could make $s(x^1)$ piecewise constant across the three spatial ``sites", (for instance $s(x^1) = s_0$ for $|x^1| <a/3$, and varying the couplings via gradients), which mimic a lattice. However, for simplicity, one could take $s$ to be homogeneous and constant.

For capturing the photonic EPs, we can then study different scales in the system. First, for $|s| < M \approx \sqrt{3/\lambda}$, which is the critical mass scale from the potential, we have the $\mathcal{PT}$-unbroken phase. In this case, real bulk solutions exist, where $\phi$ and the metric are real and the null energy condition is satisfied. Also, the boundary spectrum (QNMs of $\phi$) has real frequencies $\omega_n$, which are dual to stable photonic modes. The Dyson map which corresponds to a complexified bulk $U(1)$ gauge shift $A_M \to A_M + i \alpha \partial_M \log \eta$ maps it to a unitary theory.

Second, at the EP transition $(|s| = M)$, the eigenvalues coalesce and the QNMs branch into complex conjugates. In the bulk, the scalar decouples as there is no backreaction. The metric remains real, but the Hessian of the free energy which has a zero eigenvalue, corresponds to marginal stability. This is the holographic EP, where boundary eigenvectors (operator modes) merge, analogous to the three-resonator coalescence at critical coupling.

Third, in the $\mathcal{PT}$-broken phase ($|s|> M$), the complex bulk metric can be written as 
\begin{gather}
ds^2 = -f(z) dt^2 + dz^2/f(z) + dx_i^2, 
\end{gather}
where $f$ is complex, and the NEC is violated. Additionally, the boundary free energy $\mathcal{F} \sim |s|^4 \log |s|$ becomes complex with conjugate branches, where $\text{Im} (\mathcal{F})$ signals dissipation. The QNMs in this case have $\text{Im} (\omega) \ne 0$, with unstable modes growing or decaying exponentially due to the amplified or suppressed photonic signals post-EP.

At finite temperature (where a black brane is added: $f(z) = 1 - (z/z_h)^4$), the EP persists but shifts. In this case, the chemical potential $\mu \sim A_t(z=0)$ tunes the EP, similar to the frequency in photonics. For AdS/QCD flavor, where the chemical potential is coupled to D7-branes to model chiral symmetry, the minimal scalar suffices.
 
Note that in this case, the $4D$ strongly coupled boundary CFT is local, while the $5D$ bulk is non-local, as the gravitational degrees of freedom (metric $g_{MN}$, scalar $\phi$, and gauge field $A_M$) are described by a gauge theory where diffeomorphism invariance introduces non-local effects. The non-Hermiticity (gain/loss) is dual to complex bulk-boundary conditions, and the EP could correspond to a critical point in the bulk where QNMs coalesce, often tied to non-local features such as horizon instabilities or complex metric solutions that violate the NEC in the $\mathcal{PT}$-broken phase. Thus, the non-locality which arises from gravitational gauge redundancy and the radial smearing of interactions with the EP, would be an analogical manifestation of a non-local critical point in the QNM spectrum.

Then, to numerically find the EP, one needs to discretize the radial Sturm-Liouville  operator and build a generalized eigen-problem. The radial equation can be written as
\begin{gather}
\partial_z \left( p(z) \partial_z \psi \right)+ p(z) \lambda \psi=0,
\end{gather}
or in the operator form, $L_0 \psi + \lambda L_1 \psi =0$, with $L_1 = p(z)$ multiplication. Here, $p(\lambda) = \text{det} ( \mathcal{M} - \lambda I)$ is the characteristic polynomial.

Using our toy model, we then can build a “holographic-like" block operator $H$, which is the effective non-Hermitian Hamiltonian for the ternary coupled microring system, and is inspired by bottom-up AdS/QCD holography, and can be written as 
\begin{gather}
\begin{bmatrix}
L_0 + i \gamma L_1 & \kappa L_1 & \kappa_{13} L_1 \\
\kappa L_1 & L_0 & \kappa L_1 \\
\kappa_{13} L_1 & \kappa L_1 & L_0 - i \gamma L_1,
\end{bmatrix},
\end{gather}
where $L_0$ and $L_1$ are radial differential operators that come from discretizing the $\text{AdS}_5$ wave equation with flavor brane embedding functions. So $L_0$ and $L_1$ carry the physics in the engineered bulk geometries. Each block corresponds to one ring resonator embedded as a boundary degree of freedom, dual to a bulk flavor brane. Off-diagonal terms $(\kappa, \kappa_{13})$ encode hopping/coupling between rings, which by the holographical analogy encode the mixing between bulk fields on different branes. The diagonal terms include gain and loss $(\pm i \gamma L_1)$, which holographically represent non-Hermitian boundary conditions, meaning ingoing/outgoing flux at the AdS boundary.

Note that in the language of optics, $H$ is the effective coupled-mode Hamiltonian governing the ternary microring system, while in the holographic language, $H$ is the block operator which can encode the bulk-boundary dynamics of three coupled flavor branes. The real parts of its eigenvalues give the resonant frequencies, while the imaginary parts give the gain/loss rates. Exceptional points appear where two or more eigenvalues/eigenvectors coalesce, which in the holographic picture means the bulk modes become degenerate under the non-Hermitian boundary condition. Thus, $H$ is the “bridge” between the holographic bulk toy model and the non-Hermitian optical experiments, via $L_0$ and $L_1$, as it contains the AdS radial structure, and via $\kappa$, the microring coupling topology.

Then, in the sweeps and plots, one could define ``minsep", which is defined as the minimum eigenvalue separation of the holographic block operator $H$ as
\begin{gather}
\text{minsep} ( \gamma, \kappa) = \text{min}_{i \ne j} | \lambda_i - \lambda_j |,
\end{gather}
where $\{ \lambda_i \}$ are the complex eigenvalues of $H$. When $\text{minsep} \to 0$, an exceptional point occurs. When it is exactly zero, albeit up to numerical precision, we get higher-order EP or degeneracy.

This model is solvable numerically (e.g., via the shooting method for EoMs) or perturbatively near the EP. It also predicts universal features, like the Ferrell-Glover-Tinkham (FGT) sum rule, holding across phases, which is testable in photonic analogs via conductivity spectra.

\subsection{Ferrell-Glover-Tinkham sum rule and EP in holography}

Note that the FGT sum rule is a fundamental principle in condensed matter physics and optics, particularly in the study of superconductors and other systems with collective excitations. It relates the frequency-integrated optical conductivity to the superfluid density, or to the missing spectral weight in a material transitioning to a condensed phase, such as a superconductor.

In a superconductor, when the system transitions to the superconducting state below its critical temperature, the optical conductivity $\sigma(\omega) = \sigma_1(\omega) + i \sigma_2(\omega)$ which is a complex number, with real part $\sigma_1$ describing absorption and imaginary part $\sigma_2$ related to the phase response, exhibits a gap. The normal-state Drude peak which is centered at $\omega = 0$, would collapse, and the spectral weight is transferred to a delta function at $\omega = 0$, representing the infinite conductivity of the superfluid.

The FGT sum rule states that the integrated real part of the optical conductivity over all frequencies is conserved, so we get
$$\int_0^\infty \sigma_1(\omega) d\omega = \frac{\pi}{2} \frac{n e^2}{m},$$
where $n$ is the electron density, $e$ is the electron charge, and $m$ is the electron mass. In the superconducting state, the missing spectral weight from the gapped region would be
$$\int_0^\infty \left[ \sigma_1^{\text{normal}}(\omega) - \sigma_1^{\text{super}}(\omega) \right] d\omega = \frac{\pi}{2} \frac{n_s e^2}{m},$$
where $n_s$ is the superfluid density. This ``missing" weight is actually transferred to the delta function at $\omega = 0$, reflecting the perfect conductivity of the superfluid.

In our photonic exceptional point system with a non-Hermitian Hamiltonian, the FGT sum rule can be applied analogously as in the holographic dual, in AdS/CFT or in AdS/QCD toy models. The boundary optical conductivity, derived from the bulk gauge field $A_M$, can mimic the dissipative dynamics of the strongly coupled QFT which are dual to the photonic system. The sum rule holds across the $\mathcal{PT}$-symmetry-breaking transition, ensuring that the total spectral weight of $\sigma_1(\omega)$ remains conserved, even as modes coalesce and eigenvalues become complex.

In the $\mathcal{PT}$-unbroken phase ($|s| < M$), the conductivity has a real spectrum, akin to a normal metal. At the EP ($|s| = M$), the conductivity shows a critical redistribution of spectral weight, with modes merging. In the $\mathcal{PT}$-broken phase ($|s| > M$), complex eigenvalues lead to amplified or damped modes, and the sum rule ensures that the integrated weight remains fixed, with contributions shifting to low frequencies, which is mimicking superfluid-like behavior due to gain/loss. In the analogical holographic toy model, this is computed via the retarded Green's function of the U(1) current, $G_{JJ}(\omega) \sim \sigma(\omega)$, from the bulk gauge field's boundary response.

\begin{figure}[ht!]   
\begin{center}
\includegraphics[width=0.5\textwidth]{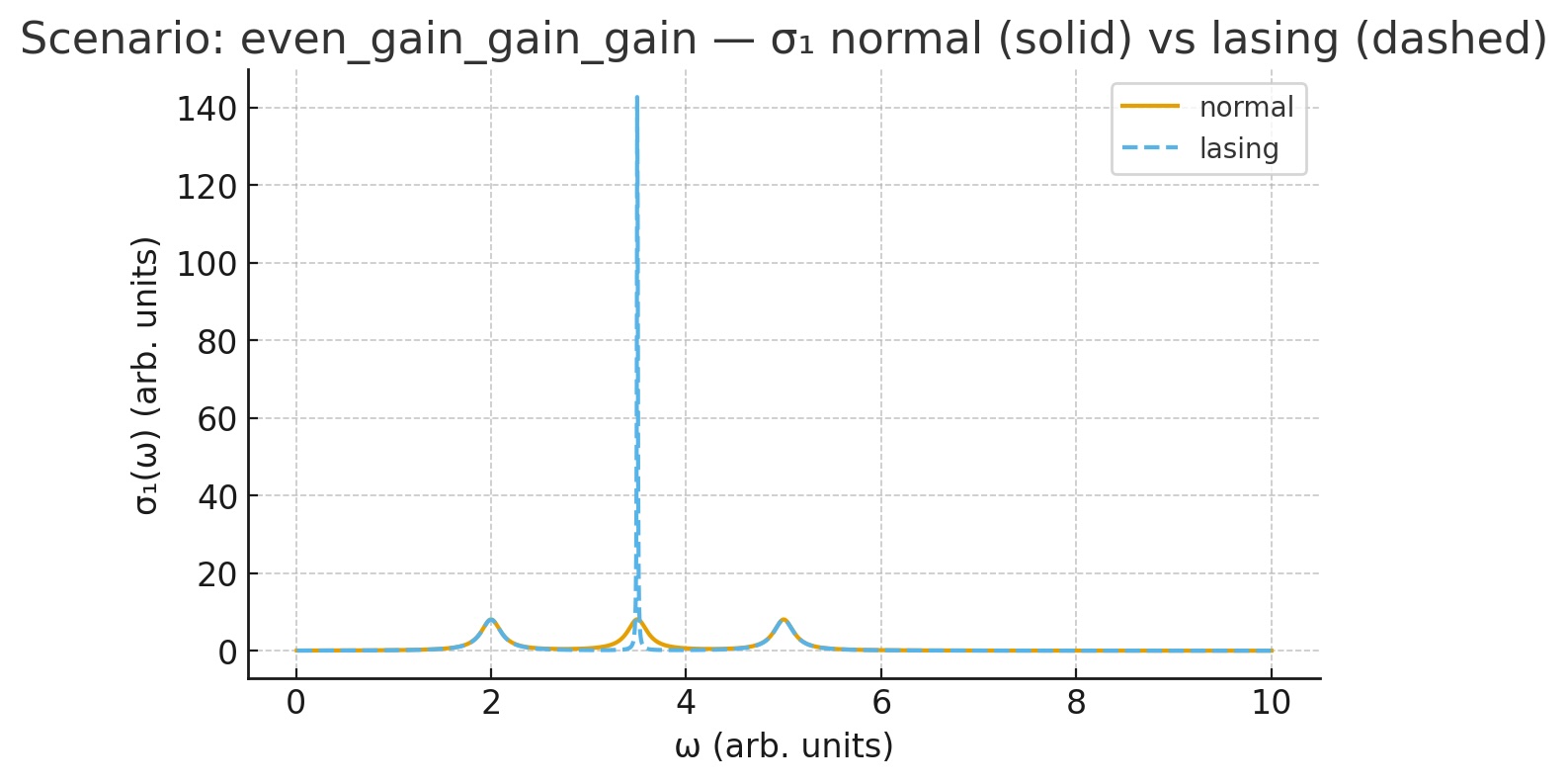} 
\includegraphics[width=0.48\textwidth]{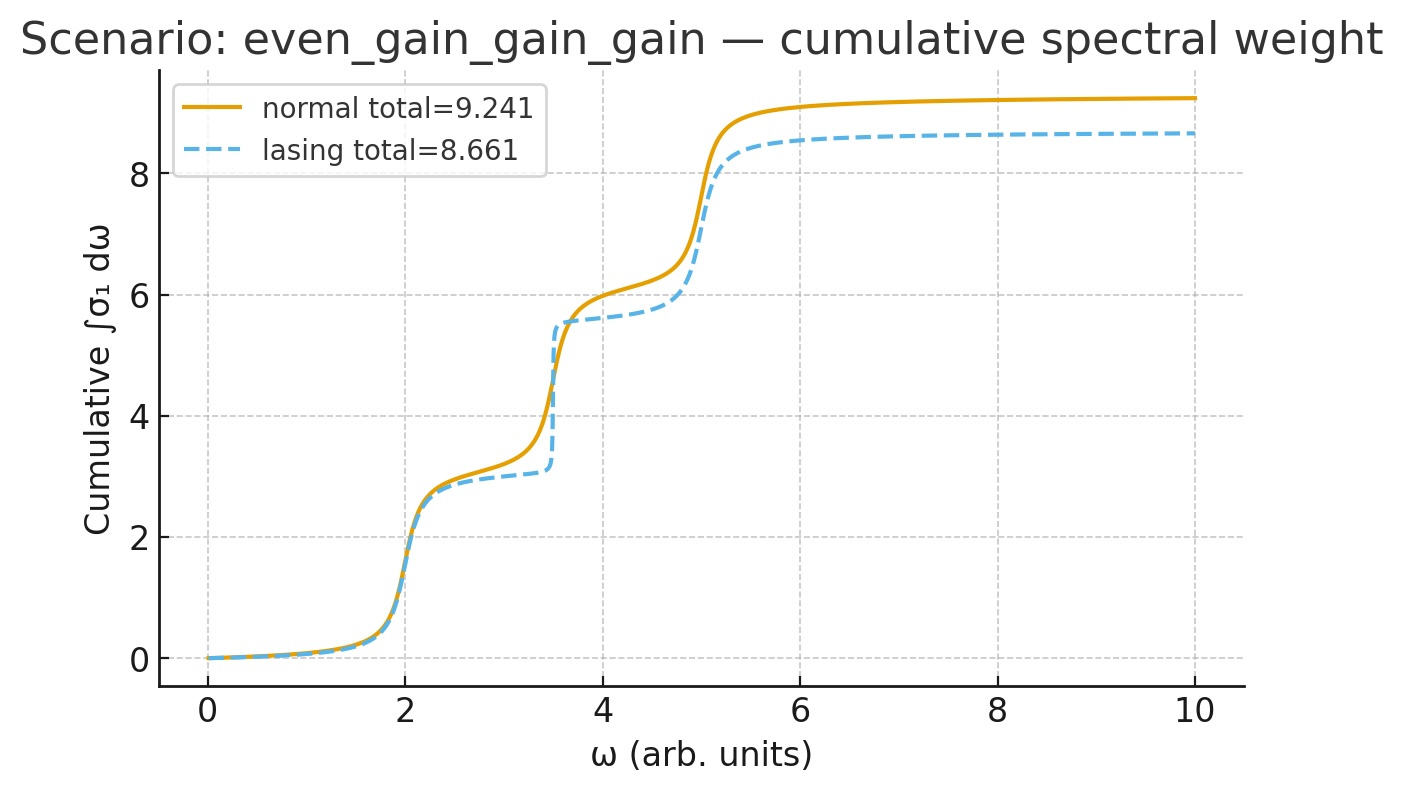} \\
\includegraphics[width=0.5\textwidth]{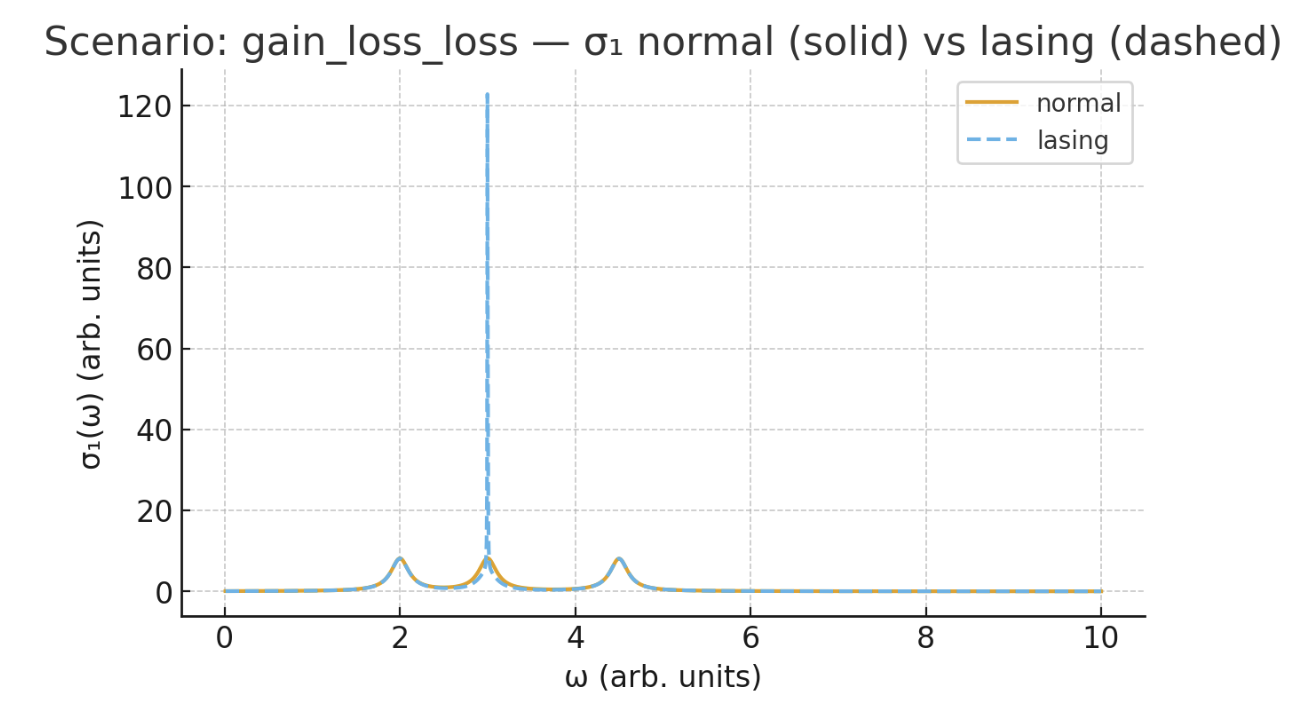} 
\includegraphics[width=0.48\textwidth]{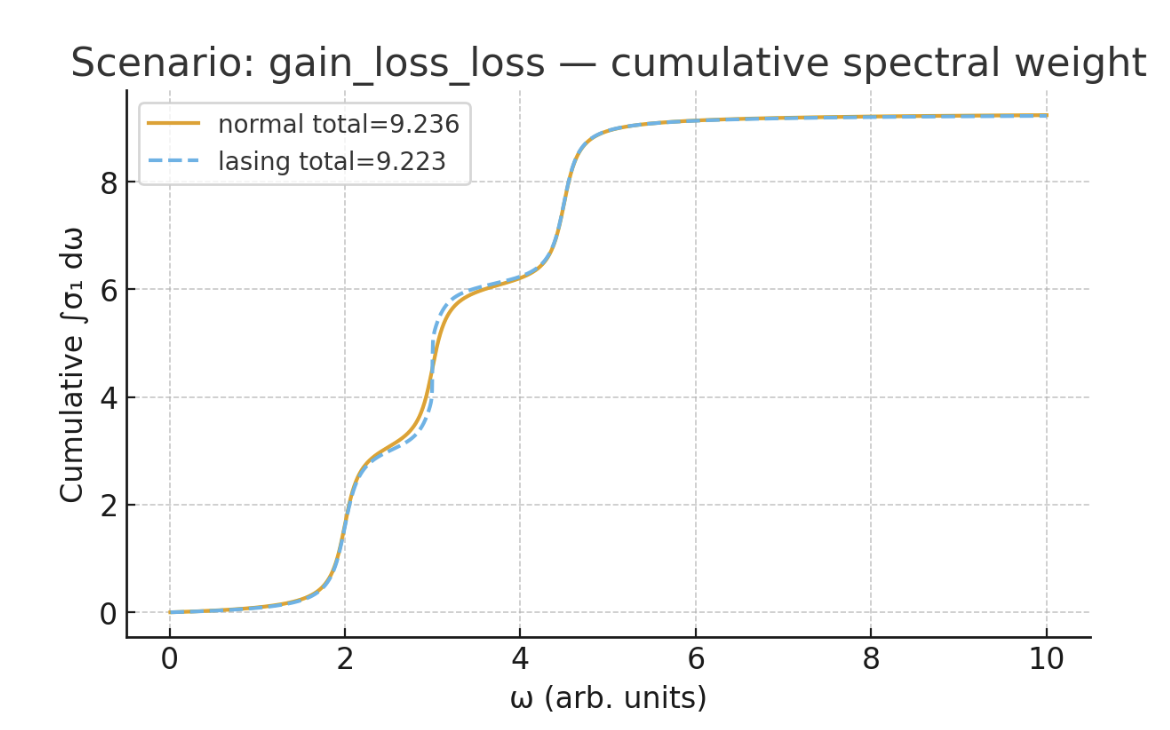} \\
\includegraphics[width=0.5\textwidth]{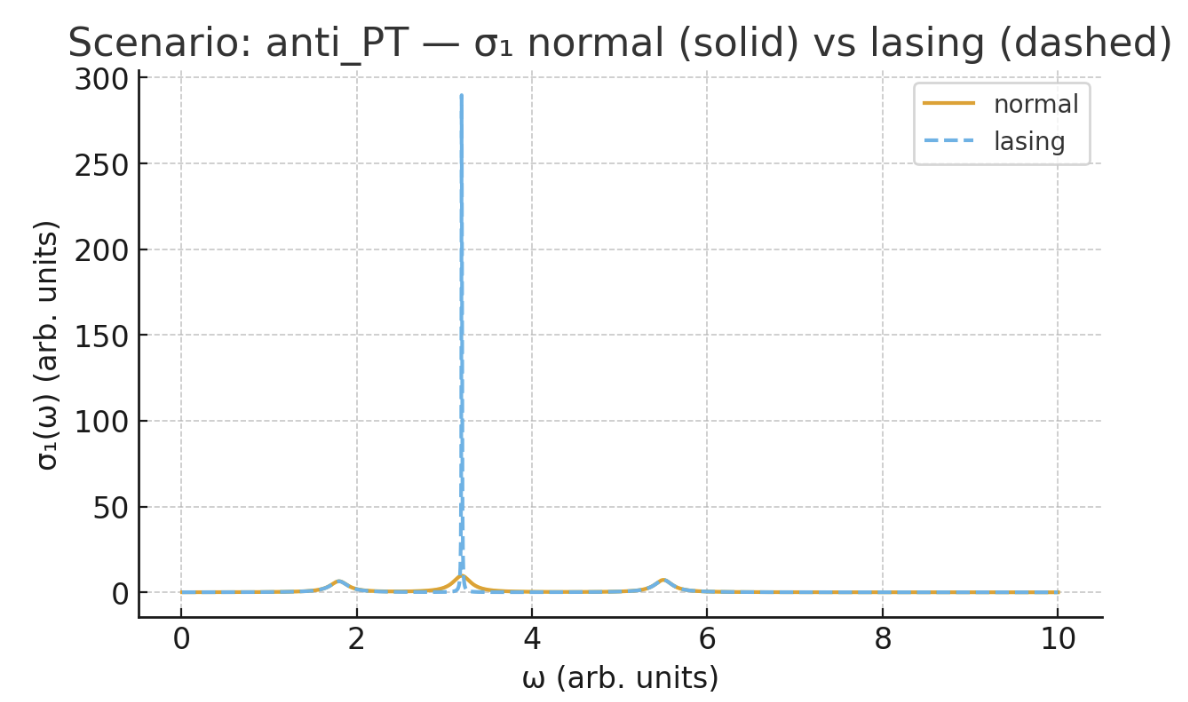} 
\includegraphics[width=0.48\textwidth]{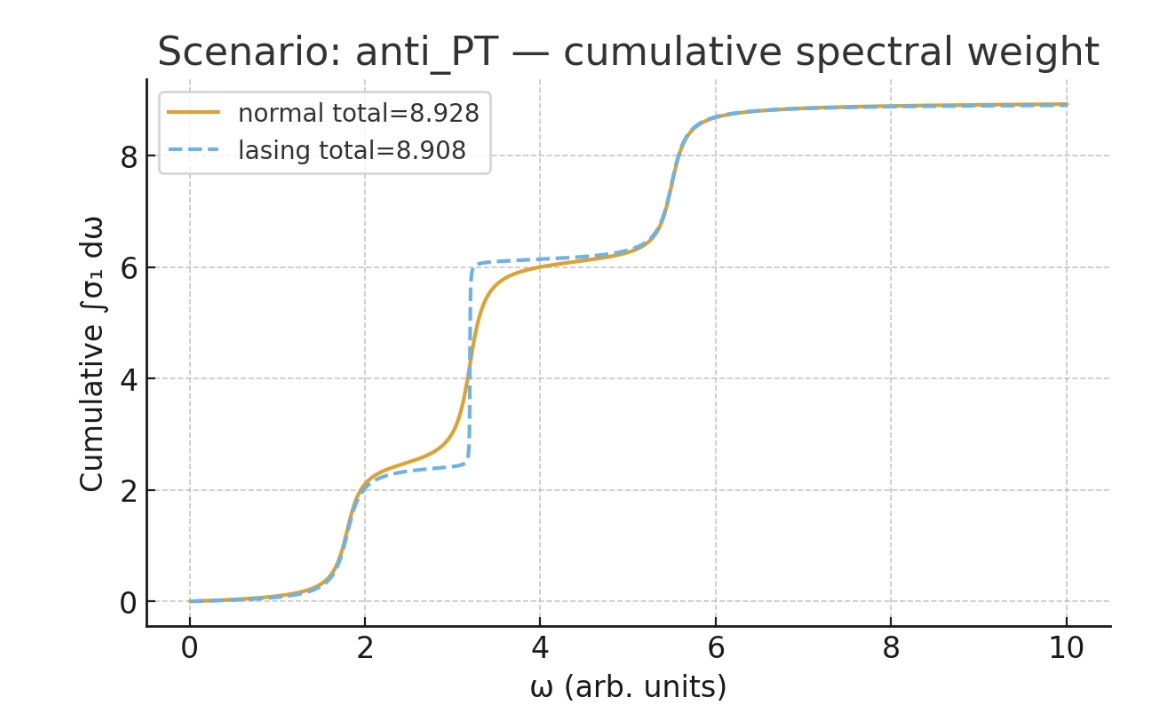} 
\caption{Ferrell-Glover-Tinkham sum rule for coupled ternary microcavities. This is a numerical demonstration of an FGT-style spectral-weight check for three toy pumping scenarios which are even gain-gain-gain, gain-loss-loss, and anti-$\mathcal{PT}$.}
\label{fig:Eigenfreq}
\end{center}
\end{figure}

When one applies the FGT sum rule, the resonance positions are then set by the real parts of the eigenvalues of the analogical holographic block operator $H$.

\begin{figure}[ht!]   
\begin{center}
\includegraphics[width=0.6\textwidth]{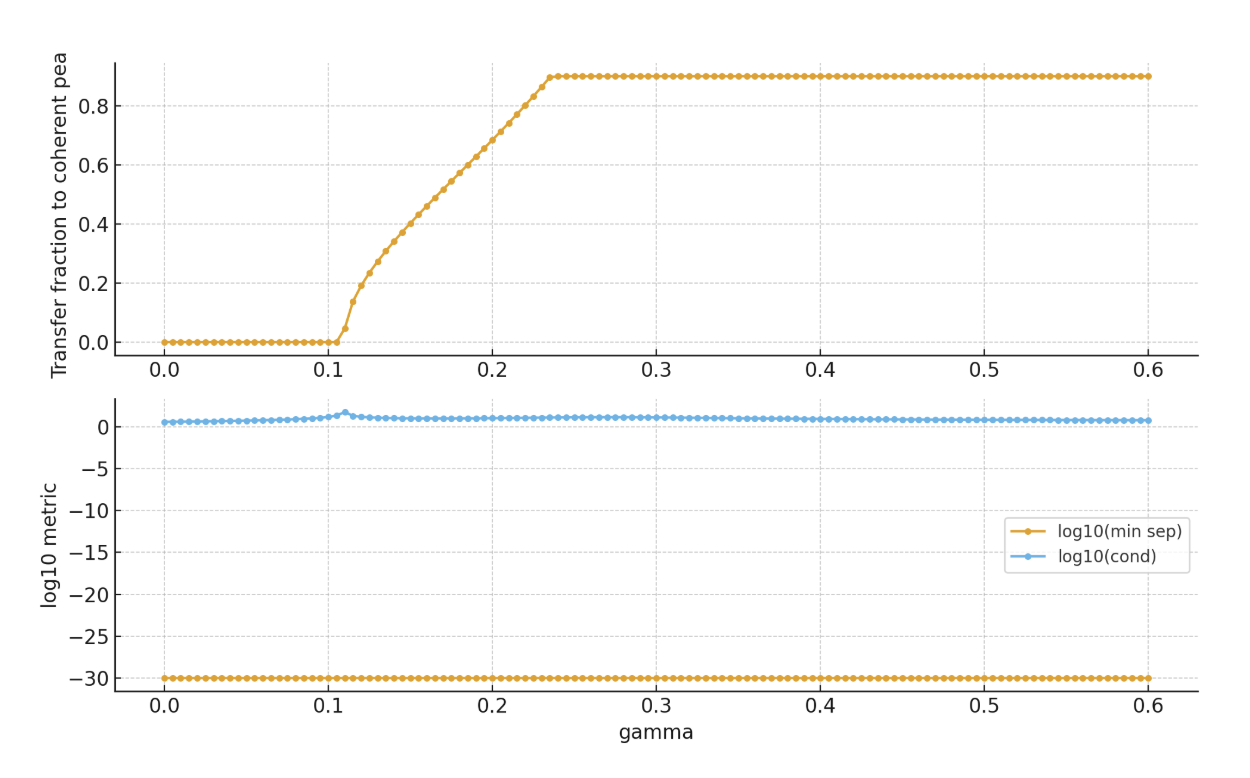}  
\caption{ A $1d$ parameter sweep in the gain/loss (pumping) parameter $\gamma$, with coupling fixed. Here $\kappa_{12} = \kappa_{23}=0.15$, $\kappa_{13}=0.06$ and $\gamma: 0 \to 0.6$, $N_z=20$.}
\label{fig:seepgamma1d}
\end{center}
\end{figure}

Now, in the above holographic model, one can run a parameter sweep by gradually increasing pumping and then produce a phase diagram showing how spectral weight transfers into the narrow, coherent peak and where EPs appear. This is shown in figures \ref{fig:seepgamma1d} and \ref{fig:spectralasing}. These results come from the holographic block operator $H$, where the real parts of $H$'s eigenvalues locate the resonances, while the imaginary parts are interpreted as gain/loss which drive the oscillator strengths, and the eigenvectors' IR localization is used to modulate oscillator strength. Here, we also defined a transfer fraction, $T(\gamma)$, which is a heuristic fraction of the total spectral area moved into a narrow coherent peak at the dominant mode frequency. It is simply defined by the relation $T = \text{min} (1.8 \times \text{max}  (\text{Im} \lambda_n, 0), 0.9)$.

In the upper part of figure \ref{fig:seepgamma1d}, one can see that at small $\gamma: 0 \to 0.10$, the transfer fraction $T$ is effectively zero, as there is no coherent peak. Around $\gamma \approx 0.11$, a threshold is crossed, and $T(\gamma)$ rises quickly as the system starts to move spectral weight into coherent/narrow peak. Then, $T(\gamma)$ grows and saturates around $\approx 0.85-0.9$ for larger $\gamma$, which means that when the gain becomes large enough, a dominant mode collects most of the spectral weight and the system becomes single-mode lasing.

In the bottom panel of figure \ref{fig:seepgamma1d}, the minimal eigenvalue separation shows a clear minimum at $\gamma = 0$, which is zero, and this corresponds to the analytic EP2 at $\gamma = 0$. Away from zero, the minsep is finite, and there is a tiny bump in the condition-number curve near $\gamma \sim 0.12$, where the transfer fraction first grows. This indicates mode hybridization/competition in the region where lasing starts to pick a single mode.

\begin{figure}[ht!]   
\begin{center}
\includegraphics[width=0.4\textwidth]{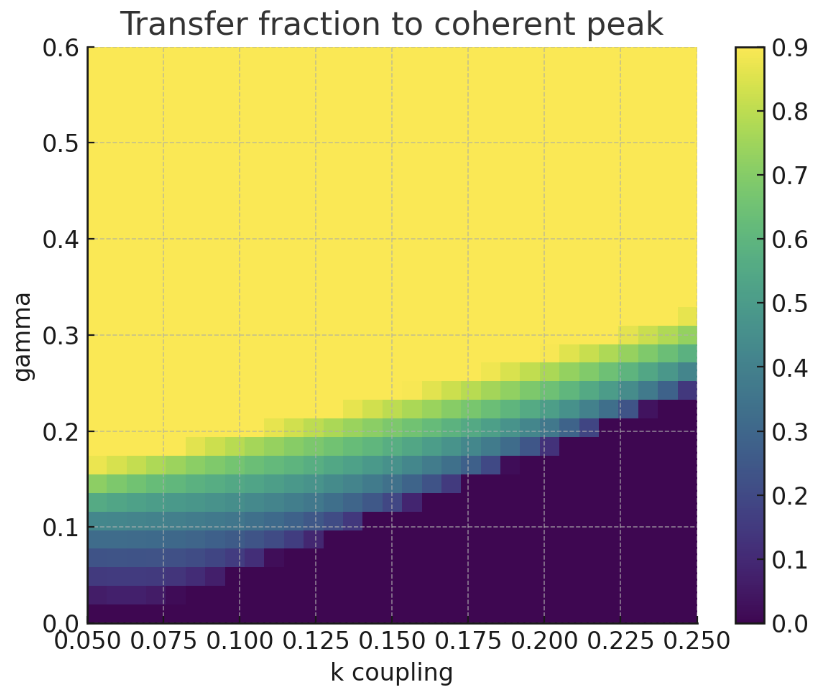} 
\includegraphics[width=0.4\textwidth]{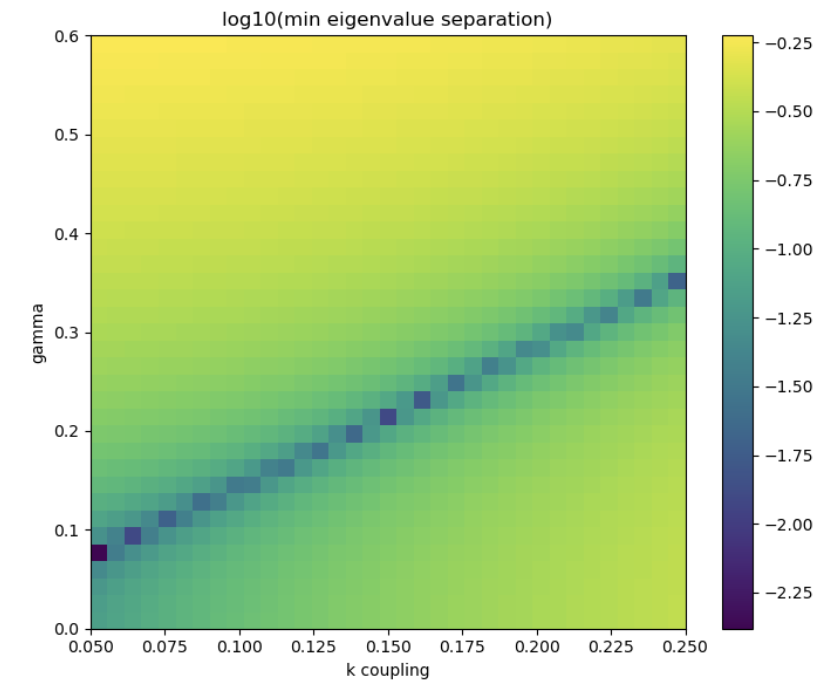} 
\caption{A coarse $2D$ parameter sweep $(31\times 31)$ in the left. This figure shows the transfer fraction of spectral weight into the coherent lasing peak as a function of the coupling $\kappa$ and pumping $\gamma$. }
\label{fig:seepgamma2d}
\end{center}
\end{figure}

\begin{figure}[ht!]   
\begin{center}
\includegraphics[width=0.85\textwidth]{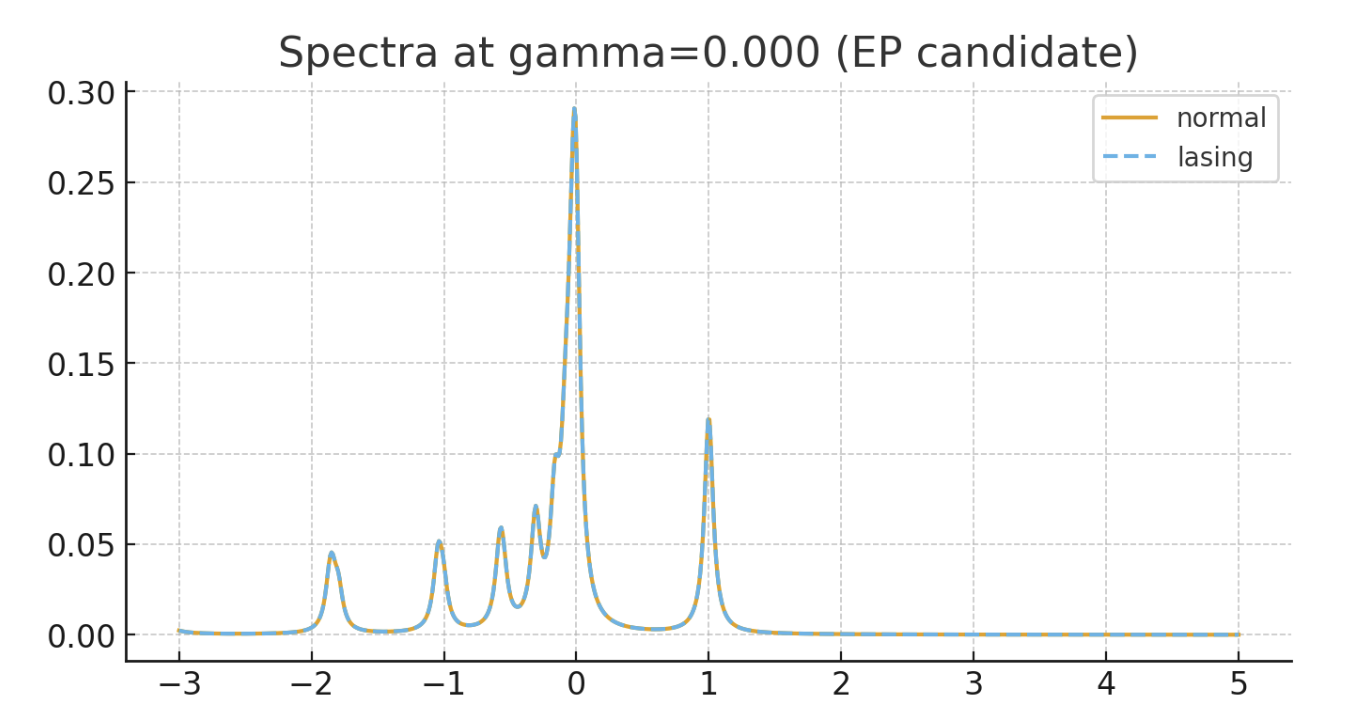} 
\caption{Example spectra at an EP candidate. This is the plot of $\sigma_1 (\omega)$, normal and lasing, at the $\gamma$ that gives the minimal eigenvalue separation. This matches the behavior in \cite{citekeyJahangiri}, showing the capability of holography to simulate photonic systems.}
\label{fig:spectralasing}
\end{center}
\end{figure}

Then, in figure \ref{fig:seepgamma2d}, the transfer fraction of spectral weight into the coherent lasing peak as a function of coupling $\kappa$ and pumping $\gamma$ is shown. One could see an EP-like bifurcation structure. Also, as one would expect, by increasing gain/loss, the system is pushing toward the coherent supermode, while larger coupling $\kappa$ would delay the onset, so the threshold line slopes upward.

In figure \ref{fig:spectralasing}, the behavior of spectra at an EP is shown. To get this result, we use the holographic model, where we find the eigenvalues and eigenvectors of the full holographic block operator $H$. One could see that the general and qualitative behavior match the experimental results of \cite{citekeyJahangiri}. This shows that holographic models based on AdS/QCD-inspired are capable of correctly simulating the general properties of open quantum systems and photonic structures, which could be used to architect better quantum information setups with applications in quantum computers.

So, the holographic model reproduces the qualitative FGT picture as there is a transfer of finite-frequency spectral weight into a narrow, coherent peak, as pumping increases past the threshold.

The lasing threshold, where the transfer fraction rises is not identical to the analytic EP2 at $\gamma=0$. In the case of asymmetric couplings and the full radial structure, lasing onset occurs at some nonzero $\gamma$ where a particular mode gains dominance. The exceptional points then occur at special parameter values, like $\gamma=0$, which would give EP2 in the symmetric case. In the full holographic block operator, they manifest as near-degeneracies, with very small minsep and large eigenvector's condition numbers.

\subsection{Inhomogeneous case: holographic lattice for three-site photonic EP}
For deeper lattices with three sites, one could extend this model to inhomogeneous $s(x^1)$, and with RG flows to a $\mathcal{PT}$-symmetric IR fixed point, where non-Hermitian ``cooling'' which is the decrease in entropy, occurs via the horizon shrinkage.

For extending the homogeneous case to the inhomogeneous boundary condition, which can capture the multi-site structure like a three-resonator system, one can make a spatial variation in the sources, which then can mimic the discrete lattice sites with inter-site couplings. In the photonic analog, the three resonators have position-dependent gain/loss ($\Gamma_j(x)$ for site $j=1,2,3$) and nearest-neighbor coupling $g$, leading to an EP at critical $g_c$ where modes coalesce in the band structure $\omega(k)$. This structure can actually holographically map to spatially modulated boundary sources $s(x^1)$, dual to a bulk scalar $\phi(z,x)$ that develops a lattice-like profile via RG flow. The strong coupling in the bulk would smear the discrete sites into a continuous modulated geometry, while the EP would emerge as a momentum-dependent branch point in the quasinormal mode spectrum, $\omega_n(k)$, where $k$ is the Bloch wavevector.

The PT-symmetry here can be preserved by making $s(x)$ real and even under parity $x \to -x$, with gain/loss imbalance parameterized locally as $x(x) = \frac{s(x) - s^*(x)}{s(x)+s^*(x)}$.
 
The action in this case is the same as in relation \ref{eq:actionEP}, but now the fields depend on $x \equiv x^1$, e.g., $\phi = \phi(z, x, t)$, $A_M = A_M(z, x, t)$. To keep the translation invariance in other directions $(x^2, x^3, t)$, one can assume the plane-wave ansatz $e^{-i \omega t + i \vec{k}\perp \cdot \vec{x}\perp}$, but one should focus on $k \equiv k_x$ along the modulation direction.

The scalar EoM then simplifies to a $2D$, $(z,x)$ problem as
\begin{gather}
- \partial_z (f(z) \partial_z \phi) + f(z) \partial^2_x \phi + \frac{\omega^2}{f(z)} \phi - i q A_t \partial_z \phi + ... = \frac{\partial V}{\partial \phi^*},
\end{gather}
 where at the boundary $z \to 0$, we have
 \begin{gather}
 \phi(z,x) \sim s(x) z^3 + \langle \mathcal{O} (x) \rangle z + ..., \nonumber\\
 \phi^* (z,x) \sim s^* (x) z^3 + \langle \mathcal{O}^\dagger (x) \rangle z + ....
 \end{gather}
Here, $s(x)$ is piecewise constant for three sites, e.g., the period is $a= 3\ell$, with
  \[
    \alpha(x)=\left \{
                \begin{array}{ll}
                  s_1 \ \ \ \ 0<x < \ell, \nonumber\\
                  s_2 \ \ \ \  \ell< x <2\ell, \nonumber\\
                  s_3 \ \ \ \  2\ell < x < 3\ell,
                \end{array}
            \right.
  \]
and satisfies the relation $s(x+a) = s(x)$. In order to keep the PT symmetry, we can take $s_1 =s_3 = \gamma+ i \Gamma/2$ (symmetric gain/loss on outer sites), $s_2 = \gamma- i \Gamma$ (lossy central site), with $\gamma$ and $\Gamma$ to be real. The couplings then emerge from the bulk diffusion with the hopping scale of $\sim 1/\ell$. The imbalance $\Gamma$ tunes the EP, while $\gamma$ sets the overall scale.

This specific boundary condition breaks the translation invariance along $x$, where the solutions are Bloch waves: $\phi(z,x) = e^{i k x} \tilde{\phi}(z,x)$, with $k$ quantized in the Brillouin zone $\lbrack -\pi/a, \pi/a \rbrack$. Then, the boundary operator $\langle \mathcal{O}(x) \rangle$ has a Fourier expansion, dual to the photonic mode amplitude on the sites.

Now, to capture the three-site photonic EP, one should determine the effective lattice coupling. In the bulk, the scalar's Laplacian $\partial_x^2 \phi$ can generate the hopping between ``sites." The effective coupling is
\begin{gather}
 g \sim \int dz f(z)  \frac{ | \phi'(z) |^2 }{ \ell^2},
 \end{gather}
 where 
 \begin{gather}
 \tilde{\phi}(z,x) = \sum_n \phi_n (z) e^{i n \pi x / \ell},
 \end{gather}

Then, in the $\mathcal{PT}$-unbroken phase, where the gain/loss satisfies the inequality $\Gamma < \Gamma_c(k)$, and by photonic analogy $\Gamma_c \sim g \sqrt{3}$, the QNMs $\omega_n(k)$ are real for each band, and the dispersion is $\omega(k) \approx \omega_0 + 2g \cos(k \ell)$. The bulk profile $\phi(z,x)$ is a smooth modulation with no instabilities. The boundary Green's function $G(x, x'; \omega) = \langle \mathcal{O} (x) \mathcal{O}^\dagger (x') \rangle$ decays exponentially for $| x - x'| > \ell$, which mimics the localized resonator modes.

At the EP transition, where $\Gamma = \Gamma_c (k)$ and at specific $k = \frac{\pi}{2\ell}$, which is the zone edge, coalescence in the band structure would happen. The two QNMs, $\omega_{\pm}(k)$, touch at $\omega_c = g \sqrt{3}$, where $\omega(k) - \omega_c \sim \sqrt{k-k_*}$. In the bulk, the radial flow develops a logarithmic divergence in $\partial_x \phi$, signaling marginal stability, as the Hessian of the on-shell action has a zero mode at $k_*$. This is the holographic EP, where the eigenvectors (Fourier modes of $\langle \mathcal{O} \rangle$) coalesce, which is dual to photonic eigenvector degeneracy.

Then, in the $\mathcal{PT}$-broken phase, $\Gamma > \Gamma_c$, the complex conjugate branches are $\omega_\pm (k) = \omega_r(k) \pm i \omega_i(k)$, with $\omega_i >0$ for amplification, where we have gain-dominated bands. In the bulk, we get instability via the complex field $\phi(z,x)$, which has growing modes along $x$ with Petrov-type shifts. Considering the full backreaction, the metric warps into a striped geometry where $g_{xx} (z,x) \sim 1+ \delta \cos (2\pi x/a)$, and the NEC is broken locally, akin to holographic superconductors with modulated order.

In this case, the FGT sum rule extends to a momentum-dependent form as
\begin{gather}
\int_0^\infty \sigma_1 (\omega, k) d\omega = \frac{\pi}{2} \frac{n(k) e^2}{m},
\end{gather}
where $n(k)$ is the $k$-dependent density. Right after the EP is formed and at low $k$, the weight transfers to $\delta(\omega)$, while $n(k)$ is conserved across the transition.

This inhomogeneous model can predict topological features that were absent in the homogeneous case, like EP-protected edge states in finite lattices, or unidirectional amplification in the broken phase. 

\begin{figure}[ht!]   
\begin{center}
\includegraphics[width=0.4\textwidth]{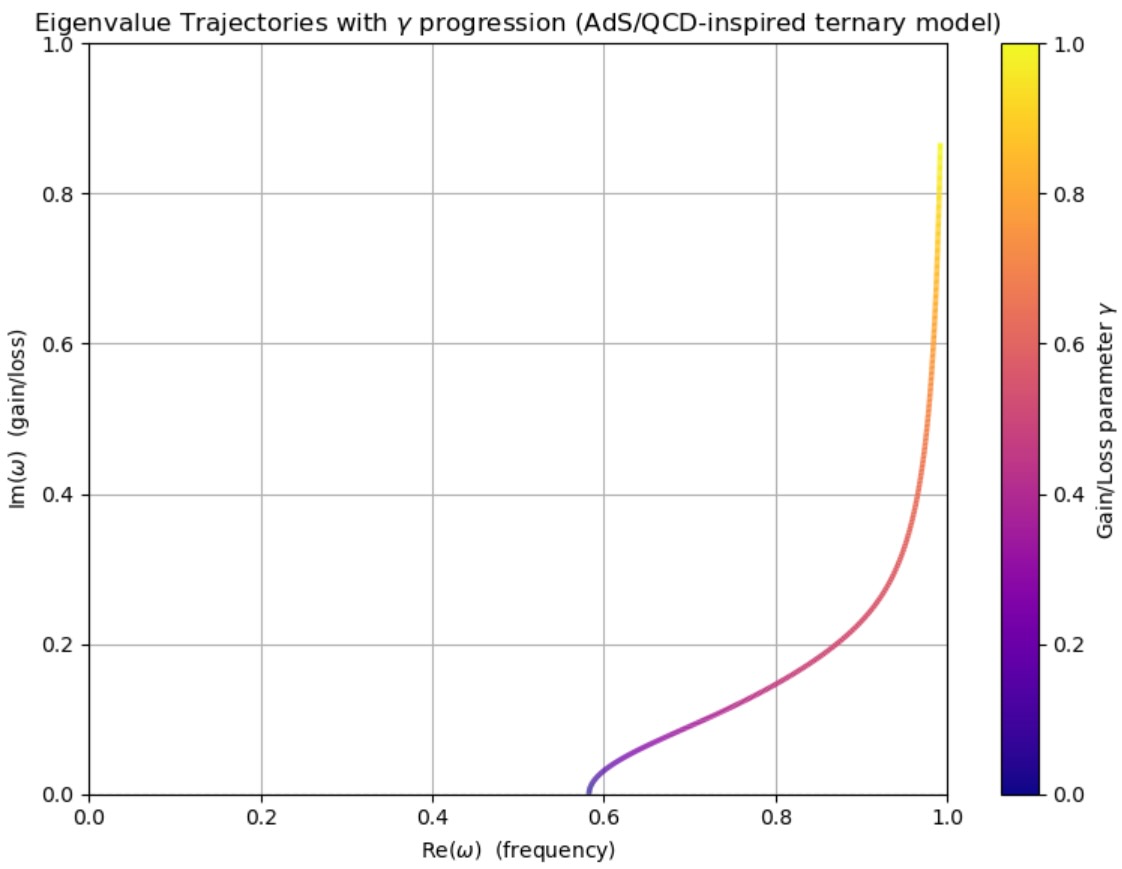} \ \ \ \ \ \ 
\includegraphics[width=0.4\textwidth]{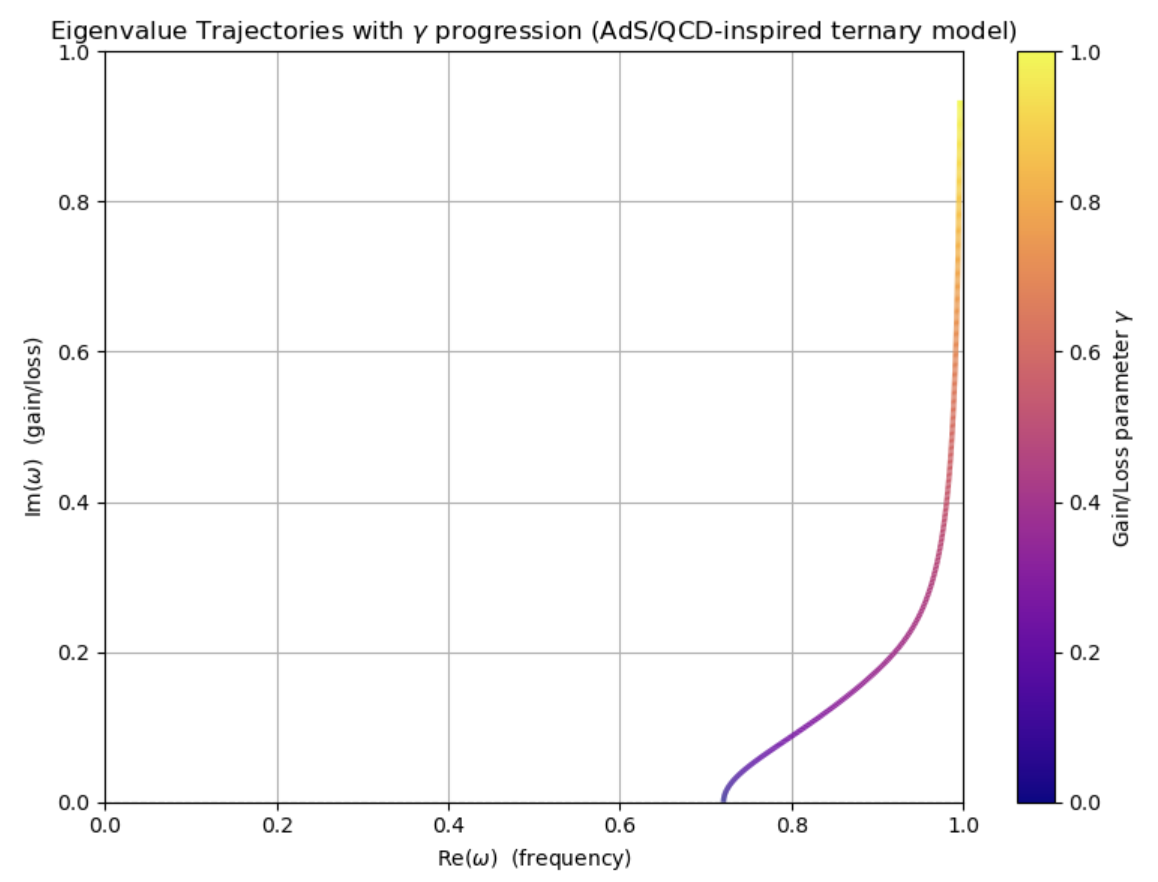} \ \  \ \ \ \ 
\includegraphics[width=0.42\textwidth]{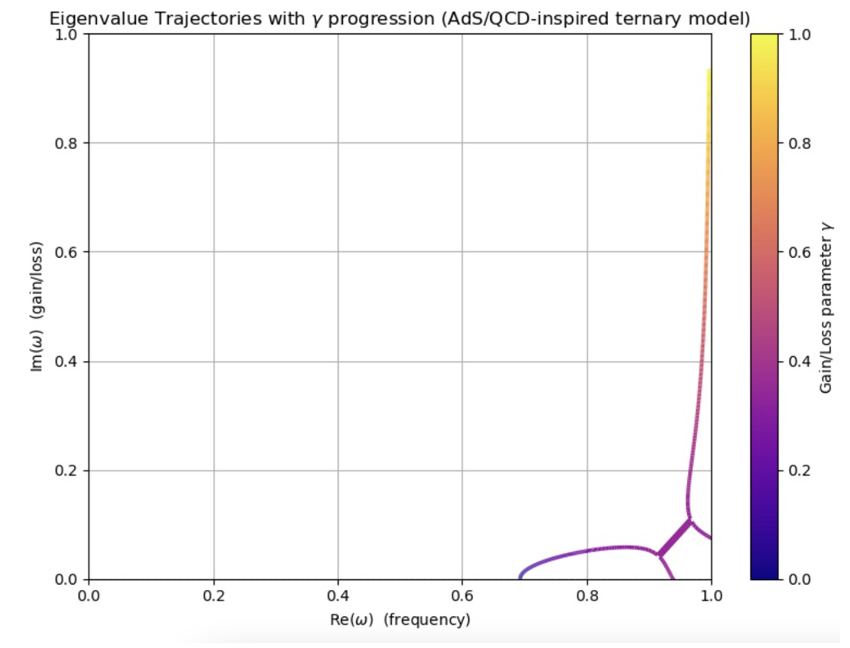}\ \ \ \  \ \
\includegraphics[width=0.41\textwidth]{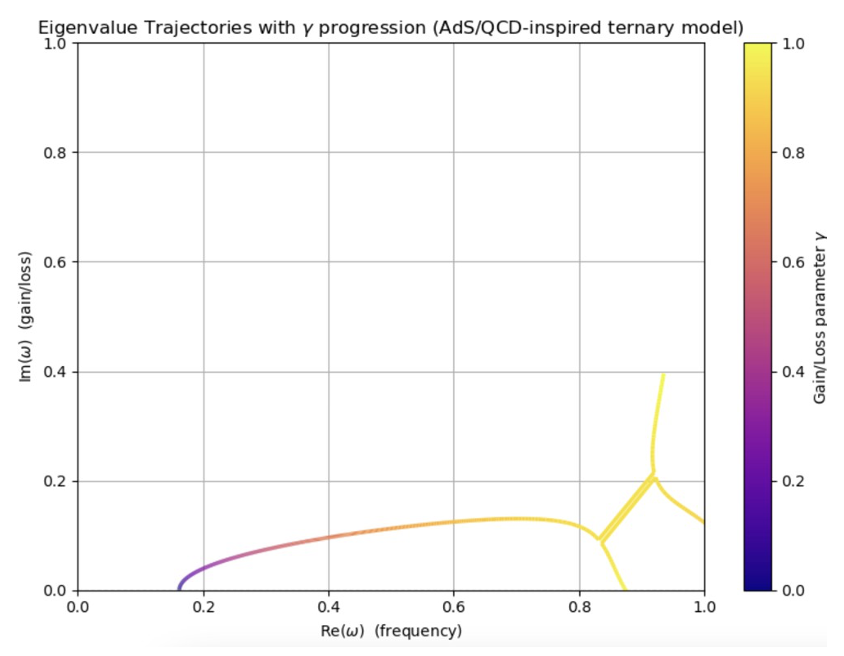}
\caption{Eigenvalue trajectories with the progression of gain/loss parameter $\gamma$ in the AdS/QCD-inspired ternary model. In the left, up $g_{12}=0.3, g_{23}=0.35$, in the right, up $g_{12}=0.25, g_{23}=0.2$, in the left, down, $g_{12}=0.25, g_{23}=0.25$, in the right, down, $g_{12}=0.65, g_{23}=0.65$.}
\label{fig:Etrajectories}
\end{center}
\end{figure}

The eigenvalue trajectories with the progression of gain/loss parameter $\gamma$ in the AdS/QCD-inspired ternary model have been shown in figure \ref{fig:Etrajectories}. It can be seen that with bigger coupling the trajectory is larger and with bigger weight for higher values of $\gamma$ as most of the trajectory would get yellower color.

In other interesting works such as \cite{Xian16}, the phase structure and phase diagram of non-Hermitian $\mathcal{PT}$-symmetric holographic model, based on stability and $\mathcal{PT}$-symmetry  has been constructed.
We could extend our work here in the same way as well and consider stability in our phase diagram to find a richer phase structure. In addition, similar to them, the square of condensate and AC conductivity for these phases could be calculated, and then we can go to the next step and check the FGT sum rule for all of these phases.

\subsection{Effects of exceptional points versus confining end-wall on chaos}

We proposed here that there could be an analogy between exceptional points in non-Hermitian photonics and the end-wall (hard-wall/IR brane) in holographic confining geometries.
As the hard-wall imposes boundary conditions, the wavefunctions vanish at $z=z_{IR}$, and this enforces a discrete spectrum of normal modes like hadron resonances. The EP also enforces a coalescence condition in parameter space, which leads to lasing threshold selection and symmetry breaking. In the confining case, the obstruction that truncates the bulk geometry would force ``level crowding" near the wall, and the nonlinear effects near the wall can induce chaotic behavior in classical string trajectories and in the quantum spectrum, leading to random matrix statistics. In the EP case, the spectra reorganize drastically at EPs, phase rigidity breaks, modes hybridize, and level-spacing statistics can move toward chaotic, non-Hermitian random matrix behavior. Mathematically, in holography, the hard-wall condition turns the Sturm-Liouville operator into a non-self-adjoint operator when dissipation or complex dilaton backgrounds are included. In photonics, the non-Hermitian coupling matrix $\mathbb{M}$  is exactly such an operator, with an effective boundary in parameter space where solutions coalesce. So the EP in photonics plays the same role as the IR wall in AdS/QCD. It is actually a geometric boundary that reorganizes the spectrum and can act as a source of chaotic dynamics.

In \cite{PhysRevE.52.4762}, it has been shown that if a regular Hamiltonian is perturbed by a term which produces chaos, the onset of chaos shifts toward larger values of the perturbation parameter if the unperturbed spectrum is degenerate. This point can also be seen using AdS/QCD and the onset of chaos there.

One can check the statistical behavior of the eigenvalues, which both follow a 
$\beta$-ensemble similar to the Random Matrix Theories (RMTs). These connections can be studied further by exploring the intrinsic properties of the quantum mechanical operators.

Noting in a more rigorous way, in \cite{1977RSPSA.356..375B, Bianchi:2022mhs, Bianchi:2023uby}, a method to diagnose chaos in Hamiltonian systems has been introduced, which is now commonly used. The random observables are the spacings between the energy levels,
\begin{gather}
\delta_n = E_{n+1} - E_n,
\end{gather}
or the \textit{ratios of successive spacings}
\begin{gather}
r_n \equiv \frac{E_{n+1}- E_n }{E_n - E_{n-1}} = \frac{\delta_{n+1} }{\delta_n}.
\end{gather}

In chaotic systems, the level spacings match the spacings of the eigenvalues of random matrices in Random Matrix Theory (RMT) and its $\beta$-ensemble, where for $N \times N$ matrices, the eigenvalues have the probability density function (PDF) in the form
\begin{gather}
P_N (\lambda_1, \lambda_2, . . . , \lambda_N) = \mathcal{C}(\beta) \times \text{exp} \left ( - \frac{\beta}{2} \sum_{i=1}^N \lambda_i^2 \right ) \prod_{1 \le i < j \le N} | \lambda_i - \lambda_j |^\beta,
\end{gather}
where for the case of a continuous $\beta$, this defines the so-called $\beta$-ensemble. This statistical behavior can also be used to model EPs. Note that, in general, EPs occur in all eigenvalue problems that depend on a system parameter.

In \cite{Bianchi:2021sug}, the authors showed that the branch points in open string scattering amplitudes in confining backgrounds follow such a $\beta$-ensemble, and in \cite{Ghodrati:2023uef}, we showed that the branch cut singularities in mutual information, or its corresponding critical distance $D_c$, also follow such behavior in various confining models. Therefore, we can show how chaos emerges in these confining setups.

Then, in \cite{Savic:2024ock}, the authors showed that the eigenphase-space distribution of the scattering of photons, gravitons, and tachyons on highly excited bosonic strings consists of both regular/Poisson and chaotic Wigner-Dyson processes, and so they are weakly chaotic. The interesting point is that for special values of momenta and scattering angles in the case of photons, they found strong chaotic behaviors in the form of random-matrix type. In fact, these specific momenta and angles are directly related to the exceptional points of the system. Another interesting result is that at these points, there is a crossover between the two regimes of scattering, one dominated by the short and the other by the long partitions of the total occupation number of the highly excited string. Also, the information entropy of the S-matrix becomes maximized, which again is a typical behavior at the EPs, as one would expect.

Based on the appearance of the periodic structures in these functions, which arise due to the boundary condition and the presence of the wall at the end of the geometry in confining backgrounds, one could imagine that similar structures could appear in the presence of exceptional points as well, since these points cause instabilities and discontinuities in the Hamiltonian and modular Hamiltonian, spectral singularities, and dramatic effects in the wave functions and scattering processes in the system. They also induce chaos. The strength of chaos then depends on the density of these EPs in the system, which can be modeled by either the hard-wall or soft-wall confining backgrounds in the holographic QCD models.

In \cite{PhysRevA.43.4159, Heiss_2012}, it has been shown that the onset of chaos becomes more apparent where there is a high density of EPs, while for mild perturbations and smaller numbers of EPs, the system remains robust. Thus, a high density of EPs is associated with a hard-wall or a clear phase transition, while low-density EPs could be associated with free AdS or soft-wall geometries. The changes in the light-propagating modes of the quark-gluon plasma versus the confined phase, i.e., gain or loss regions, can also be considered to strengthen this connection. The Berry phase shift in AdS/QCD has been briefly discussed in our previous work \cite{Ghodrati:2020vzm}, which could then be extended. The applications of EPs could also be studied in the context of the fluid/gravity correspondence, as it has been connected to instabilities in the Reynolds number \cite{GOLUBITSKY1988362}.

Another important factor is that exceptional points can enhance the sensitivity by increasing the eigenfrequency splitting. This bifurcation effect has been used in studying cavity QED (cQED), strong-field QED, and QCD. In \cite{Qu_2023}, it has been shown that collisions in a QED plasma can produce pairs whose frequency becomes large as they slow down or reverse direction. This is similar to the behavior of frequency shifting around EPs, where the sensitivity enhances as $\Delta \omega_{\text{EPN}} \propto (\kappa/\epsilon)^{2/N}$ or $\Delta \omega \sim \epsilon^{1/N}$. Thus, if we find that the eigenfrequency splitting in the QCD models increases by the order of the exceptional point, we can support our conjecture. Another important piece of evidence for the connection is the existence of topological chirality in both photonic EPs and QCD structures, as well as in specific setups of cQEDs and microcavity lasers.

Therefore, we get hints to relate the spectral degeneracies of QCD with the exceptional points of PT-symmetric systems, and we can then check if the real and imaginary parts, as well as the associated eigenvectors, coincide. Thus, exceptional points can act as new ``tools'' in AdS/CFT dictionaries.
 
Also, note that the specific dual structures of EPs in holographic models could be constructed by going one dimension higher. The dual of such particles moving on the boundary CFT would be strings hanging in the bulk, so one could expect that the dual of these exceptional points in the boundary quantum system would be ``exceptional lines,'' or exceptional Wilson lines, in the higher-dimensional bulk system.

In addition, the resolvent, or the Green's function, for the Sturm–Liouville operator $G(\lambda) = (\mathcal{L}- \lambda)^{-1}$, near $\lambda \approx \lambda_n$, where $\lambda_n$ is a simple eigenvalue, would be
\begin{gather}
G(\lambda) \sim \frac{ \ket{\psi_n^{(R)} } \bra{\psi_n ^{(L)} }  }{\lambda- \lambda_n}, \ \ \text{simple pole},
\end{gather} 
where $\ket{\psi_n^{(R)}}$ and $\bra{\psi_n ^{(L)}}$ are distinct right/left eigenfunctions. On the other hand, at an EP of order $m$, the resolvent for the system would develop a higher-order pole as $G(\lambda) \sim \sum_{k=1}^m \frac{A_k}{(\lambda- \lambda_{EP})^k} $, which has a nontrivial Jordan structure yielding nonanalytic and strongly enhanced responses. So both the IR wall and the EP are singularities of $G$.

All in all, the end-wall of confining geometries has interesting connections to exceptional points, and new holographic correspondences could be imagined using them.

\subsection{Holographic phase rigidity and Petermann factors around EPs}

An important concept in designing qubits in topological and photonic quantum computers is the Petermann factor, which is especially useful for studying the ``imperfect'' exceptional points in experiments \cite{PhysRevResearch.5.033042}, and therefore, by the relations that we built in the previous section, we could bring it to the study of QCD and specifically to the experimental QCD research.

In non-Hermitian systems where $\lbrack \hat{H}, \hat{H}^\dagger \rbrack \ne 0$, which are also non-normal, the right eigenstates $\ket{R_l}$ should be distinguished from the left eigenstates $\ket{L_l}$, and their difference quantifies the strength of the non-normal effect, which can be measured by the Petermann factor between a pair of eigenstates as
\begin{gather}
K_l : = \frac{\langle R_l \ket{R_l}  \langle{L_l} \ket{L_l} }{ |\langle L_l \ket{R_l}  |^2 }, \ \ \ \ \ \ K_l \geqslant 1.
\end{gather}

This factor, which has also been called the excess-noise factor, quantifies the linewidth broadening, which, for instance, could come from the quantum excess noise in lasers.

The phase rigidity then can be defined as
\begin{gather}
r_l := \frac{|\langle L_l \ket{R_l} |}{\sqrt{\langle R_l \ket{R_l} \langle L_l \ket{L_l}  } }, \ \ \ \ 0 \leqslant r_l=1/\sqrt{K_l} \leqslant 1,
\end{gather}
which quantifies the complexness of the wave functions in $\mathcal{PT}$-broken symmetry systems. Note that around an EP, the Petermann factor diverges and the phase rigidities vanish. Also, around the EPs, as the vector space becomes severely skewed, we have a losing dimensionality effect, which is related to the divergence of the Petermann factor.

So the Petermann factor can be used in studying QCD phase diagrams and the search for the critical points. In the Quark-Gluon-Plasma (QGP) phase, there are resonant states and unstable quasi-particles which have complex eigenvalues. Near the chiral phase transition, deconfinement transition, or around the critical point, the Petermann factor sould increase sharply as there is an enhanced sensitivity.

At an EP of order $n$, the spectral response strength can be written as
\begin{gather}
\xi : = || \hat{G}_n ||_2,
\end{gather}
where the Green's function near the exceptional points is expanded as \cite{PhysRevE.61.929}, 
\begin{gather}
\hat{G} (E) = \frac{\mathds{1}}{E - E_{\text{EP} } }+\frac{\hat{G}_2}{(E-E_{\text{EP}} )^2} + ... +\frac{\hat{G}_n}{(E-E_{\text{EP} } )^n}.
\end{gather}

For the soft-wall AdS/QCD model, the Green's function can be written as
\begin{gather}
G(p, z, z') = \sum_{n=1}^\infty \frac{\phi_n(z) \phi_n(z')}{(-p^2) - m_n^2}.
\end{gather}

The similar behavior of this Green's function, and also Green's functions in other holographic confining models could further strengthen our duality.

For the above holographic model that we built in section \ref{sec:Holoconf}, we can find the Petermann factor and phase rigidities near the exceptional points based on the parameters of the metric or action, such as the coupling or dilaton field. The Petermann factors and phase rigidities near exceptional points are defined in \cite{Wiersig:2023xoq}.

 \begin{figure}[ht!]
 \centering
  \includegraphics[width=14cm] {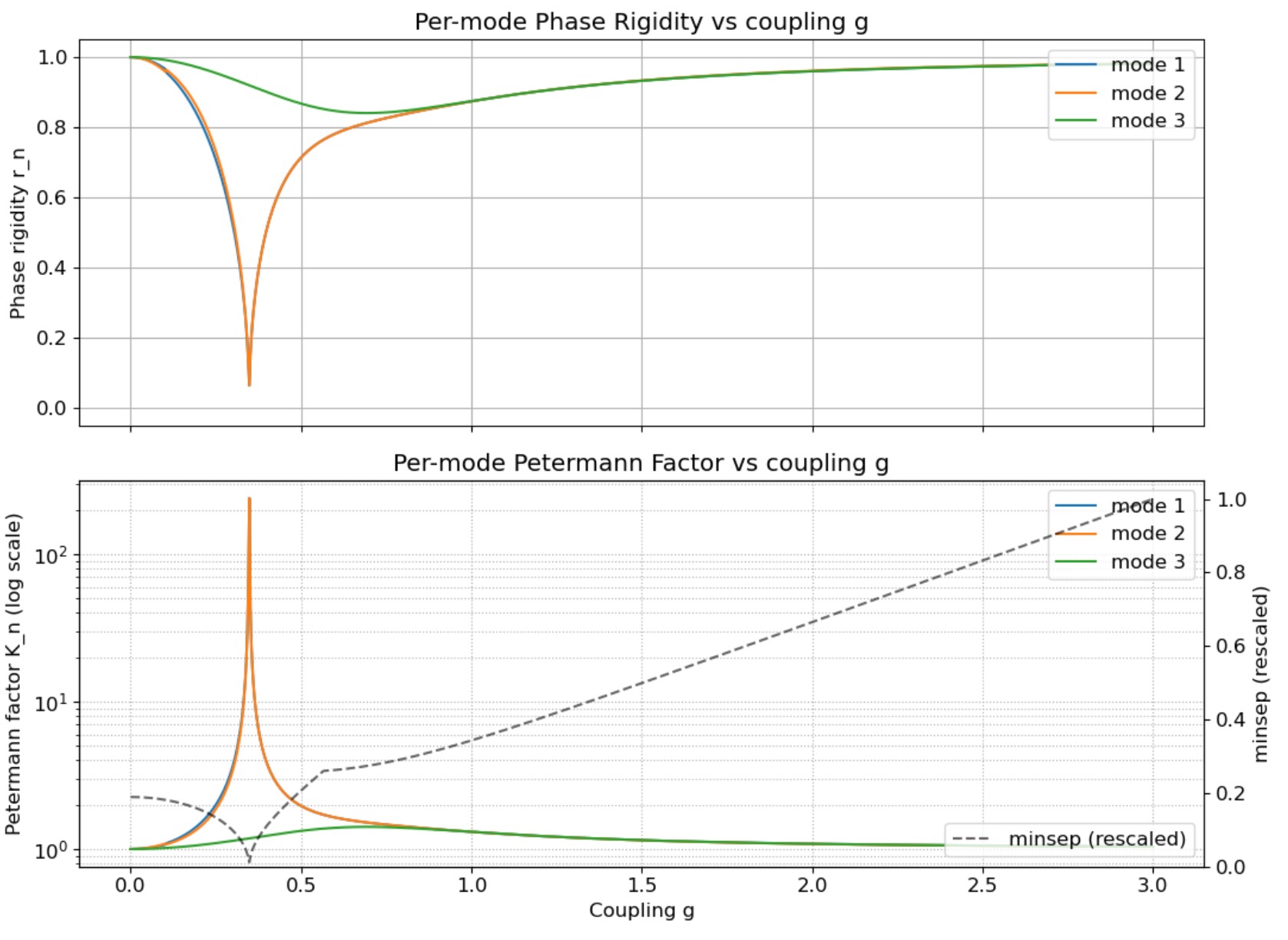}  
  \caption{Petermann factor and phase rigidities as a function of the coupling $g$, while other parameters are fixed, for the holographic model based on AdS/QCD constructed in our section \ref{sec:Holoconf}.}
 \label{fig:monodromy}
\end{figure}
 
From figure \ref{fig:monodromy}, one can see that, for small $g$, modes are weakly coupled and we have $r \approx 1$ and $K \approx 1$. As $g$ increases, eigenmodes start coalescing. Then, near the exceptional point $g = g_c$, $r \to 0$ and $K \to \infty$, which reproduces the mode condensation or chaotic transition observed in holographic models. The mode depicted by orange shows a dip around $g \approx 0.4-0.5$, which is a hallmark of an EP. Note that the “coupling” $g$ can correspond to boundary deformation strength, brane-brane coupling, or IR wall transparency. The EP then marks a transition between confined and quasi-open modes, akin to resonant mode merging at the AdS hard wall. Divergent $K$ and vanishing $r$ can thus signal non-Hermitian mixing between bulk normal modes or complexified quasinormal frequencies.
 
  \begin{figure}[ht!]
 \centering
  \includegraphics[width=14cm] {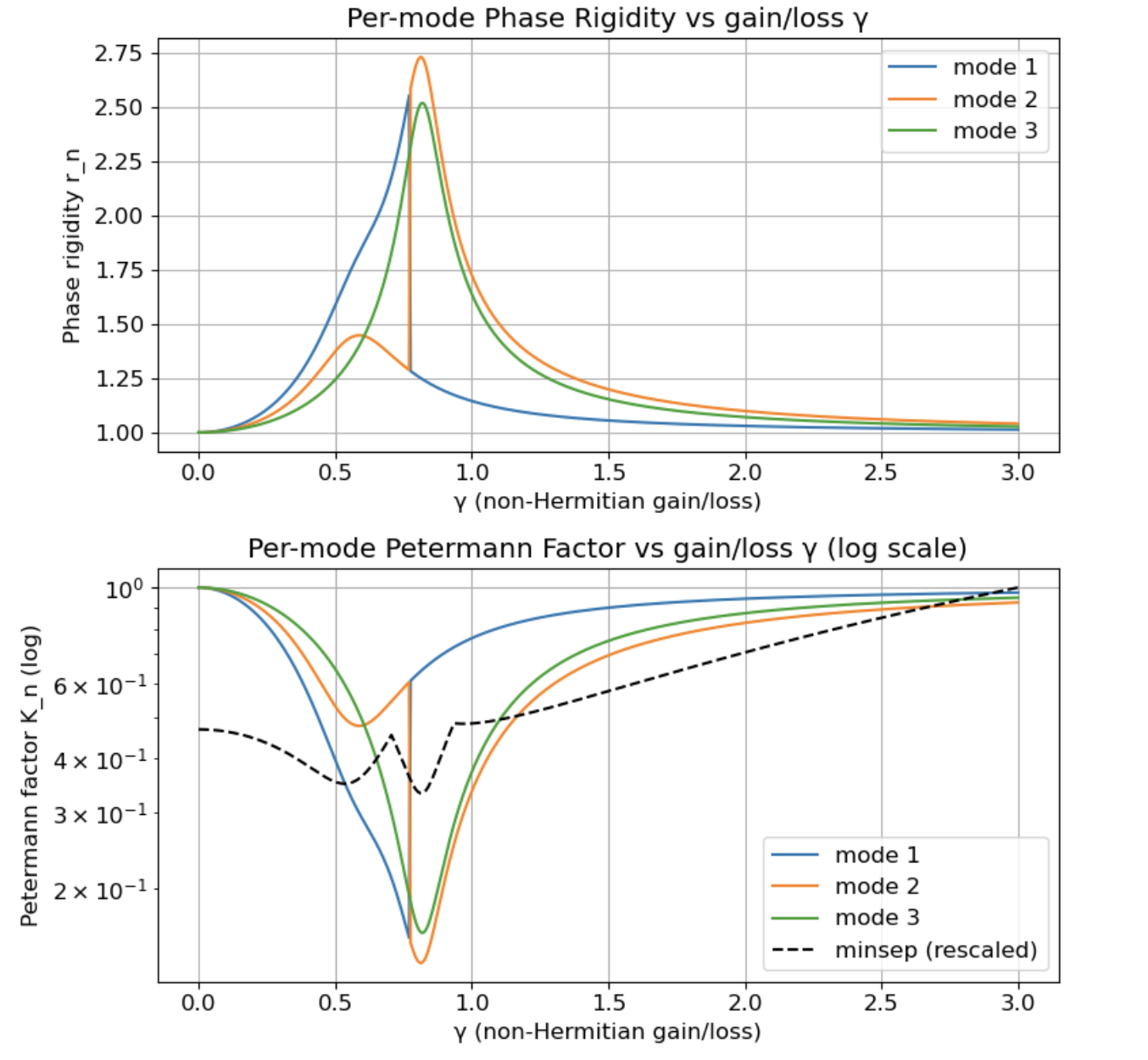}  
  \caption{Petermann factor and phase rigidities as a function of gain/loss $\gamma$, while other parameters are fixed, for the holographic model based on AdS/QCD constructed in our section \ref{sec:Holoconf}.}
 \label{fig:monodromy2}
\end{figure}
 
Next, in figure \ref{fig:monodromy2}, the Petermann factor and phase rigidities as a function of gain/loss $\gamma$ are shown while other parameters are fixed. Here the “dip-spike” structure around $\gamma \approx 0.8-1.0$ is where the EP occurs. The minseps which are the dashed black curve, reaches a minimum indicates coalescing eigenvalues and phase rigidity $r_n$ varies strongly near the EP.

\section{EPs, photonic molecules and quark bound states}\label{sec:Kaons}

An interesting model that one can use is the picture of the bound states for a ternary, parity-time-symmetric photonic laser molecule. For instance, mesons, which consist of a quark-antiquark pair, can model the EP2 structures, and baryons, which consist of three color quarks, can be associated with EP3. However, note that this section is only an ``\textbf{analogy}" and unlike previous section, we don't make any holography argument here and only are building a toy model.

So as noted in \cite{citekeyJahangiri}, by using the ternary systems and then breaking the PT symmetry and perturbing the anti-PT symmetry, a pure single-mode spectrum can be achieved. This is similar to the confinement/deconfinement phase transition. So by an analogy, we can relate these two matrices as
\begin{gather} \label{eq: quarkH}
\Phi  = \begin{pmatrix}
\frac{\pi^0}{\sqrt{2}} + \frac{\eta}{\sqrt{6} } & \pi^+ & K^+\\
\pi^- & -\frac{\pi^0}{\sqrt{2}}+\frac{\eta}{\sqrt{6} } & K^0 \\
K^- & \bar{K}^0 & -\frac{2\eta}{\sqrt{6}}
\end{pmatrix}  \iff  H=  \begin{pmatrix}
m_\pi - i \gamma_\pi/2 & g_{\pi K} & g_{\pi \eta} \\
g^*_{\pi K} & m_K - i \gamma_K/2 & g_{K\eta} \\
g^*_{\pi \eta} & g^*_{K \eta} & m_\eta- i \gamma_\eta/2
\end{pmatrix}.
\end{gather}

The $H$ matrix on the left is non-Hermitian, and the mesons are considered to be unstable and decay. For instance, pions decay into leptons or Kaons into pions. Here, $m_i$ denote the masses, $\gamma_i$ are the decay widths, and $g_{ij}$ are the interaction strengths between mesons. Therefore, considering the Kaon decay as an open non-Hermitian quantum system, another way of connecting the behavior of exceptional points and QCD dynamics could be envisioned, which we will discuss in section \ref{sec:Kaons}.

First, note that the flavor $SU(3)$ current in terms of the vector currents of $u$-, $d$-, and $s$-quarks can be written as \cite{Klingl:1996by}
\begin{gather}
J_\mu^{em} = \frac{2}{3} \bar{u} \gamma_\mu u - \frac{1}{3} \lbrack \bar{d} \gamma_\mu d + \bar{s} \gamma_\mu s \rbrack,
\end{gather}
which can then be rewritten as \cite{Klingl:1996by}
\begin{gather}
J_\mu^{em} = \frac{1}{\sqrt{2}} J_\mu^{(\rho)} + \frac{1}{3\sqrt{2}} J_\mu^{(\omega)}- \frac{1}{3}J_\mu^{(\phi)}.
\end{gather}
Here, $\rho$ is related to the neutral $\rho^0$-meson.

The decay of the vector meson into the $e^+ e^-$ pair can be described by the matrix elements of
\begin{gather}
\langle 0 | J_\mu^{em} | V \rangle = - \frac{m_V^2}{g_V} \epsilon_\mu^{(V)},
\end{gather}
where $m_V$ denotes the vector meson mass, $\epsilon_\mu^{(V)}$ is a polarization vector, and $g_V$ is the vector meson coupling, which is related to the $e^+ e^-$ decay widths \cite{Klingl:1996by}. This meson coupling ($g$) can be related to the couplings between the resonators in the photonic molecule, i.e., $\kappa$, of the coupled cavity system \cite{cite-key}.

The three-flavor QCD in the presence of a nonzero $\theta$-angle can be written in terms of the chiral Lagrangian as
\begin{gather}
\mathcal{L}= \frac{F_\pi^2}{4} \text{Tr} \lbrace \nabla_\mu \Sigma \nabla^\mu \Sigma^\dagger \rbrace + \frac{F_\pi^2}{2} G \text{Tr} \lbrace M \Sigma + M^\dagger \Sigma^\dagger \rbrace - \frac{a F_\pi^2}{4}\left ( \theta - \frac{i}{2} \text{Tr} \lbrace \log \Sigma - \log \Sigma^\dagger \rbrace \right )^2,
\end{gather}
where $F_\pi$ is the pion decay constant, $F_\pi^2 G$ corresponds to the chiral condensate at zero $\theta$-angle, $F_\pi^2 a/2$ is the topological susceptibility of the underlying Yang-Mills theory, and $M= \text{diag} (m_u, m_d, m_s)$ is the quark mass matrix.

The matrix field $\Sigma \in U(3)$ can be written as
\begin{gather}
\Sigma = U \Sigma_0 U, \ \ \ \ \ U=exp \left ( \frac{i \Phi}{\sqrt{2} F_\pi} \right), \ \ \ \ \ \Phi= \Pi^a T^a + \frac{S}{\sqrt{3}},
\end{gather}
where $\Sigma_0$ maximizes the static Lagrangian and $T^a$ are the $SU(3)$ generators.
Here, $\Sigma$ describes the pseudo-scalar meson octet
\begin{gather} \label{eq: quarkH}
\Pi^a T^a  = \begin{pmatrix}
\frac{\pi^0}{\sqrt{2}}+\frac{\eta}{\sqrt{6}} & \pi^+ &K^+\\
\pi^- & -\frac{\pi^0}{\sqrt{2}}+\frac{\eta}{\sqrt{6}} & K^0 \\
K^- & \bar{K}^0 & -\frac{2\eta}{\sqrt{6}}
\end{pmatrix}.
\end{gather}

For $\eta = 0$, the effective chiral Lagrangian in the flavor $SU(3)$ sector, which includes the $u$, $d$, and $s$ quarks, can be written as
\begin{gather} \label{eq: quarkH}
\Pi^a T^a  = \sqrt{2} \begin{pmatrix}
\frac{\pi^0}{\sqrt{2}} & K^+ & \pi^+\\
K^- & 0 & \bar{K}^0 \\
\pi^- & K^0 & \frac{-\pi^0}{\sqrt{2}}
\end{pmatrix}.
\end{gather}
and the singlet field $S$ is related to the physical $\eta'$ state.

The matrix $\Pi^a T^a$ should be compared with the Hamiltonian of a scanning electron micrograph (SEM) structure of \cite{cite-key}. Then, similar to \cite{cite-key}, one can find the modal field evolution which satisfies $i , d\mathbf{V}/dt = H \mathbf{V}$, where $\mathbf{V} = (a, b, c)^T$ is the modal column vector, and $a$, $b$, and $c$ are the field amplitudes in the amplifying, neutral, and lossy cavities, correspondingly.

 The non-Hermitian Hamiltonian is
\begin{gather} 
H  = \begin{pmatrix}
i g+\epsilon & \kappa & 0\\
\kappa & 0 & \kappa \\
0 & \kappa & - i g
\end{pmatrix}.
\end{gather}
So if the pion fields $\pi^+$ and $\pi^-$ are taken to be zero, the Kaon field can be associated with the coupling $\kappa$, and $\pi^0$ corresponds to the gain $g$. Then we can build the duality.

Note that the structures are $\pi^+ : u\bar{d}, \pi^0 : (u \bar{u} - d \bar{d})/\sqrt{2}, \pi^- : d \bar{u}, K^+ : u \bar{s}, K^0 : d \bar{s}, K^- : s \bar{u}$. Therefore, from this observation, we can deduce that \textbf{having the strange quark is directly connected to having an EP in the QCD system}. Note that the strange quark has isospin $I_3 = 0$, while the up and down quarks have isospin values of $+ \frac{1}{2}$ and $- \frac{1}{2}$, respectively. So we can speculate that \textbf{the presence of isospin might prevent the formation of an exceptional point}. The presence of the strange quark also increases the lifetime of particles because, due to the Cabibbo angle, the window for the decay of the strange quark into the up quark is very small, which causes similar effects on the structures of EPs as well. In the $s\bar{s}$ pairs that hadronize into $\Lambda$ and $\bar{\Lambda}$ hyperon pairs, all the spin is also carried by the strange quark.

Note also that \textbf{pions} are unstable, so it makes sense to simulate them with the gain $i g$ and loss $-i g$ sectors. Kaons, however, are more stable and can therefore be modeled by the sectors containing the coupling $\kappa$ and the eigenstate splittings in the Hamiltonian of the open quantum system.

With zero disturbance $(\epsilon = 0)$, the modal field $\mathbf{V}$ should have a harmonic dependence of the form $e^{-i \omega_n t}$. Then, by solving the matrix equation, the evolution of eigenfrequencies for quark bound states and the behavior of the system around exceptional points can be studied. Assuming all the $K$ fields equal $\kappa$, and using the relation $\det(H_0 - \omega_n I) = 0$, we get
\begin{gather}
\omega _n^3 -\left(4 \kappa ^2+\pi _0^2+2 \pi ^- \pi ^+\right) \omega _n-2 \sqrt{2} \left(\pi ^-+\pi ^+\right) \kappa ^2=0,
\end{gather}
which has three solutions as
\begin{gather}
\omega_0 \to -\frac{4 \sqrt[3]{3} \kappa ^2+\sqrt[3]{3} \left(\pi _0^2+2 \pi ^- \pi ^+\right)}{3^{2/3} \sqrt[3]{\frac{1}{6} \sqrt{5832 \left(\pi ^-+\pi ^+\right)^2 \kappa ^4-108 \left(4 \kappa ^2+\pi _0^2+2 \pi ^- \pi ^+\right){}^3}-9 \sqrt{2} \left(\pi ^-+\pi ^+\right) \kappa ^2}}+ \nonumber\\
+\frac{\left(\frac{1}{6} \sqrt{5832 \left(\pi ^-+\pi ^+\right)^2 \kappa ^4-108 \left(4 \kappa ^2+\pi _0^2+2 \pi ^- \pi ^+\right){}^3}-9 \sqrt{2} \left(\pi ^-+\pi ^+\right) \kappa ^2\right){}^{2/3}}{3^{2/3} \sqrt[3]{\frac{1}{6} \sqrt{5832 \left(\pi ^-+\pi ^+\right)^2 \kappa ^4-108 \left(4 \kappa ^2+\pi _0^2+2 \pi ^- \pi ^+\right){}^3}-9 \sqrt{2} \left(\pi ^-+\pi ^+\right) \kappa ^2}}, \nonumber\\
\omega_1 \to \frac{ \sqrt[3]{-3} \left(4 \kappa ^2+\pi _0^2+2 \pi ^- \pi ^+\right)}{3^{2/3} \sqrt[3]{\frac{1}{6} \sqrt{5832 \left(\pi ^-+\pi ^+\right)^2 \kappa ^4-108 \left(4 \kappa ^2+\pi _0^2+2 \pi ^- \pi ^+\right){}^3}-9 \sqrt{2} \left(\pi ^-+\pi ^+\right) \kappa ^2}}+ \nonumber\\
\frac{\left(1-i \sqrt{3}\right) \left(\frac{1}{6} \sqrt{5832 \left(\pi ^-+\pi ^+\right)^2 \kappa ^4-108 \left(4 \kappa ^2+\pi _0^2+2 \pi ^- \pi ^+\right){}^3}-9 \sqrt{2} \left(\pi ^-+\pi ^+\right) \kappa ^2\right){}^{2/3}}{2\ 3^{2/3} \sqrt[3]{\frac{1}{6} \sqrt{5832 \left(\pi ^-+\pi ^+\right)^2 \kappa ^4-108 \left(4 \kappa ^2+\pi _0^2+2 \pi ^- \pi ^+\right){}^3}-9 \sqrt{2} \left(\pi ^-+\pi ^+\right) \kappa ^2}}, \nonumber\\
\omega_{-1} \to \frac{\left(1+i \sqrt{3}\right) \left(\frac{1}{6} \sqrt{5832 \left(\pi ^-+\pi ^+\right)^2 \kappa ^4-108 \left(4 \kappa ^2+\pi _0^2+2 \pi ^- \pi ^+\right){}^3}-9 \sqrt{2} \left(\pi ^-+\pi ^+\right) \kappa ^2\right){}^{2/3}}{2\ 3^{2/3} \sqrt[3]{\frac{1}{6} \sqrt{5832 \left(\pi ^-+\pi ^+\right)^2 \kappa ^4-108 \left(4 \kappa ^2+\pi _0^2+2 \pi ^- \pi ^+\right){}^3}-9 \sqrt{2} \left(\pi ^-+\pi ^+\right) \kappa ^2}}+ \nonumber\\
\frac{\left(4 \kappa ^2+\pi _0^2+2 \pi ^- \pi ^+\right) \text{Root}\left[\text{$\#$1}^3+24\&,2,0\right]}{2\ 3^{2/3} \sqrt[3]{\frac{1}{6} \sqrt{5832 \left(\pi ^-+\pi ^+\right)^2 \kappa ^4-108 \left(4 \kappa ^2+\pi _0^2+2 \pi ^- \pi ^+\right){}^3}-9 \sqrt{2} \left(\pi ^-+\pi ^+\right) \kappa ^2}}.
\label{eq:solOmega}
\end{gather}

One of these solutions is real, while the other two are imaginary. Each of these fields could then be perturbed separately to produce novel behaviors and phase structures. The perturbation of each field above corresponds to perturbing (i) the gain cavity, (ii) the neutral cavity, or (iii) the gain and neutral cavities, as in \cite{cite-key}. Thus, in this more general case, the bifurcations in the eigenvalues can be obtained in a similar way.

If we now assume $\pi^+ = \pi^- =0$, we get
\begin{gather}
\omega _0\to 0 , \  \omega _1\to -\sqrt{4 \kappa ^2+\pi _0^2} , \ \omega _{-1}\to \sqrt{4 \kappa ^2+\pi _0^2},
\end{gather}
which is similar to the result of \cite{cite-key}.

\begin{figure}[ht!]
 \centering
  \includegraphics[width=5cm] {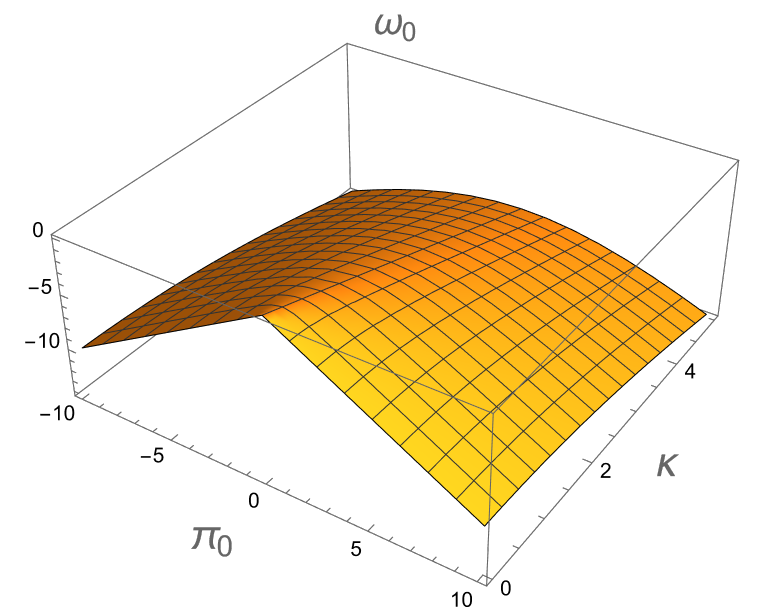} 
    \includegraphics[width=5cm] {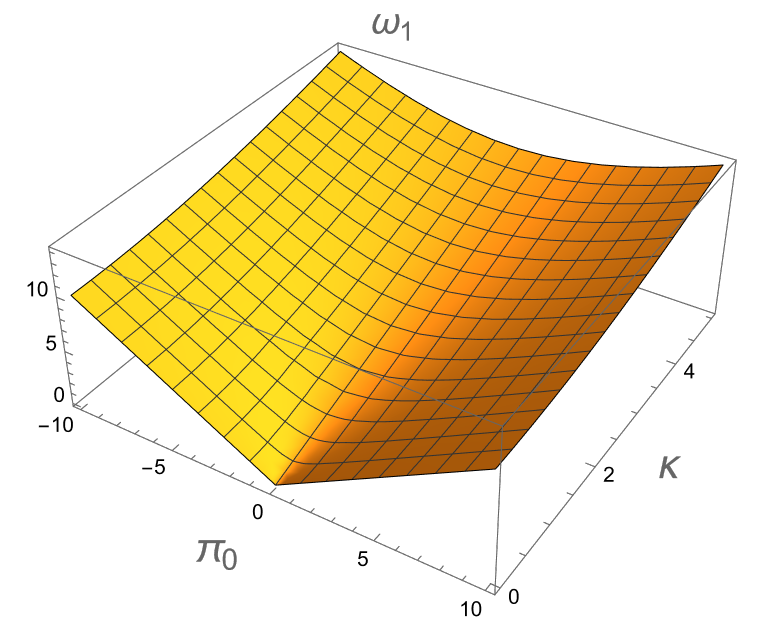} 
        \includegraphics[width=5.4cm] {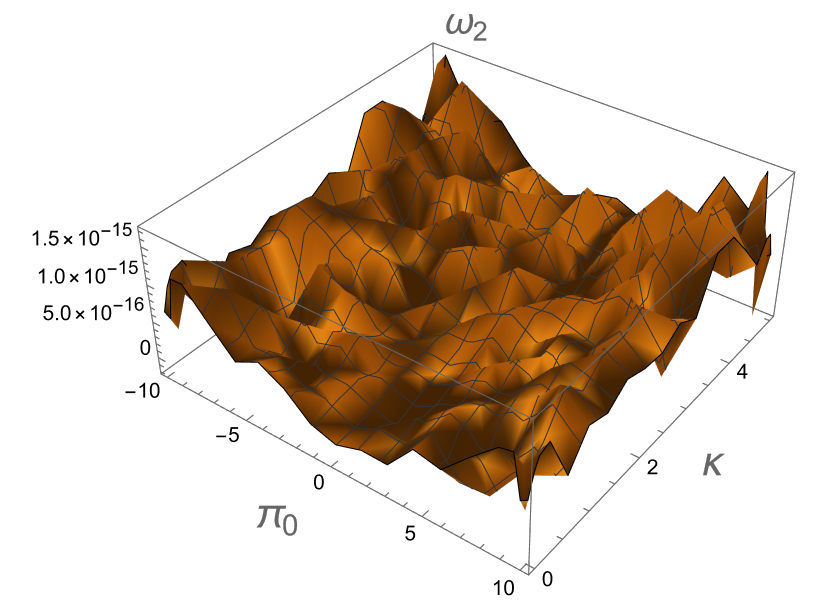} 
  \caption{The eigenfrequency solutions of \eqref{eq:solOmega} for the case of $\pi^+ = \pi^- = 0$.}
 \label{fig:Omegas}
\end{figure}

\begin{figure}[ht!]
 \centering
  \includegraphics[width=5cm] {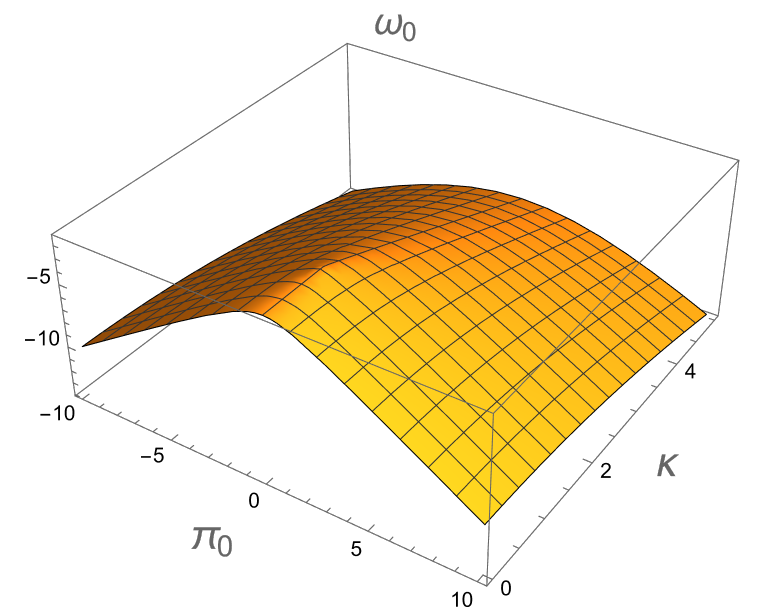} 
    \includegraphics[width=5cm] {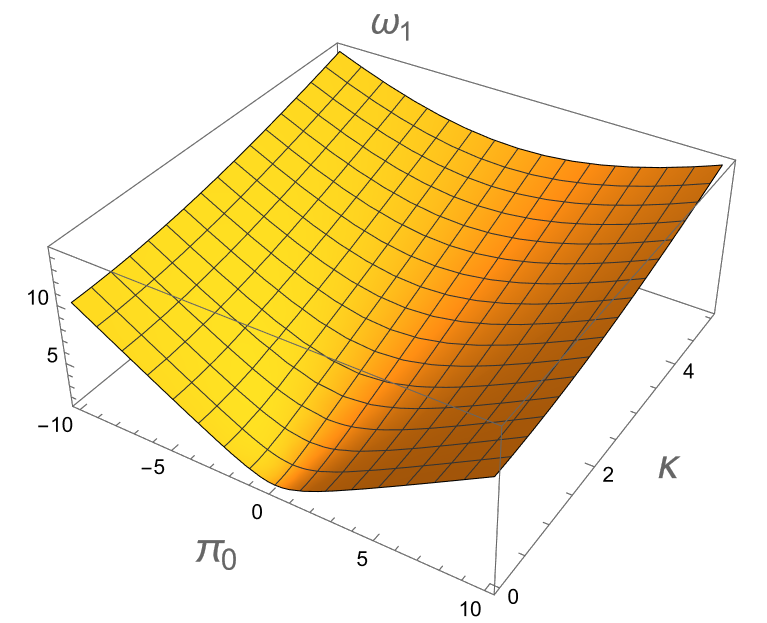} 
        \includegraphics[width=4.6cm] {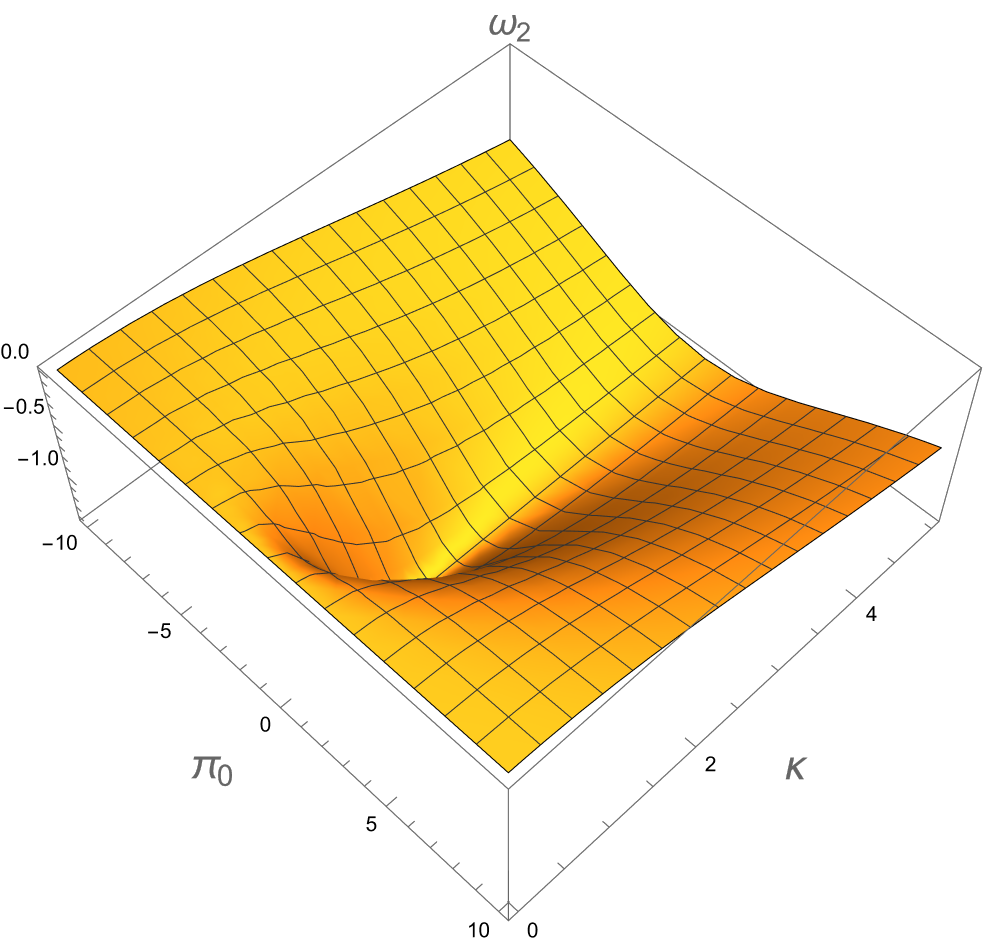} 
  \caption{The eigenfrequency solutions of \eqref{eq:solOmega} for the case of $\pi^+ = \pi^- = 1$. }
 \label{fig:Omegas1}
\end{figure}

\begin{figure}[ht!]
 \centering
  \includegraphics[width=5cm] {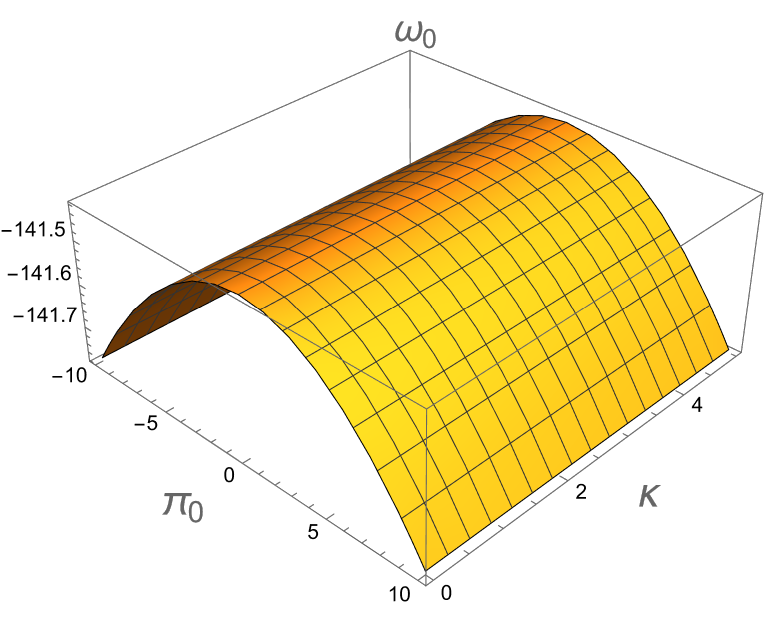} 
    \includegraphics[width=5cm] {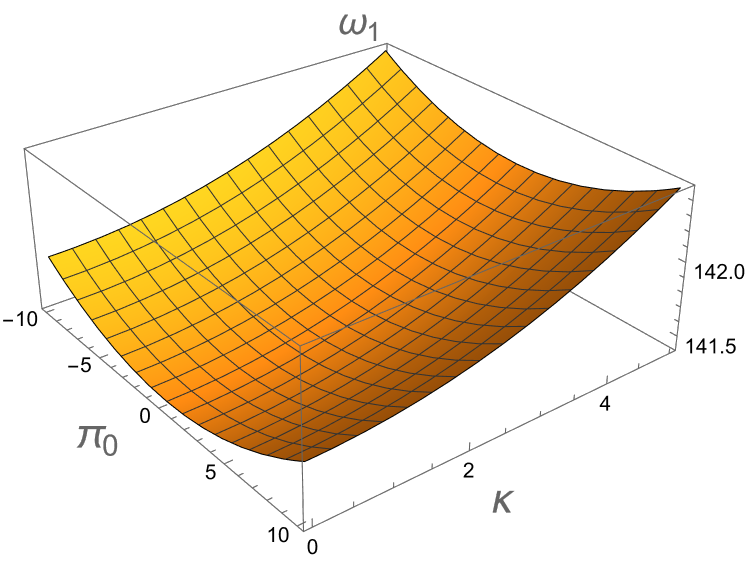} 
        \includegraphics[width=4.6cm] {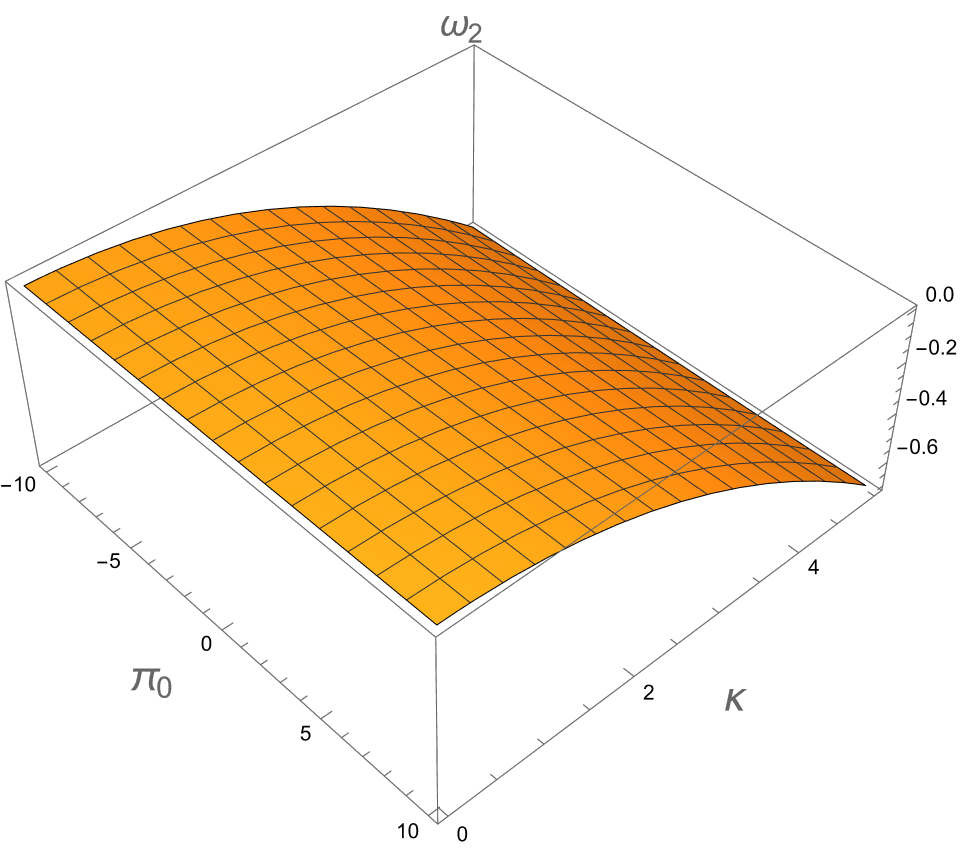} 
  \caption{The eigenfrequency solutions of \eqref{eq:solOmega} for the case of $\pi^+ = \pi^- = 100$. }
 \label{fig:Omegas2}
\end{figure}

We can then study these solutions (relation \ref{eq:solOmega}) further, as their behaviors are shown in figure \ref{fig:Omegas}. The case of $\pi^+ = \pi^- = 1$ is shown in figure \ref{fig:Omegas1}, and the case of $\pi^+ = \pi^- = 100$ is shown in figure \ref{fig:Omegas2}

First, one can note that increasing the coupling $\kappa$ clearly distinguishes the frequencies, as can be seen by comparing the behaviors of $\omega_0$ and $\omega_1$. This is similar to the result of \cite{cite-key}, where, for instance, in a two-ring system, the sensitivity enhancement follows the square root of the coupling factor. On the other hand, these results could point to new methods of measuring fields or couplings in QCD experiments using the frequency splittings.

It can be seen that by turning on the fields $\pi^+$ or $\pi^-$ and increasing them, the behaviors of all the eigenfrequencies, especially $\omega_2$, become smoother. In the limit of very large values of $\pi^+$ and $\pi^-$ relative to the couplings $\kappa$ and the field $\pi_0$, which acts as gain here, the eigenfrequencies become very smooth; yet the conclusion that increasing the gain distinguishes the eigenfrequencies remains valid.

To further study the bifurcation properties in our QCD system, we can perturb the Hamiltonian by modifying the gain cavity $\epsilon_1$, the neutral cavity $\epsilon_2$, and the loss cavity $\epsilon_3$. Therefore, the determinant of the following matrix should be set to zero.
\begin{gather}
\text{det}  \begin{pmatrix}
\begin{array}{ccc}
 \frac{-\omega _n+\kappa  \epsilon _1+\pi _0}{\sqrt{2} \kappa } & 1 & \frac{\pi ^+}{\kappa } \\
 1 & \frac{\kappa  \epsilon _2-\omega _n}{\sqrt{2} \kappa } & 1 \\
 \frac{\pi ^-}{\kappa } & 1 & \frac{-\omega _n+\kappa  \epsilon _3-\pi _0}{\sqrt{2} \kappa } \\
\end{array}
\end{pmatrix} =0,
\end{gather}
which leads to the equation
\begin{gather}
\omega _n^3 - \omega _n^2 \kappa (  \epsilon_1+  \epsilon_2+  \epsilon_3)+ \omega _n \left(\kappa ^2 \epsilon _1 \epsilon _2+\kappa ^2 \epsilon _1 \epsilon _3+\kappa ^2 \epsilon _2 \epsilon _3-\pi _0 \kappa  \epsilon _1+\pi _0 \kappa  \epsilon _3-\pi _0^2+2 \pi ^- \pi ^+\right) + \nonumber\\
\kappa ^3 \left(-\epsilon _2 \epsilon _3 \epsilon _1-2 \epsilon _1+2 \epsilon _3\right)+\kappa ^2 \left(\pi _0 \epsilon _1 \epsilon _2-\pi _0 \epsilon _2 \epsilon _3-4 \pi _0+2 \sqrt{2} \pi ^-+2 \sqrt{2} \pi ^+\right)+ \kappa  \epsilon _2 \left(\pi _0^2-2 \pi ^- \pi ^+\right) =0.
\label{eq:omegaequ}
\end{gather}

Note that in this case the fields $\pi^+$ and $\pi^-$ are still present, which can make the behavior a bit more complicated. Similar to \cite{cite-key}, we can neglect higher-order perturbations. Then, for the case of perturbing only the gain cavity, we assume $\epsilon_1 = \epsilon$, $\epsilon_2 = \epsilon_3 = 0$, and by taking $\omega \sim c_1 \epsilon^{1/3} + c_2 \epsilon^{2/3}$, we obtain the following three solutions
\begin{gather}
\omega_{-1} \sim  \sqrt[3]{-2} \kappa \epsilon^{1/3} -\frac{(-1)^{2/3}\pi_0  }{3 \sqrt[3]{2}}  \epsilon^{2/3}, \nonumber\\
\omega_{0} \sim  -\sqrt[3]{2} \kappa \epsilon^{1/3} -\frac{ \pi_0  }{3 \sqrt[3]{2}}  \epsilon^{2/3}, \nonumber\\
\omega_{1} \sim  -(-1)^{2/3} \sqrt[3]{2} \kappa \epsilon^{1/3}+  \frac{1}{3} \sqrt[3]{-\frac{1}{2}} \pi_0 \epsilon^{2/3}
\label{eq:perteq}
\end{gather}
 
\begin{figure}[ht!]
 \centering
  \includegraphics[width=4.5cm] {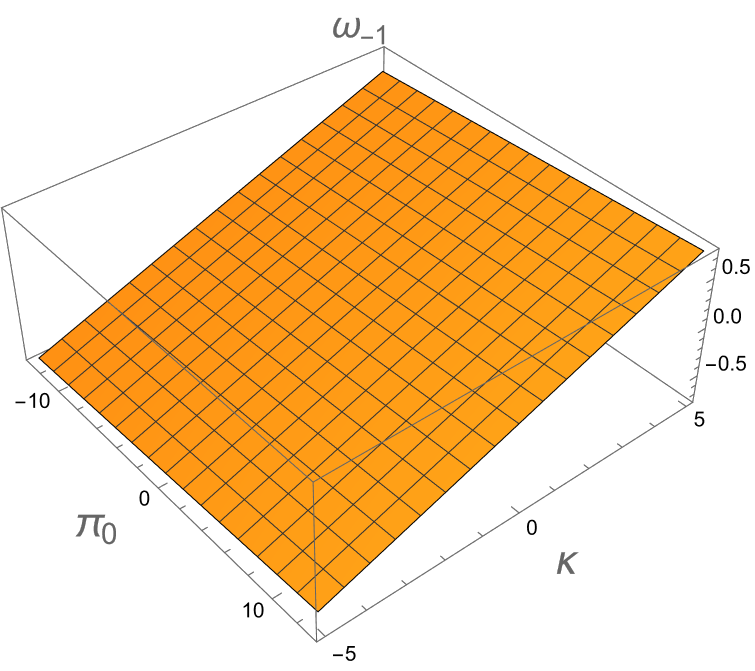} 
    \includegraphics[width=4.5cm] {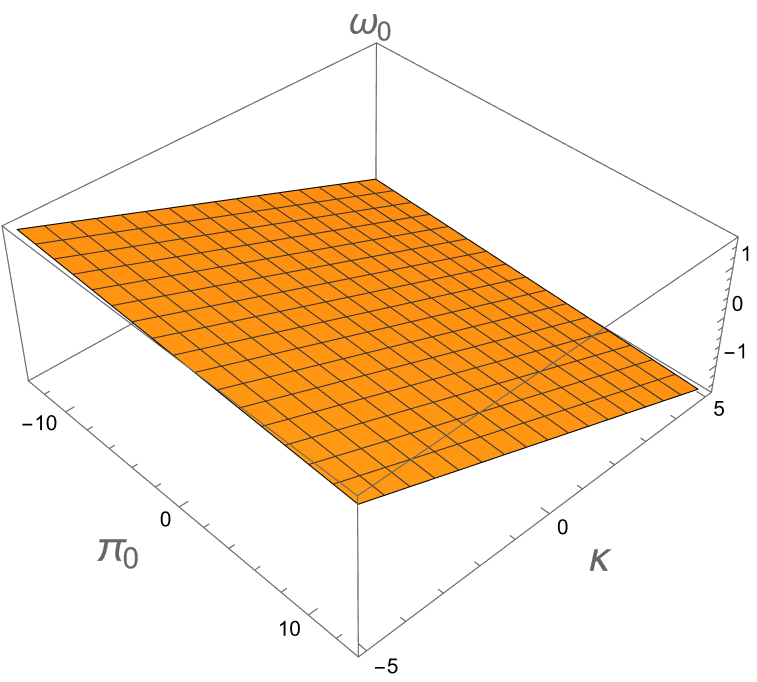} 
        \includegraphics[width=4.5cm] {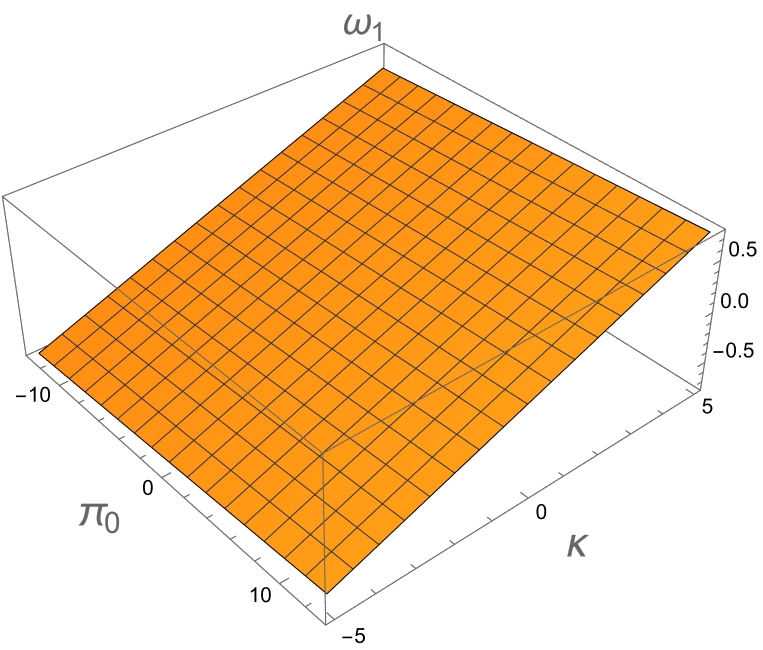} 
  \caption{The eigenfrequency solutions of \ref{eq:perteq} for the case $\epsilon = 0.01$. }
 \label{fig:pertOmegas}
\end{figure}

Again, it can be seen that by increasing the coupling $\kappa$, the splitting between the real parts of $\omega_{-1}$ and $\omega_0$ increases, as shown in figure \ref{fig:pertOmegas}. However, this is not the case for the real parts of $\omega_{-1}$ and $\omega_1$.

In the case of perturbing only the neutral cavity, similar to \cite{cite-key}, we should set $\epsilon_2 = \epsilon$ and $\epsilon_1 = \epsilon_3 = 0$, for which relation \ref{eq:omegaequ} leads to 
\begin{gather}
\omega _n^3-\kappa  \epsilon  \omega _n^2+\left(2 \pi ^- \pi ^+-\pi _0^2\right) \omega_n+\pi _0^2 \kappa  \epsilon -2 \pi ^- \pi ^+ \kappa  \epsilon + 2 \left(\sqrt{2} \left(\pi ^-+\pi ^+\right)-2 \pi _0\right) \kappa^2 =0. 
\end{gather}

Taking $\omega \sim c_1 \epsilon^{1/3} + c_2 \epsilon^{2/3}$, we then get
\begin{flalign}
-4 \pi _0 \kappa ^2+2 \sqrt{2} \pi ^+ \kappa ^2+2 \sqrt{2} \kappa ^2 \pi _n+ \left(2 \pi ^- \pi ^+ c_2-\pi _0^2 c_2\right) \epsilon ^{2/3}+ \left(2 \pi ^- \pi ^+ c_1-\pi _0^2 c_1\right) +  \sqrt[3]{\epsilon }+& \nonumber\\  \epsilon  \left(c_1^3+\pi _0^2 \kappa -2 \pi ^- \pi ^+ \kappa \right)+ 3 c_1^2 c_2 \epsilon ^{4/3}+ \epsilon ^{5/3} \left(3 c_1 c_2^2-c_1^2 \kappa \right) + \epsilon ^2 \left(c_2^3-2 c_1 c_2 \kappa \right)-c_2^2 \kappa  \epsilon ^{7/3}&=0,
\end{flalign}
which, due to the presence of the $\pi^{\pm}$ fields and unlike the case of \cite{cite-key}, has more complicated solutions. These fields change the behavior around exceptional points significantly. The case where the perturbations of both the gain and neutral cavities are turned on would be a combination of the two previous scenarios.

Now, for the sake of comparison with \cite{cite-key}, if we assume from the beginning the equation
\begin{gather}
\text{det}  \begin{pmatrix}
\begin{array}{ccc}
 \frac{-\omega _n+\kappa  \epsilon _1+\pi _0}{\sqrt{2} \kappa } & 1 & \frac{\pi ^+}{\kappa } \\
 1 & \frac{\kappa  \epsilon _2-\omega _n}{\sqrt{2} \kappa } & 1 \\
 \frac{\pi ^-}{\kappa } & 1 & \frac{-\omega _n+\kappa  \epsilon _3-\pi _0}{\sqrt{2} \kappa } \\
\end{array}
\end{pmatrix} =0,
\end{gather}
and then, by taking $\pi^+ = \pi^- = 0$, we can find some equations that are more similar to the results of \cite{cite-key}. Here, we assume $\tilde{\omega} = \omega_n / (\sqrt{2} \omega_n)$, but for simplicity we can again replace $\tilde{\omega}_n$ with $\omega_n$. So we get
\begin{gather}\label{eq:omegaep}
 \omega _n^3-\kappa  \omega _n^2 \left(\epsilon _1+\epsilon _2+\epsilon_3\right)-\omega_n(-\pi _0 \kappa  \epsilon _3+\pi _0 \kappa  \epsilon _1+\pi _0^2+2)+ \kappa  \left(\pi_0^2 \epsilon _2+\epsilon _1+\epsilon _3\right)=0.
\end{gather}

Then, for the case of perturbing the gain cavity, we take $\epsilon_1 = \epsilon$ and $\epsilon_2 = \epsilon_3 = 0$, and we then find
\begin{gather}
\text{eq}_g: \omega _n^3-\kappa  \epsilon  \omega _n^2-\omega _n \left(\pi _0 \kappa  \epsilon +\pi _0^2+2\right)+\kappa  \epsilon=0.
\end{gather}

Next, using a Newton-Puiseux series and taking $\omega_n = c_1 \epsilon^{1/3} + c_2 \epsilon^{2/3}$, we get
\begin{gather}
\epsilon ^{4/3} \left(3 c_1^2 c_2-\pi _0 c_1 \kappa \right)+\epsilon ^{5/3} \left(c_1^2 (-\kappa )-\pi _0 c_2 \kappa +3 c_2^2 c_1\right)+\left(-c_2^2\right) \kappa  \epsilon ^{7/3}-\left(\pi _0^2+2\right) c_2 \epsilon ^{2/3}+\nonumber\\
\epsilon ^2 \left(c_2^3-2 c_1 c_2 \kappa \right)+\epsilon  \left(c_1^3+\kappa \right)+\left(\pi _0^2+2\right) \left(-c_1\right) \sqrt[3]{\epsilon }=0,
\end{gather}
where again the presence of $\pi_0$ changes the behavior significantly, as the lower powers are related to this field and not to the higher-order perturbation terms.

\begin{figure}[ht!]
 \centering
  \includegraphics[width=6cm] {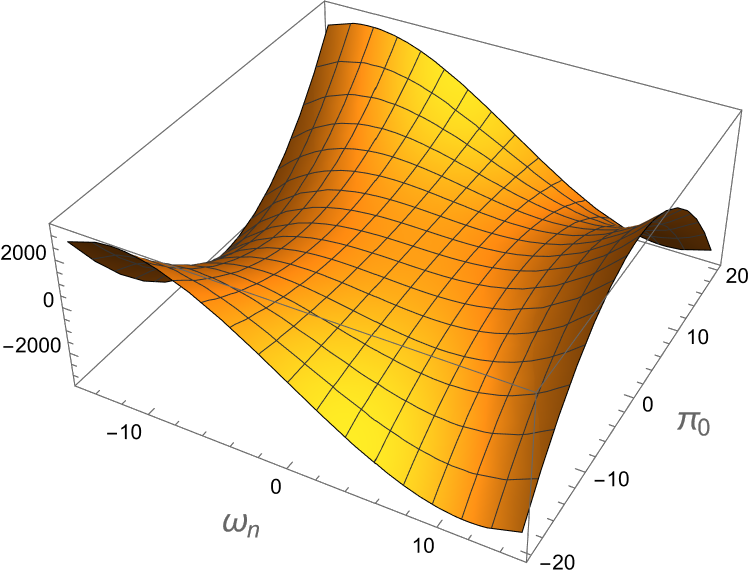}  \ \  \ \  \ 
    \includegraphics[width=6cm] {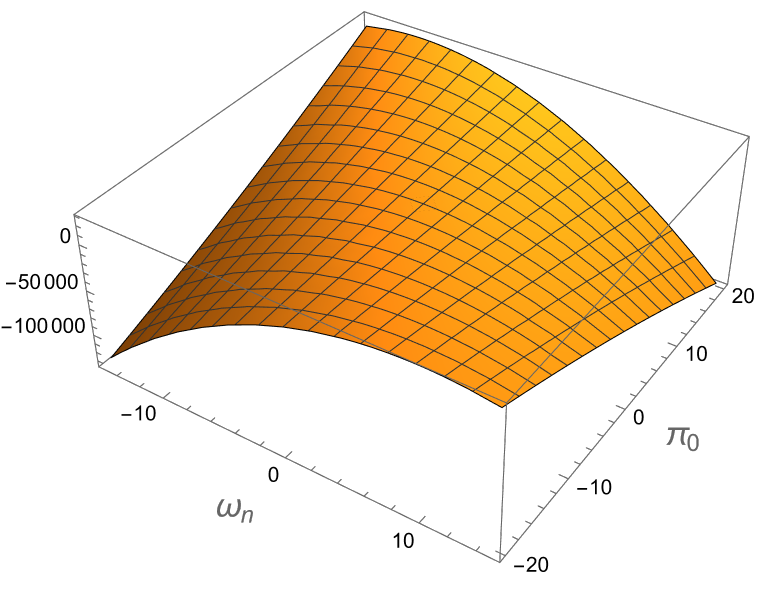} 
  \caption{The behavior of $\text{eq}_g$ versus $\omega_n$ and $\pi_0$ when perturbing the gain cavity only. Here, we set $\epsilon=0.01$ and $\kappa=1$ on the left and $\epsilon=0.5$ and $\kappa=500$ on the right. One can see that increasing $\kappa$ and $\epsilon$ makes this plot smoother. The increase in $\kappa$ has a bigger effect. }
 \label{fig:pomega}
\end{figure}

To study the equation $\text{eq}_g$ further, we can plot its behavior, which is shown in figure \ref{fig:pomega}. One can see that increasing $\kappa$ and $\epsilon$ can have a detuning effect.

Then, in the case of neutral cavity perturbation, where $\epsilon_2 = \epsilon$ and $\epsilon_1 = \epsilon_3 = 0$, equation \ref{eq:omegaep} becomes
\begin{gather}
\text{eq}_n: \omega _n^3-\left(\pi _0^2+2\right) \omega _n-\kappa  \epsilon  \omega _n^2+\pi _0^2 \kappa  \epsilon=0.
\end{gather}

 \begin{figure}[ht!]
 \centering
  \includegraphics[width=6cm] {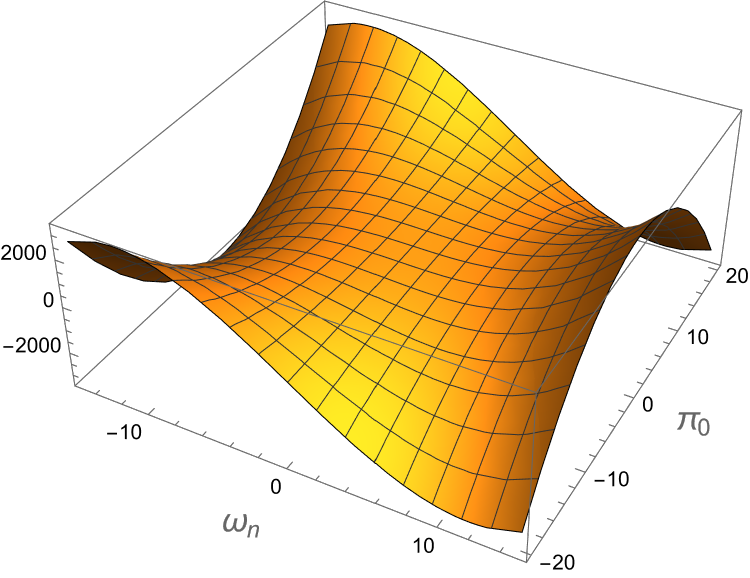}  \ \  \ \  \ 
    \includegraphics[width=6cm] {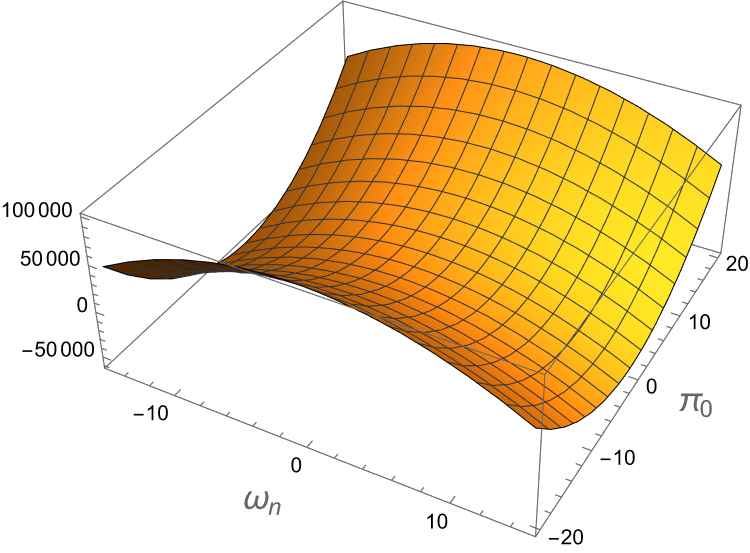} 
  \caption{The behavior of $\text{eq}_n$ versus $\omega_n$ and $\pi_0$ in the case of perturbing the neutral cavity only. Here, we set $\epsilon = 0.01$ and $\kappa = 1$ on the left and $\epsilon = 0.5$ and $\kappa = 500$ on the right. One can see that increasing $\kappa$ and $\epsilon$ makes this plot smoother. The increase of $\kappa$ has a bigger effect here as well.}
 \label{fig:p2omega}
\end{figure}

Comparing figures \ref{fig:pomega} and \ref{fig:p2omega}, one can see that for small $\epsilon$ and $\kappa$, the behavior is very similar, but increasing these two parameters makes the behavior distinct. When increasing the perturbation $\epsilon$ and coupling $\kappa$ in the gain-cavity case, the tuning setups become more intricate, but it remains easy in the neutral-cavity case, as there is still a minimum there. This, of course, matches our intuition as well.

In the case of perturbing both the gain and neutral cavities, where $\epsilon_1 = \epsilon$ and $\epsilon_2 = \alpha \epsilon$ ($\alpha < 1$), the equation \ref{eq:omegaep} becomes
\begin{gather}
\text{eq}_{p, n}: \omega _n^3 -\omega _n^2 (\alpha  \kappa  \epsilon +\kappa  \epsilon )-\omega _n \left(\pi _0 \kappa  \epsilon +\pi _0^2+2\right)+\kappa  \epsilon+ \pi _0^2 \alpha  \kappa  \epsilon=0.
\end{gather}

Note that this is similar to the argument of \cite{Liu:2024qnh}, which mentions that the quantum gravity fluctuations of the bulk extremal $\text{AdS}_5$ are the source of confinement in super-Yang-Mills theory.

 \begin{figure}[ht!]
 \centering
  \includegraphics[width=5.5cm] {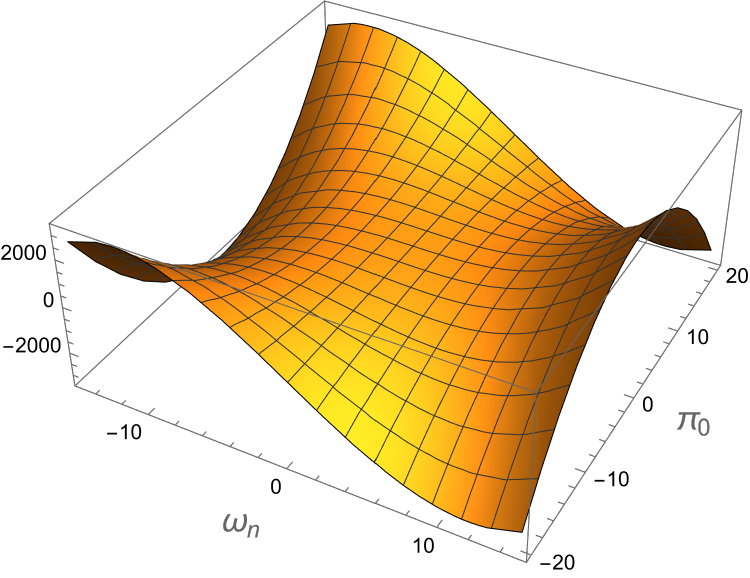}  \ \  \ \  \ 
    \includegraphics[width=5.5cm] {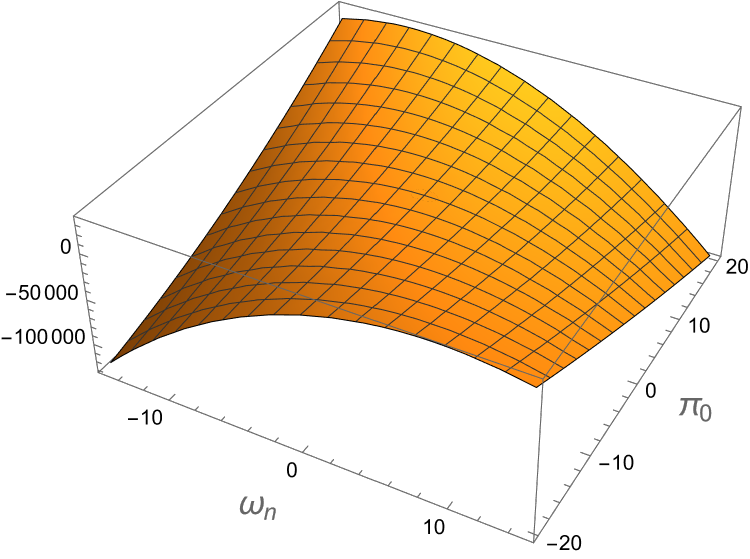} 
        \includegraphics[width=5.5cm] {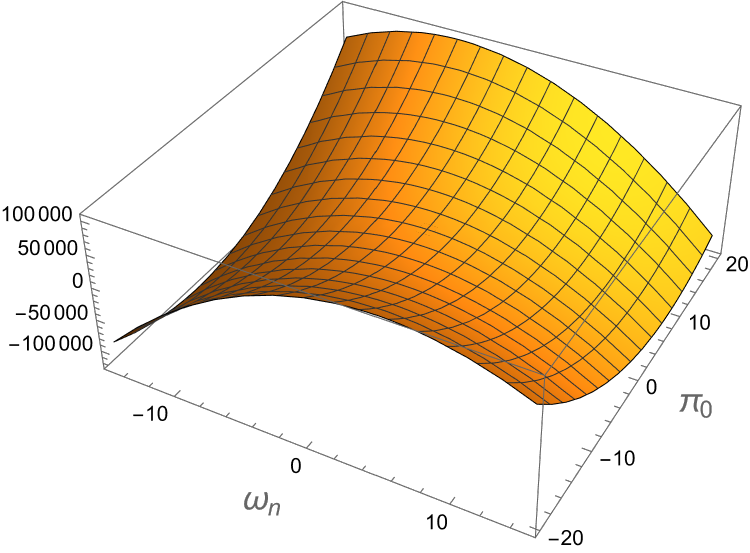} 
  \caption{The behavior of $\text{eq}_{g,n}$ versus $\omega_n$ and $\pi_0$, in perturbing the neutral cavity only, is shown for $\epsilon=0.01$, $\alpha=0.1$, and $\kappa=1$ on the left; $\epsilon=0.5$, $\alpha=0.1$, and $\kappa=500$ in the middle; and $\epsilon=0.5$, $\alpha=0.9$, and $\kappa=500$ on the right. One can see that increasing $\kappa$ and $\epsilon$ makes the plot smoother, with $\kappa$ having the larger effect.}
 \label{fig:p2omega}
\end{figure}

One can see that increasing $\alpha$ makes the behavior of the system closer to the case of perturbing the neutral cavity, while decreasing it makes the behavior more similar to the gain cavity case, which makes sense since $\epsilon_2$ here corresponds to perturbing the neutral cavity. The same procedure can be applied to the neutral cavity and to the neutral + gain cavities, yielding results comparable to \cite{cite-key}. For instance, the presence of $\epsilon_2$ also reduces the sensitivity.

So, the gain and loss processes in photonic molecules can be modeled by the Lagrangian of the weak interaction and QCD models. Insights from exceptional points can provide further interesting information about the eigenfrequencies of quark-bound states and suggest new ways of studying fundamental particles.

In some cases, the coupling $\kappa$ can play the role of a virtual loss \cite{Hodaei:2014iws}. It would be interesting to investigate further why the corresponding Kaon could play such a role as well. Our proposition is that Kaons have a distinguished role in CP violation and parity-symmetry breaking, and the presence of a strange quark in their bound state can cause exotic effects, such as neutral particle oscillations. Also, they have two weak eigenstates: one long-lived neutral Kaon ($K_L$), which decays into three pions, and one short-lived neutral Kaon ($K_S$), which decays into two pions. Due to oscillation and the time dependence of the decay, there is a mass splitting between $K_S$ and $K_L$, which corresponds to \textbf{eigenstate frequency splitting around EPs}.

It is worth noting that the Einstein-Podolsky-Rosen \cite{PhysRev.47.777} experimental setup has been constructed using Kaons and photons. The test of entanglement has been performed both with photons, using time-energy correlations through optical fibers, and with massive K-mesons, using strangeness correlations. The similarities and differences between these two setups were discussed in \cite{Gisin_2001}. Actually, in \cite{Gisin_2001}, it was shown specifically that there are similarities between the decays related to Kaons and the polarization-dependent losses in optical fibers. They showed that both the strangeness mixing in neutral Kaons and the birefringence in the optical fibers cause rotations of Poincaré vectors of the quantum states of Kaons and photons, similar to the behaviors around EPs. Therefore, they also demonstrated the similarities between these two systems. In optical fibers, polarization-dependent losses (PDL), and, on the other hand, Kaon decay, both produce similar stretches of the state space \cite{Gisin_2001}.
 
The PDL can be formulated using the evolution operator $T$ as \cite{Gisin_2001}
 \begin{gather}
 T =  \begin{pmatrix}
\sqrt{T_{\text{max}} } & 0 \\
0 & \sqrt{T_{\text{min}} }  
\end{pmatrix} =
  \begin{pmatrix}
e^{-\frac{\alpha_{\text{max}  } z }{2} } & 0 \\
0 & e^{-\frac{\alpha_{\text{min}  }z }{2} }
\end{pmatrix}, 
\end{gather}
as the transmission coefficients (for the intensity) are related to the fiber length $z$.

On the other hand, the time evolution operator $U(t)$ for Kaons can be written as
\begin{gather}
U(t) =  \begin{pmatrix}
e^{-(i m_S + \frac{\gamma_S}{2} )t} & 0 \\
0 & e^{-(i m_L + \frac{\gamma_L}{2} )t} \\
\end{pmatrix}
= e^{-i \frac{m_S + m_L}{2} t} . \ e^{-i \frac{m_S-m_L}{2} t \sigma_3 } . \begin{pmatrix}
e^{\gamma_S t /2 } & 0 \\
0 & e^{\gamma_L t /2} \\
\end{pmatrix},
\end{gather}
where $m_L$ ($m_S$) is the mass of the long (short) eigenstate, and $\gamma_L$ ($\gamma_S$) is the width of the long (short) eigenstate.  
In the above relation, the first term is just a global phase factor, the second term represents the strangeness mixing, which corresponds to birefringence in optics, and the third term is related to the decay of unstable Kaons, where $K_S$ decays faster than $K_L$. Thus, on a Poincaré sphere, the precession would be toward $K_L$ as it remains longer.

Therefore, one could imagine that, in addition to photons, Kaons can also be implemented in quantum computers, as they are massive and can be detected and controlled more easily. The problem with experiments involving Kaons, however, is their intrinsic decay, which the experimenter cannot control, while the PDL in photon systems can be controlled. However, the strangeness mixing of Kaons, which is again intrinsic and analogous to optical birefringence, has the advantage that it allows the experimenter to effectively rotate the analyzer just by delaying the measurement \cite{Gisin_2001}. This effect can also be useful in designing quantum computers.

Also, in a recent interesting work from the STAR collaboration \cite{STAR:2025njp}, using the Relativistic Heavy-Ion Collider (RHIC) and spin entanglement, QCD confinement has been probed. Their strange quark-antiquark pairs are initially spin-entangled and then undergo quark confinement to form $\Lambda$ hyperons. When the angles between the hyperon pairs increase—which, due to decoherence of the quantum system, could actually prevent the formation of exceptional points—the correlation vanishes. Before transitioning from quark confinement to final-state hadrons, the quark condensate contains similar numbers of virtual up $(u)$, down $(d)$, and strange $(s)$ quark pairs, leading to degeneracy and the formation of exceptional points in the system. Also, the quantum number here is $J^{PC} = 0^{++}$, making the spins of the pairs all parallel and keeping them in spin-triplet states, again consistent with the formation of EPs.

The fact that they could introduce a method using the spin correlation of $\Lambda \bar{\Lambda}$ hyperon pairs and observe the degree of (de)coherence of the entangled $s\bar{s}$ pairs transitioning into hadrons-showing that entanglement entropy can probe confinement and QCD processes-again confirms our conjecture about the interplay between entanglement patterns, exceptional points in open quantum systems, and QCD. It would then be interesting to classify EPs based on each model of confining geometries, i.e., hard-wall, soft-wall, AdS-soliton, Sakai-Sugimoto, Witten model, etc.

Also, another specific and interesting regime of QCD is the “lasing regime.” The holographic dual of this regime in the AdS/QCD setup could also be constructed. For instance, soft-wall models with a background dilaton field and linear instability ($\text{Im}>0$), mode selection, gain saturation, and then mapping to the microring toy model could simulate such behaviors for us, including enhanced coherence or collective excitations \cite{Cao_2022}. The pion dynamics with chiral symmetry breaking could also describe collective behavior in lasing regimes \cite{PhysRevA.98.063837}. The spectral evolution of the transitions between the lasing and non-lasing regimes, which can be tracked by spectrometers, can then be compared with each other.

\section{Exceptional point and timelike entanglement entropy}\label{sec:EPtime}

Recently, there has been significant interest in defining and using entanglement measures in time \cite{Doi:2023zaf, Milekhin:2025ycm, Xu:2024yvf}, which quantifies quantum correlations across timelike boundaries. In particular, \cite{Glorioso:2024xan} defined mutual information in time, \cite{Nunez:2025gxq} and \cite{Ghodrati:2023uef} used it for holographic confining geometries, \cite{Guo:2025pru} used it for probing the interior of black holes, \cite{Ikeda:2025gju} applied it in quantum energy teleportation, among other works.

The new timelike entanglement entropy (TEE), being complex-valued, can serve as a powerful tool to probe exceptional points in non-Hermitian systems such as our photonic setup. Additionally, the imaginary part of pseudo-entropy is linked to the emergence of the time coordinate in holography and cosmology.

TEE is defined via the modular Hamiltonian or replica trick on a timelike interval $\lbrack t_1, t_2 \rbrack$ in the boundary theory, often yielding a complex value
\begin{gather}
S_T= - \text{Tr} (\rho_T \log \rho_T) = S_T^R + i S_T^I,
\end{gather}
where $\rho_T$ is the reduced density matrix for the timelike subsystem. The real part, $S_T^R$, captures the ``entanglement-like" temporal correlations, while the imaginary part, $S_T^I$, is related to commutators of twist operators or emergent time structures, signaling dissipation or out-of-time-order correlations in open quantum systems. In holography, TEE is computed via the Ryu-Takayanagi prescription generalized to timelike strips. There, the extremal surfaces in the bulk are anchored to a timelike boundary interval, often requiring analytic continuation to complex coordinates for well-defined saddles. Recent works \cite{Doi:2023zaf,Narayan:2023ebn, Xu:2024yvf,Nunez:2025ppd} show that multiple such surfaces exist, with the physical one selected by dominance principles, and the imaginary part tied to bulk horizon effects or pseudo-entropies.

As noted EPs mark a transition where the non-Hermitian Hamiltonian's spectrum shifts from real $\mathcal{PT}$-unbroken and stable modes to complex conjugate pairs with $\mathcal{PT}$-broken and amplification or decay. This manifests in temporal dynamics as exponential growth or decay, which TEE is uniquely suited to capture. Note that in the $\mathcal{PT}$-broken phase, our photonic system's modes have $\text{Im} \ne 0$, leading to non-unitary time evolution. The imaginary part of TEE, which can probe these temporal commutators, is given by
\begin{gather}
S_T^I \propto \int dt \ \langle [ A(t), B(0) ] \rangle,
\end{gather}
where $A$ and $B$ are local operators and $S_T^I$ is sensitive to dissipation and non-Hermitian effects.

Then, at $\Gamma = \Gamma_c$, TEE exhibits a branch point or cusp, where the real part shows a temporal area-law behavior, while the imaginary part branches into complex conjugates, which mirrors the coalescence of eigenvalues. Unlike the FGT sum rule or OTOCs, TEE is local in time and complex from the start, making it ideal for open quantum systems. For $\text{AdS}_{d+1}$, the extremal surface satisfies the relation,
\begin{gather}
\frac{d}{dz} \left ( \frac{f(z) x'(z)}{\sqrt{1-f(z) x'(z)^2} } \right) =0,
\end{gather}
where by continuing into the complex time direction, the solutions of this equation would merge or bifurcate, as for instance, two complex saddles could coalesce. The coalescence condition is
\begin{gather}
\frac{\partial S_{\text{eff} } }{\partial z_*}=0, \ \ \ \ \ \frac{\partial^2 S_{\text{eff} } }{\partial z_*^2}=0.
\end{gather}
It is the exceptional point condition for the effective potential $S_{\text{eff}}$. Thus, the branch point in the complex extremal surface manifold corresponds to an EP in the spectral problem.

In photonics, it can be measured via interferometric setups tracking temporal correlations in coupled resonators. For instance, in works such as \cite{cite-keyHales2025}, to experimentally witness and control entanglement in light-driven quantum materials via ultrafast spectroscopic measurements using inelastic X-ray scattering, a systematic approach to quantify the time-dependent quantum Fisher information has been proposed.

In the holographic model we discussed ($\text{AdS}_5$ with complex scalar $\phi$ sourced by $s$ for gain/loss), TEE for a timelike strip $\lbrack t_L, t_R \rbrack$ can be computed as
\begin{gather}
S_T = S_T^R + i S_T^I, \quad S_T^R \sim \frac{\beta \log \beta}{\epsilon^2}, \quad S_T^I \sim \beta \ \text{Im}  \omega_{\text{QNM}} \sim \beta \sqrt{\Gamma- \Gamma_c } \Theta(\Gamma-\Gamma_c),
\end{gather}
where $\beta = t_R - t_L$ is the boundary length, $s = \gamma + i \  \Gamma$ is the source, $\epsilon$ is a UV cutoff, and $\Theta$ is the step function signaling $\mathcal{PT}$ breaking. So when $\Gamma < \Gamma_c$, $S_T^I \approx 0$, as there would be no dissipation in this case. At $\Gamma = \Gamma_c$, we get an EP, as there is coalescence where multiple bulk surfaces compete and $S_T$ develops a $\sqrt{\Gamma-\Gamma_c}$ branch point. When $\Gamma > \Gamma_c$, we get a $\mathcal{PT}$-broken phase where $S_T^I \ne 0 \sim |\text{Im}(\omega)| \beta$, where for positive imaginary part there is amplification and for negative part there is decay. There is a discontinuity in the imaginary part which signals the transition.

For an inhomogeneous lattice with the scalar field $\phi(z,x) = e^{ikx} \tilde{\phi} (z,x)$, with periodic source $s(x+3\ell) = s(x)$, TEE is
\begin{gather}
S_T(k) \sim \frac{\beta \log \beta}{\epsilon^2} + i \beta \sqrt{\Gamma^2 - g^2_{\text{eff} }(k) } \ \Theta\Big(\Gamma - g_{\text{eff} }(k)  \Big),
\end{gather}
where $g_{\text{eff}} = 2g \cos(k \ell) $ and the dispersion relation at the EP is
\begin{gather}
\omega(k) \approx \omega_0 + 2 g \cos(k \ell) + i \sqrt{\Gamma^2 - g_{\text{eff}}^2 (k) },
\end{gather}
with coalescence at $k_* = \pi/(2\ell)$ and $\Gamma_c = g \sqrt{3}$.

This interesting connection could then be extended. In \cite{Milekhin:2025ycm}, a matrix called $T_{AB}$, which is the analogue of the density matrix, is defined. The trace over a certain Hilbert space, i.e., $\text{Tr} T_{AB}^n$, and the analytical continuation of $\text{Tr}(\rho_{AB}^n)$, can be used to compute time-separated correlation functions of twist operators. We can assume that this $T_{AB}$ defines an open quantum matrix with gain and loss, similar to the non-Hermitian systems, and the singularities discussed in \cite{Milekhin:2025ycm} would correspond to the exceptional points of this matrix.

Here $A$ and $B$ define the entire system before and after the evolution and can be written as
\begin{gather}
T_{AB}= J ( \rho \otimes 1_B),
\end{gather}
where the operator ordering is important, making this matrix non-Hermitian.

The matrix $T_{AB}$, when contracted with the operators $\mathcal{O}_A$ and $\mathcal{O}_B$, builds the Wightman correlation function as
\begin{gather}
\text{Tr} ( T_{AB} ( \mathcal{O}_A \otimes \mathbf{1}_B ) ( \mathbf{1}_A \otimes \mathcal{O}_B )) = \langle \mathcal{O}_A(0) \mathcal{O}_B (t) \rangle,
\end{gather}
and $T^\dagger_{AB}$ produces the opposite ordering as
\begin{gather}
\text{Tr} ( T^\dagger_{AB} \mathcal{O}_A \otimes \mathcal{O}_B) = \langle \mathcal{O}_B (t) \mathcal{O}_A (0) \rangle.
\end{gather}

In \cite{Milekhin:2025ycm}, it has been shown that, similar to conventional density matrices which do not have eigenvalues larger than one, the singular values of the matrix $T_{AB}$ are no larger than one, and all eigenvalues lie inside the unit disk:
\begin{gather}
||T_{AB}|| \le 1.
\end{gather}
Therefore, in this scenario, the probability of having many exceptional points is very high. Additionally, it has been shown that $\text{Tr} (T_{AB} T^\dagger_{AB})$ is very sensitive to lattice effects, similar to the case of higher-order exceptional points. Also, as pointed out in \cite{Milekhin:2025ycm}, the lattice value of $\text{Tr} (T_{AB} T^\dagger_{AB})$ is much larger than $\text{Tr} (T^2_{AB})$ by a factor of $\sim e^{1/a}$, which becomes exponentially large as the lattice spacing $a$ becomes very small. This is similar to the bifurcation properties discussed in \cite{cite-key}, where the sensitivity is enhanced at higher-order exceptional points, as
\begin{gather}
\omega_n = e^{-i (2n+1) \pi/3} \kappa^{2/3} \epsilon^{1/3} + \frac{i \sqrt{2} }{3} e^{i(2n+1) \pi/3} \kappa^{1/3} \epsilon^{2/3}, \ \ \ \ \ n \in \lbrace -1,0,1 \rbrace, \nonumber\\
\Delta \omega_{\text{EP}3} = 3 \kappa^{2/3} \sqrt[3] \epsilon /2.
\end{gather}

In addition, in \cite{Milekhin:2025ycm}, it has been suggested that the entanglement p-imagitivity, $|| T_{AB} - T^\dagger_{AB}||_2$, should obey the Lieb-Robinson bounds, where the precise formulation of the bound depends on locality. In the language of exceptional points, this actually would be related to $\epsilon$. Also, the lattice connectivity \cite{Hastings:2005pr, Nachtergaele_2006} would correspond to the coupling constant $\kappa$ in the EP terms. Therefore, it can be seen that photonic setups for studying EPs could also be useful in exploring spacetime entanglement and quantum correlations. Other phenomena, such as dichroism, Dirac points, or photonic crystals, could have interesting relations with entanglement in time. Additionally, as mentioned in \cite{Milekhin:2025ycm}, the tensor $T_{AB}$ could be viewed as a kind of Kirkwood-Dirac distribution, $Q(x,p)$, which is a quasiprobability distribution and can be written as
\begin{gather}
Q(x,p) = \frac{1}{2\pi \hbar} \bra{x} \hat{\rho} \ket{p} \langle p | x \rangle.
\end{gather} 

The biorthogonal density matrix, $\rho_{ij}^{KD} = \langle \phi_i^L | \rho | \phi_j^R \rangle$, is actually a Kirkwood-Dirac–type distribution.

In general, this $Q(a,b)$ is complex and can be used to define the complex timelike entanglement entropy as
\begin{gather}
S= - \sum_{a,b} Q(a,b) \log Q(a,b).
\end{gather}

On the other hand, in non-Hermitian systems, such as our microrings, the eigenmodes are biorthogonal, i.e., $\langle \phi_i^L | \phi_j^R \rangle = \delta_{ij}$, and the density or the correlation between left and right eigenstates is complex, similar to the KD distribution. At exceptional points, left and right eigenvectors coalesce, and the KD distribution becomes singular or highly non-classical. Thus, the KD distribution can be seen as a natural analytic tool to describe EPs. Through the KD distribution, we can also see the connections between EPs and TEE.

\begin{figure}[ht!]
 \centering
  \includegraphics[width=9cm] {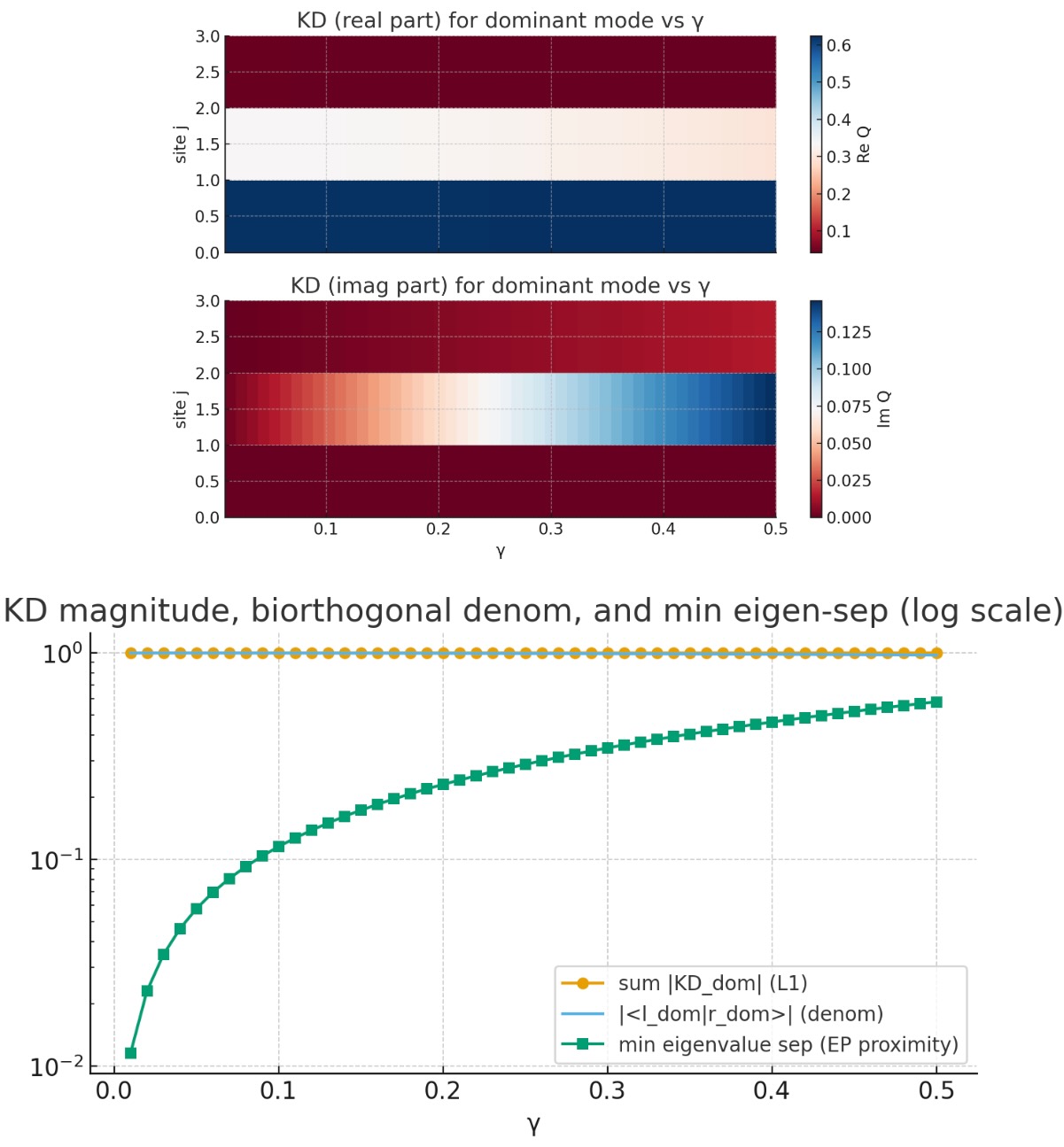} 
    \includegraphics[width=7.4cm] {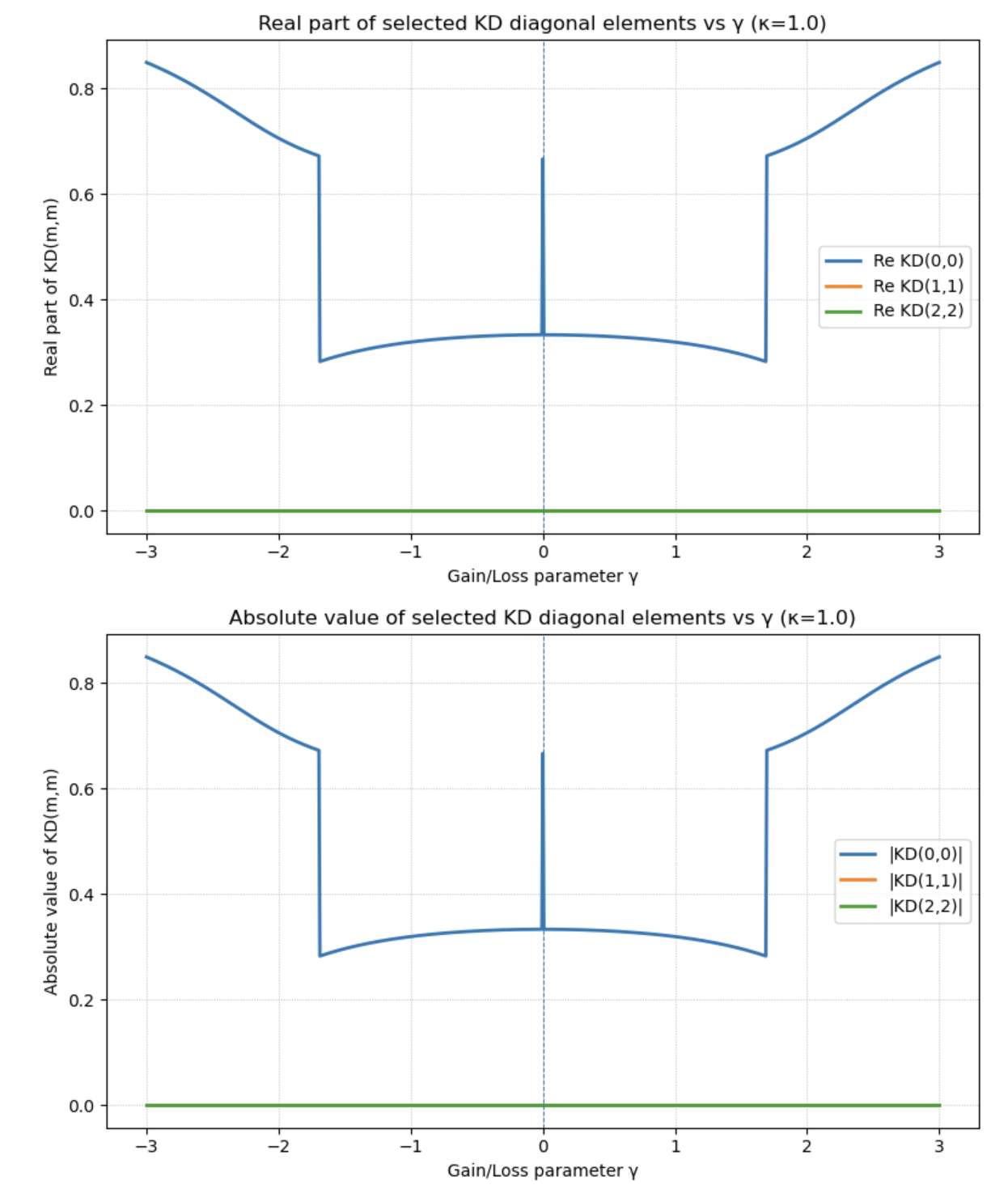} 
  \caption{The behavior of the KD distribution versus the gain/loss parameter $\gamma$ in a coupled ternary system of microrings. }
 \label{fig:KDmicroring}
\end{figure}

For a three-coupled ternary system of microrings, the behavior of the imaginary and real parts of the KD distribution is shown in figure \ref{fig:KDmicroring}. The EP occurs at $\gamma=0$, and the increase in KD magnitude versus $\gamma$ and the separation of modes are apparent. 

In \cite{Milekhin:2025ycm}, the following inequality has been proposed,
\begin{gather}
|| T_{AB} - T^\dagger_{AB}||_2 \le \text{dim} \mathcal{H}_A C |A| |B|  e^{- \mu d(A,B)} \left( e^{v |t|}-1 \right),
\end{gather}
where $|| T_{AB} - T^\dagger_{AB}||_2$ is the 2-imagitivity, and $C$, $v$, and $\mu$ only depend on the structure of the lattice and the Hamiltonian. The corresponding figure 2 of \cite{Milekhin:2025ycm} for the Ising chain would be related to Fig.~4b of \cite{cite-key}, as the intensity is related to $|\langle [ Y_1, Y_6 ] \rangle|$, where the Pauli operator $Y$ acts on the first and sixth qubits at different times. The peaks in both cases are related to the exceptional points.

Also, based on Lieb-Robinson bounds, both the commutator and $|| T - T^\dagger ||_2$ of \cite{Milekhin:2025ycm}, and the intensity in \cite{cite-key}, start increasing from zero after a long elapsed ``time'' in a similar pattern. In addition, as the correlators of examples such as free fermions obey Wick's theorem, the reduced density matrices are Gaussian. Similarly, $T$ is a Gaussian operator since the time-separated correlation functions also obey Wick's theorem. This is also true in the case of a ternary parity-time-symmetric system at a third-order exceptional point, as shown in figure 4b of \cite{cite-key}. Furthermore, in the microcavities, the above inequality implies that imagitivity decays exponentially with spatial separation, but in the parameter space $(n, \chi)$, and tuning toward EPs would increase imagitivity.

The microscopic definition of $T$ can be extracted from the two-point correlation functions. In \cite{Time-asymmetric}, the time-asymmetric loop around an exceptional point has been studied and in \cite{MissingDimension}, the revealing of the missing dimension of the Hilbert space, known as the Jordan vector, at the exceptional point has been studied. There, the radiation field becomes fully decoupled from the eigenstates of the environment with opposite chirality compared to the coalesced eigenstate, which leads to vortex emissions in the far field and anomalous wave-matter interactions. This is similar to probing inside of a black hole or hitting the end-wall of confining geometries in the holographic dual of timelike entanglement entropy.

As pointed out in \cite{Milekhin:2025ycm}, the singular values related to matrix product operators (MPOs) and the matrix $T$, which is the analogue of a density matrix for timelike intervals, can be examined by partitioning the system into halves. By increasing the number of time-separated intervals while keeping each interval small to the point of a single qubit, one can characterize the singular values via the time-separated correlations and analyze the bond dimension $\chi$. The photonic exceptional points then can be used to measure the strength of time-separated correlations.

\begin{figure}[ht!]
 \centering
  \includegraphics[width=5.5cm] {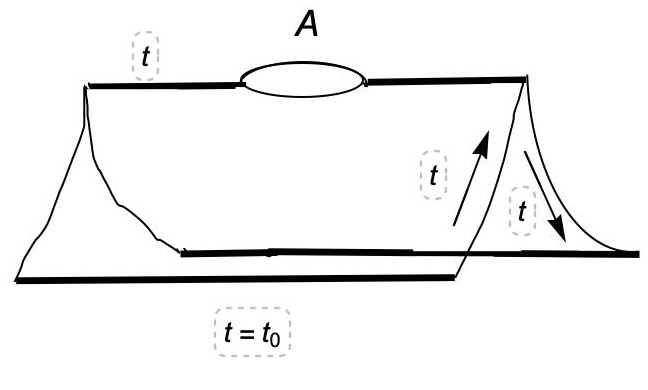}   \ \ \ \  \ \ \ \ 
    \includegraphics[width=5.5cm] {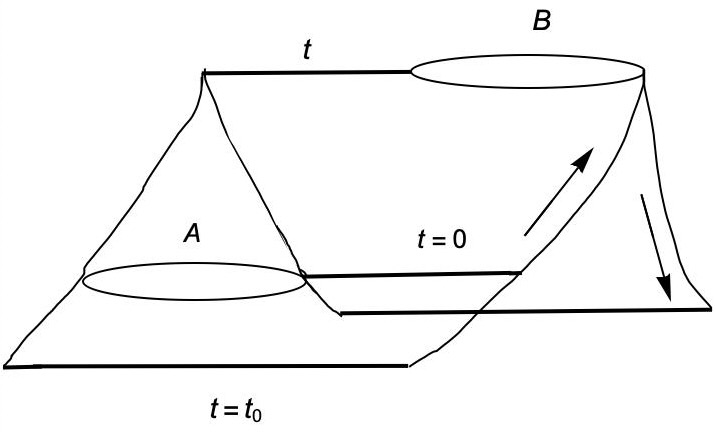}  
  \caption{The Schwinger-Keldysh (SK) representation of $\rho_A(t) := \text{tr} \ {\bar{A}}  \ \rho(t)$ is shown on the left \cite{Dong:2016hjy}, and the SK representation of $T_{AB}$ is shown on the right. Using the KD distribution, we connect them to EPs and TEE. }
 \label{fig:Keldysh}
\end{figure}

One would therefore expect that, similar to entanglement in time and timelike pseudoentropy, the dynamics around exceptional points could also be written in terms of the correlation functions of time-separated twist operators. This is indeed related to two twist operators with the ``topological defect lines'' attach to them. These topological defect lines connect different replica copies, similar to the EPs, as shown in figure \ref{fig:Keldysh}.

When the subsystems $A_i$ are causally disconnected, the reduced density transition operator $T_{A_0 A_1 \dots A_i \dots A_{N-1}}$ is Hermitian, and so it reduces to the usual density matrices. However, if any two subsystems $A_i$ and $A_j$ are causally connected, then the \textit{reduced spacetime density matrix} $T_{A_0 A_1 \dots A_i \dots A_{N-1}}$ is non-Hermitian and the spectrum is complex-valued. One should note that, similar to the case of EPs, the form of $T_{A_0 A_1 \dots A_i \dots A_{N-1}}$ depends on the choice of subsystems as well as the Hamiltonian or evolution operator.

As for the experimental setup, for a ring cavity that operates at an exceptional point with $\mathcal{PT}$-symmetric refractive index modulation \cite{MissingDimension},
\begin{gather}
n(\varphi) = n_0 + \delta n_R \cos (2l \varphi) - i \delta n_I \sin(2l\varphi),
\end{gather}
where $n_0$ is the background refractive index, and $\delta n_R$ and $\delta n_I$ are the real and imaginary parts of the refractive index modulation, respectively. The coupled-mode equations of the system can be written as
 \begin{gather}
 \frac{d}{dt} 	
 \begin{pmatrix}
a_{\text{CW}}\\
a_{\text{CCW}}
\end{pmatrix} =  
\begin{pmatrix}
i\omega-\gamma_{tot} & \chi_{ab} \\
\chi_{ab}  & i\omega- \gamma_{tot}
\end{pmatrix}   \begin{pmatrix}
a_{\text{CW}}\\
a_{\text{CCW}}
\end{pmatrix} \equiv i \mathcal{H} \begin{pmatrix}
a_{\text{CW}}\\
a_{\text{CCW}}
\end{pmatrix}.
\end{gather}
 
The clockwise and anticlockwise amplitudes, $a_{\text{CW}}$ and $a_{\text{CCW}}$ are complex. Also, $\omega$ is the resonant frequency, and $\gamma_{\text{tot}}$ is the total decay rate of the unperturbed counterpropagating whispering-gallery mode (WGM) in the non-Hermitian Hamiltonian $\mathcal{H}$. Using this setup, one could design boundary-region-like partitions—such as temporal gratings, time-like windows, or transition matrices between states—to probe pseudo-entropy in these open systems and test how these complex information measures have measurable signatures in EP systems. In both cases, analytic continuation, branch-cut topology, and complex-valued observables are employed. Moreover, the “missing dimension” in wave radiation at an EP is analogous to TEE producing an imaginary (timelike) component, which can be interpreted as an emergent time direction from entanglement/pseudo-entropy.

Then, in \cite{Guo:2024lrr,Xu:2024yvf}, it has been proposed that the timelike entanglement entropy can be uniquely determined by a linear combination of the spacelike entanglement entropy and its first-order temporal derivative. The imaginary component of the timelike entanglement entropy is supposed to be derived from the non-commutativity between the twist operator and its first-order temporal derivative, as

\begin{flalign*}
S(t,x; t',x') &= \frac{1}{4}  \left ( S(0,-u; 0,-u') + S(0,-u; 0, v') + S(0, v; 0, -u')+S(0,v;0,v') \right) \nonumber\\ &
+ \frac{1}{4} \int_{-u'}^{v'} \bar{x'} \partial_{t'} S(0, -u; 0, \bar{x}') + \frac{1}{4} \int_{-u'}^{v'} d\bar{x}' \partial_{t'} S(0, v; 0, \bar{x}') \nonumber\\
&+ \frac{1}{4} \int_{-u}^v d \bar{x} \partial_t S(0, \bar{x}; 0, -u') + \frac{1}{4} \int_{-u}^v d\bar{x} \partial_t S(0,\bar{x} ; 0, v') \nonumber\\
&+ \frac{1}{4} \int_{-u}^v d\bar{x} \int_{-u'}^{v'} d\bar{x}' \int_{-u'}^{v'} d\bar{x}' \partial_t \partial_{t'} S(0, \bar{x}; 0 , \bar{x}').
\end{flalign*}

For an interval between $(t,x)$ and $(t',x')$ in Minkowski spacetime, where the two points are timelike separated, the timelike Rényi entropy is
\begin{gather}
S_n(t,x ; t',x') := \frac{\log \text{tr} (\rho_0,A)^n \big |_{\tau \to it+\epsilon, \tau' \to it'+ \epsilon'} }{1-n} \nonumber\\
=\frac{2h_n}{n-1} \log \Big \lbrack \Delta s^2 +2i (\epsilon-\epsilon') (t-t') \Big \rbrack, \nonumber\\
\Delta s^2 = -(t-t')^2 + (x-x')^2.
\end{gather}

The imaginary part of this entanglement acts like a phase around the exceptional points, where, as the time difference $(t-t')$ increases, this phase also increases. The universal imaginary part of the timelike entanglement entropy is a constant, $\frac{i \pi c}{6}$, whose significance with respect to exceptional points could be studied further.

 \begin{figure}[ht!]
 \centering
 \begin{subfigure}{0.32\textwidth}
     \includegraphics[width=5cm] {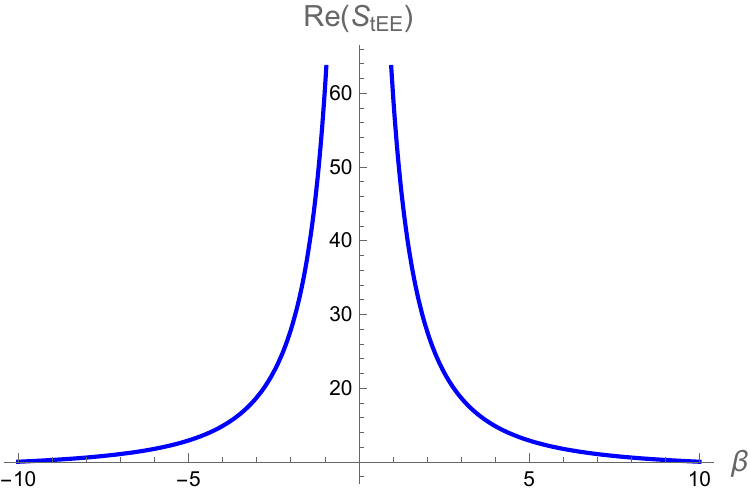} 
      \caption{} \label{fig:SEa}
     \end{subfigure}
      \begin{subfigure}{0.32\textwidth}
  \includegraphics[width=5cm] {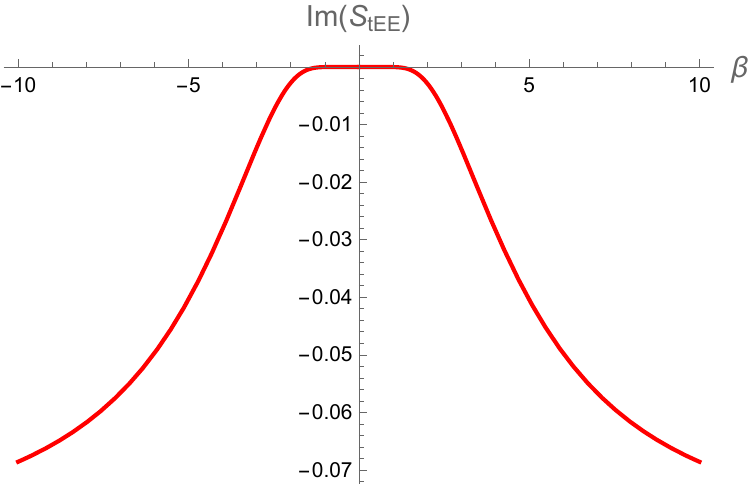}   
   \caption{} \label{fig:SEb}
    \end{subfigure}
     \begin{subfigure}{0.32\textwidth}
        \includegraphics[width=5cm] {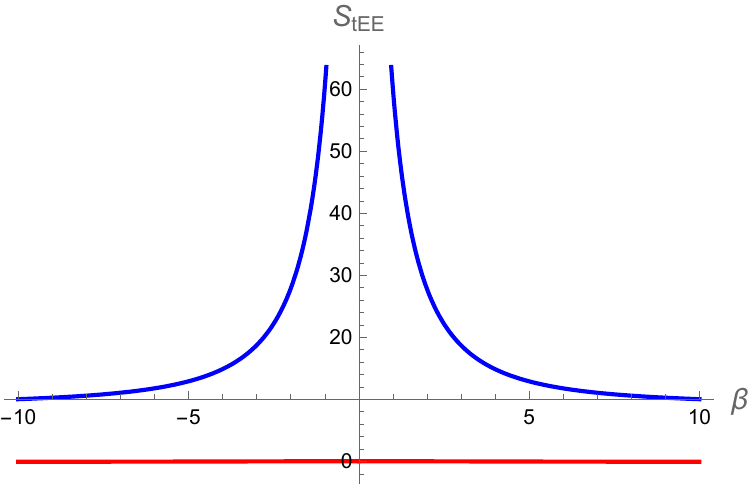}  
           \caption{} \label{fig:SEc}
         \end{subfigure}
  \caption{The real and imaginary parts of the timelike entanglement entropy, compared in magnitude relative to each other.}
 \label{fig:ImRetEE}
\end{figure}

In figure \ref{fig:ImRetEE}, the real and imaginary parts of timelike entanglement entropy are shown. One can see that decreasing $\beta$, or increasing temperature, increases the real part of the entanglement entropy to the point that it can diverge, as expected. The imaginary part of the entanglement, although much smaller, behaves in the opposite way. This makes the encircling of the exceptional point easier, which is important for quantum computation. Note that the imaginary part encodes contextuality and non-commutativity, which can manifest at lower temperatures. It also reflects non-unitarity, phase coherence, and analytic continuation effects, precisely the kinds of behavior that appear near exceptional points. Moreover, it acts as a holographic order parameter for the degree of non-orthogonality of eigenmodes and diverges near an EP.

This behavior is also consistent with the results of \cite{Park:2022agr}, where multiple EPs were generated in two-level systems using strong imaginary coupling in a non-Hermitian Hamiltonian. By encircling these multiple EPs, their topological structures were examined. There, it was shown that the imaginary parts of eigenvalues bifurcate around two critical points.

In figures 4 and 5 of \cite{Park:2022agr}, the behavior of eigenvalue trajectories for double EPs is shown. Due to strong imaginary coupling $g$, a new form of topological structure, $f(z) = \sqrt{(z-z_1)(z-z_2)}$, was found for double EPs. There, the real parts of eigenvalues show avoided crossing while the imaginary parts show mode crossing. Consequently, the branch cuts for real and imaginary parts differ as the real part has a single branch cut connecting the two EPs, while the imaginary part has two branch cuts extending outward to infinity. Similarly, the real part of TEE typically has a finite branch cut related to the replica singularities or the “turning point” between two analytic sheets, while its imaginary part may have semi-infinite branch cuts connected to the Wick-rotated, timelike regions of the correlation function, extending to infinity in complex time.

For entanglement in time \cite{Milekhin:2025ycm}, and in holographic scenarios, when choosing between various extremal surfaces between two intervals in time, one can select the minimal real part. This approach seems to work when the endpoints of $A$ and $B$ are close to null separation. However, in more complicated situations, for instance, in QCD models or at finite temperature, non-physical results might arise, requiring a microscopic analysis of timelike entanglement entropy. We propose that the topological branch cuts in photonic exceptional points in optics and quantum information could offer a path to examine these structures further.

One can treat the “past” interval $AB$ as the input matrix legs and the “future” subsystem $CD$ as the output legs. In \cite{Milekhin:2025ycm}, it is mentioned that from the MPO representation, only $2 \chi$ singular values would be nonzero. They also found that, similar to our results in \cite{Ghodrati:2023uef, Ghodrati:2021ozc}, the singular values follow an exponential decay, as $T$ can be represented by an MPO.

In \cite{Milekhin:2025ycm}, they used the notion of 2‑imagitivity, which captures the quantum correlations that propagate through time, unlike standard mutual information. For this measure, \cite{Milekhin:2025ycm} found the following inequalities,
\begin{gather}
\frac{| \langle \lbrack \mathcal{O}_A (0), \mathcal{O}_B(t)\rbrack \rangle  |}{|| \mathcal{O}_A||_2 ||\mathcal{O}_B||_2} \le ||T_{AB} - T^\dagger_{AB}||_2,
\end{gather}
and 
\begin{gather}
\frac{1}{\text{dim} \mathcal{H}_A}  ||T_{AB} - T^\dagger_{AB}||_2 \le \frac{| \langle \lbrack \mathcal{O}_A (0), \mathcal{O}_B(t)\rbrack \rangle  |}{|| \mathcal{O}_A||_2 ||\mathcal{O}_B||_2}
\end{gather}
where $\mathcal{O}_A$ and $\mathcal{O}_B$ are bosonic operators. Based on the connections we found, and since 2‑imagitivity is a measure of the non-Hermitian character of the spacetime density matrix, these two inequalities could also place constraints on the distances between branch cuts of EPs, similar to those in \cite{Park:2022agr}. Near EPs, commutators are amplified. For example, in optomechanical sensors, EPs detect deformed commutators with quantum-noise-limited bounds. The distance $\Delta$ between double EPs must be large enough to allow measurable commutator growth  but small enough to avoid exponential suppression. For instance, if imagitivity scales as $1/\sqrt{\Delta}$ near EPs which is due to square-root branching, the inequalities constrain $\Delta$ to values where commutators remain finite but detectable, e.g., $\Delta \gtrsim 1 / (\dim \mathcal{H}_A \cdot C)$.

In an interesting recent work, \cite{Hashimoto:2026kjy}, the emergence of spacetime of $\text{AdS}_2$ black holes using Lindblad harmonic oscillator and Lindblad SYK models has been explained. They also showed that the Lindblad time evolution of the harmonic oscillator coupled to a bath with jump operators, which also form  a Fokker- Planck equation, create a time trajectory in the Wasserstein space which then produce a geometry similar to the event horizon of a $2d$ black hole. As these two systems are open quantum systems, the emergence of EPs and their connections to this emergent \textit{``time"} could be connected to our calculation here and be further studied.

Finally, in this regard, there could be interesting connections between the timelike tube theorem \cite{Strohmaier:2023opz} and EPs. The timelike envelope $E(U)$ includes all points that one can reach by starting with a timelike curve in $U$ and slightly deforming it while keeping endpoints fixed, so $A(U) = A(E(U))$, where $A(U)$ is the algebra of observables for a region $U$. For this specific theorem to hold, the system should be Hermitian and real-analytic, so that the spacetime geometry could be extended to complex numbers in a smooth way. Interestingly, behaviors like those in PT-symmetric theories suggest that timelike tubes could exhibit “encircling” effects as  traversing parameters around an EP might switch system branches, altering causality or entanglement in $E(U)$, similar to how EPs enable chiral transport or enhanced sensing. Then, $E(U)$ is built from deforming timelike curves with fixed endpoints.

\begin{figure}[ht!]
 \centering
     \includegraphics[width=5cm] {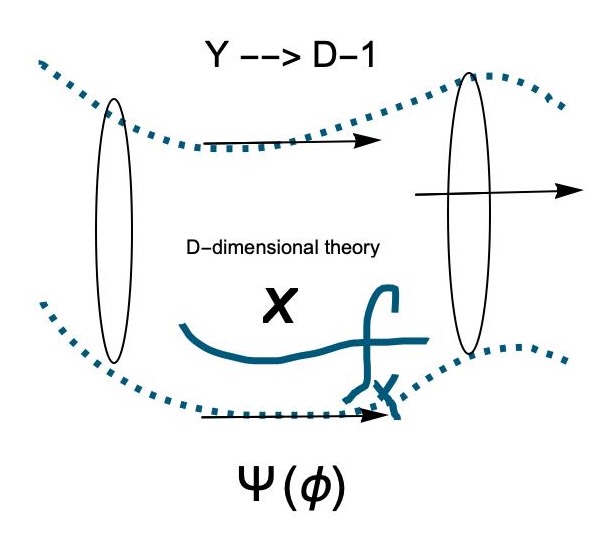} 
  \caption{Schematic of the timelike tube $E(U)$. Introducing EPs and non-Hermiticity into the system could shrink or distort $E(U)$, cause multiple “branches” of solutions to emerge, or make the envelope boundaries ill-defined beyond null geodesics.}
 \label{fig:Timetube}
\end{figure}

In the timelike tube case, Fig.~\ref{fig:Timetube}, one considers the analytic continuation of field solutions in the complexified spacetime and the consequent uniqueness of continuation from $U$ to $E(U)$. For EPs, analytic continuation of eigenvalues and eigenvectors of a non-Hermitian operator in complex parameter space shows that branch points (EPs) are singularities of this analytic continuation. Both are controlled by the same type of object which is a determinant, a boundary-value functional, or equivalently the poles or zeros of the resolvent or Green’s function. This suggests a mathematical mapping that is worth studying further.

In a non-Hermitian setup with EPs, the effective dynamics like wave equations or propagators could become non-unitary or exhibit amplified or divergent behaviors near the EP. This might shrink or distort $E(U)$, as deformations could encounter branch points where uniqueness breaks, for instance, multiple “branches” of solutions emerge, or the envelope boundaries become ill-defined beyond null geodesics. In extreme cases, $E(U)$ could fragment or fail to form a connected region, especially if EPs introduce instabilities in high-energy modes. Thus, introducing EPs would likely invalidate or require major modifications to the timelike tube theorem, as the core assumptions of analyticity and Hermitian structure break down. The envelopes $E(U)$ would become non-unique, distorted, or limited in extent.

Another interesting point worths to mention here is the recent consideration of an ``observer" in a form of a clock \cite{Maldacena:2024spf,Yang:2025lme}, a charged black hole in de Sitter space \cite{Chen:2025jqm}, or two coupled SYK systems \cite{Zheng:2025apa,Liu:2026mox}, which can mimic the properties of an open quantum system and to explain properties such as complex phase space in the one-loop gravitational path integral. These observers, by continuous monitoring can enhance chaos in the systems similar to EPs and can cancel the complex phases.

 \section{EPs, $\theta$-vacuum of QCD and winding numbers}

In this section, we further study the connections between EPs and various other structures in QCD. Specifically, based on the winding number in QCD, we define a winding number for EPs and compare their structures. We also study the connections between the ``functional renormalisation” (fRG) method and the emergence of different topological sectors based on the Chern-Simons number, EPs, and complexified time-dependent entanglement entropy.

The topological structures and EP encircling can be related to the QCD instantons. The EPs exhibit nontrivial topology in parameter space, such as Riemann surfaces or branch cuts. QCD, on the other hand, also has topologically nontrivial vacua, such as instantons or non-perturbative tunneling and $\theta$-vacuum structures. So EP encircling can simulate such topological phase changes, which is the aim of this section.

In \cite{Lenz:1994du}, the system is analyzed using the adiabatic Born-Oppenheimer approximation where fast fermion modes are ``integrated out" first, yielding an effective potential $U(a)$ for slow gauge variables $a_p$. EPs occur where pairs of fermion energy levels, where one is left-handed and one is right-handed, cross zero energy, e.g., at $g L a = 2 \pi (n + \frac{1}{2})$ for $SU(2)$, where $g$ is the coupling, $L$ the spatial size, and $n$ an integer. These points divide the configuration space into sectors with different fermion occupations in the Dirac sea. At EPs, the adiabatic approximation breaks down because the Dirac sea restructures, causing spectral flow, which is a topological shift in fermion modes, and then non-adiabatic transitions occur in the system.

The $\theta$-angle parameterizes superpositions of these vacua with $n$ the topological charge. EPs mark boundaries between topological sectors, enabling tunneling between wells via instanton-like paths. This resolves vacuum selection and generates condensates. EPs enhance tunneling between vacua which have overlap $\sim \exp(-\pi^{3/2} gL)$, boosting condensates and explaining why adjoint fermions break continuous axial symmetry while preserving discrete $Z_N$, conflicting with topology to yield the $\theta$-structure. So EPs mathematically bridge non-Hermitian spectral singularities to QCD topology, as they mark sector boundaries, induce spectral flow via anomalies and index theorems, shape multi-well potentials, and enable the $\theta$-superposition through instanton-like tunneling.

\begin{figure}[ht!]
 \centering
  \includegraphics[width=8.4cm] {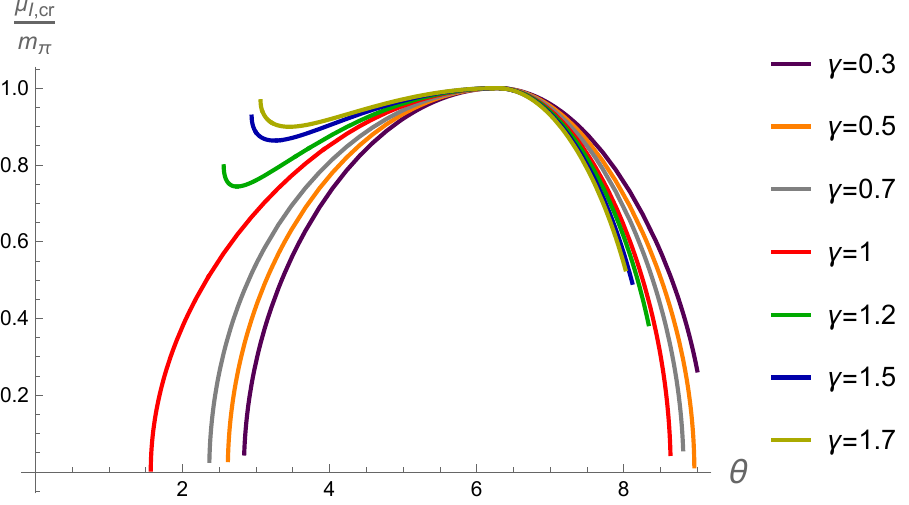}  
  \caption{For various quark mass ratios $\gamma \equiv m/m_s = \frac{m^2_\pi}{2 m^2_K - m^2_\pi}$, the behavior of the critical isospin chemical potential as a function of $\theta$ in QCD is shown. }
 \label{fig:quarkEPisospin}
\end{figure}

In \cite{Bersini:2025yvt}, the $\theta$-angle physics of QCD under pressure, with its strange and isospin phase diagrams, has been discussed. There, it was shown that for degenerate quark masses, one could induce a superfluid transition by varying $\theta$ while keeping the isospin $\mu_I$ and strangeness $\mu_s$ fixed. This is actually related to our work here, which connects exceptional points with degeneracy to QCD phase transitions. Around $\theta=\pi$, there is a novel parity-preserving superfluid phase. At $\theta=0$, there are two different superfluid phases, namely the Pion and Kaon phases, whose emergence can be characterized by the onset of Pion and Kaon condensation, and also by the emergence of exceptional points in the system. The transition between the Pion and Kaon phases does not depend on the $\theta$ angle except in the limit of degenerate quark masses. This behavior as a function of $\theta$ and the divergence around $\theta=\pi$ is shown in figure \ref{fig:quarkEPisospin}.

As in \cite{Vonk:2019kwv}, the $\theta$-vacuum of Quantum Chromodynamics, in terms of the winding number, can be defined as
 \begin{gather}
 \mathcal{L}^\theta_{\text{QCD}} = - \theta w(x),
 \end{gather}
 where here
 \begin{gather}
 w(x)= \frac{g^2}{16 \pi^2} \text{Tr} \lbrack G_{\mu \nu} \tilde{G}^{\mu \nu} \rbrack
 \end{gather}
 is the winding number density, and its integral $\int d^4 x \, w(x) = \nu$ is the winding number of the field configuration. Here, $G_{\mu\nu}$ is the QCD field strength tensor, and $\tilde{G}_{\mu\nu}= \frac{1}{2} \epsilon_{\mu\nu \rho \sigma} G^{\rho \sigma}$ is its dual.  
The vacuum structure of QCD then depends on the winding number. If we have a complex source for the $\theta$-term or the quark chemical potentials, a non-Hermitian AdS/QCD system could be modeled.

For the case of EPs, one can also define the winding number as in \cite{Yoshida:2024xua}, where, for instance, in the case of EP3, the winding number $W_3$ is
 \begin{gather}
 W_3 = \frac{\epsilon^{ijkl} }{2 \pi^2} \int d^3 \mathbf{p} f_{ijkl}, \ \ \ \ \ \ \ \ \ 
 f_{ijkl} = n_i \partial_1 n_j \partial_2 n_k \partial_3 n_l,
 \end{gather}
Here, $\mathbf{n} = \mathbf{R} / \sqrt{\mathbf{R} \cdot \mathbf{R}}$, where $\mathbf{R}$ is the radius of a three-dimensional sphere surrounding the EP3 and $\mathbf{p} = (p_1,p_2,p_3)^T$ is a point on this sphere. So, $\epsilon^{ijkl} f_{ijkl} = \epsilon^{ijkl} n_i \partial_1 n_j \partial_2 n_k \partial_3 n_l$ corresponds to $\text{Tr} \lbrack G_{\mu\nu} \tilde{G}^{\mu \nu}\rbrack$.

For EPn in $n-1$ dimensions, we have
 \begin{gather}
 W_{n-2} = \frac{\epsilon^{i_1 . . . i_{n-1} } }{A_{n-2} } \int d^{n-2} \mathbf{p} f_{i_1 . . . i_{n-1}}, \ \ \ \ \ \ \
  f_{i_1 . . . i_{n-1}} = n_{i_1} \partial_1 n_{i_2} \partial_2 n_{i_3} . . . \partial_{n-2} n_{i_{n-1} }, 
 \end{gather}
This can then be extended to the case of QCD.
 
To consider the interplay between Abelian and non-Abelian topology and for further extensions, as in \cite{Yoshida:2024xua}, it has been proposed that non-Abelian contributions are expected to emerge when additional bands are added, i.e., $\text{EP}_n$ in $m$-band models with $m>n$. For these classification schemes, the braiding of complex eigenvalues around $\text{EP}_n$ should be accounted for. Note that if symmetries are present and symmetry-protected $\text{EP}_n$ exist, the effective dimension reduces from $2n-2$ to $n-1$.

 The Euclidean action of QCD with the $\theta$-term over the spacetime volume $V$ is
 \begin{gather}
 S_E= \int_V d^4 x \left ( \frac{1}{2} \text{Tr} \lbrack G_{\mu\nu} G_{\mu\nu}\rbrack - i \theta w(x) + \bar{q} ( - i \gamma_\mu \mathcal{D}_\mu + \mathcal{M} ) q \right) \nonumber\\
 = S_G - i \theta \nu + \int_V d^4 x \ \bar{q}  ( - i \gamma_\mu \mathcal{D}_\mu +\mathcal{M}) \ q,
 \end{gather} 
 where $q$ is the quark field, $\mathcal{D}_\mu$ is the QCD gauge covariant derivative, $\mathcal{M}$ is the quark mass matrix and 
 \begin{gather}
 S_G = \int_V d^4 x \frac{1}{2} \text{Tr} \lbrack G_{\mu\nu} G_{\mu\nu} \rbrack,
 \end{gather}
 is the gluon action.
 
Next, one could postulate a corresponding Euclidean action for photonic systems with exceptional points which for the case of EP3 in 3-dimensions we propose the following terms
\begin{flalign}
S_E(EPs)& =  S_{GEPs}- i \theta W_3 + \int d^3  \mathbf{p} \ \text{Tr} \lbrack e^{-\beta H(\mathbf{p}) } \rbrack, \nonumber\\
S_{GEPs}& = \int d^3 \mathbf{p} \frac{1}{2} \text{Tr} \lbrack n_i \partial_1 n_j \partial_2 n_k \partial_3 n_l \rbrack, \nonumber\\ 
 W_3 &= \frac{1}{2\pi^2} \int d^3 \mathbf{p} \epsilon^{ijkl} n_i \partial_1 n_j \partial_3 n_l.
\end{flalign}
 
This action could help to better understand the interplay between topologies and the emergence of exceptional points. Then, one could connect the topological susceptibility in QCD with the topological structures of non-Hermitian EPs.

As in \cite{Vonk:2019kwv}, using the Euclidean action with the $\theta$-term in QCD, one can write the partition function as
\begin{gather}
Z(\theta) = \int \lbrack D A_\mu \rbrack \ \text{exp} ( -S_G + i \theta \nu)\ \text{det} ( - i \gamma_\mu \mathcal{D}_\mu + \mathcal{M}) \equiv  \sum_{\nu= - \infty}^{+\infty} e^{i \theta \nu} Z_\nu,
\end{gather}
which is a path integral over the gauge-field configurations characterized by the winding number $\nu$. Similarly, for the case of exceptional points in 3D, one can define
\begin{gather}
Z_{W_3} := \int d^3 \mathbf{p} \ \text{exp} (- i S_{GEPs}) \ \text{Tr}_B \lbrack e^{- \beta H(\mathbf{p}) } \rbrack,
\end{gather}
where $\text{Tr}_B$ here denotes the biorthogonal trace
\begin{gather}
\text{Tr}_B \lbrack O \rbrack = \sum_n \bra{\psi_n^L} O \ket{\psi_n^R},
\end{gather}
and then similar to the case of QCD, one can have a probabilistic interpretation for 
\begin{gather}
p_{W_3} = \frac{Z_{W_3} }{Z(\theta=0)},
\end{gather}  
which is the probability to find an exceptional point with winding number $W_3$. Then, by considering the $n^{\text{th}}$ derivative of $Z(\theta)$ at $\theta=0$, one gets
\begin{gather}
\langle W_3^n \rangle_{\theta=0} = \frac{1}{i^n Z(\theta=0)} \left \lbrack \frac{\partial^n Z}{\partial \theta^n}\right \rbrack_{\theta=0},
\end{gather}
and similarly here $Z$ could be regarded as the moment-generating function of the distribution of the winding number $W_3$ for the exceptional points. Next, the topological susceptibility for EPns can be defined as
\begin{gather}
\chi_{\text{top}} := \frac{\langle W_3^2 \rangle}{V}.
\end{gather}

This quantity is important in studying the topological effects of the emergence of EPs in various systems. For instance, in \cite{PhysRevB.108.L060506}, it was shown that non-Hermiticity enhances the properties of Majorana zero modes and the edge states of topological superconductors. In fact, in \cite{Bersini:2025yvt}, it has been shown that by considering the strange quark mass heavier than the degenerate up and down quark masses, the properties of QCD at $\theta \sim \pi$ can be derived. In this region, two vacuum states coexist and become degenerate at $\theta = \pi$. So the point at $\theta = \pi$ is an exceptional point for this system, and it causes the amplification of CP-violating effects near $\theta \sim \pi$, which is the “dash line” where vacua destabilize. In works such as \cite{Arean:2016hcs}, the CP-odd sector and $\theta$ dynamics in holographic QCD have been studied.

Also, in \cite{Guan:2025xoj}, the $\theta$-vacuum from “functional renormalisation” (fRG) has been studied. There, the topological properties of a quantum mechanical system with a $U(1)$ symmetry (quantum rotor model), such as vacuum energy structure and topological susceptibility, have been analyzed. The $\theta$-term in this model creates topological effects, like splitting the vacuum into  ``sectors" based on Chern-Simons numbers, which then lead to interesting behavior in the system, such as changes in the energy and effective potential based on the parameter $\theta$. Specifically, they used the complexification of the flow equation and the symmetry ($U(1) \to \mathbb{C}$), because the “spinning” in the model can only be described by a complex field, which has both magnitude and direction.

The action of their model could be written as
\begin{gather}\label{eq:Sthetaaction}
S\lbrack \varphi \rbrack = \int_\tau \left \lbrack \frac{m}{2} \dot \varphi^* \dot \varphi - \frac{\theta}{4\pi} (\varphi^*  \dot \varphi-  \dot \varphi^* \varphi) + V(\varphi^* \varphi) \right \rbrack,
\end{gather} 
which for a complex field $\varphi = r e^{i \vartheta} $ would lead to
\begin{gather}
S= \int_\tau \left \lbrack \frac{m}{2} (\dot r^2+r^2 \dot \vartheta^2) - i \frac{\theta}{2\pi} r^2 \dot\vartheta+\frac{g}{4}(r^2-1)^2 \right\rbrack.
\end{gather}

For the case of $g \to \infty$, the `$\theta$'-term becomes topological and has the value $i \theta \nu$, where $\nu$ is a winding number and $\nu \in \mathbb{Z}$.

The energy level then depends on $\theta$ as
\begin{gather}
E_n ( g \to \infty, \theta) = \frac{1}{2m} \left ( n -\frac{\theta}{2\pi} \right ),
\end{gather}
and the ground-state energy is periodic in $\theta$, indicating the topological nature of the model.

The zero-point energy here would be proportional to $\sqrt{g}$ as
\begin{gather}
E^{(0)}(g) = \sqrt{ \frac{g}{2}},
\end{gather}
where $g$ is the coupling parameter in the potential
\begin{gather}
V(\varphi^* \varphi) = \frac{g}{4} ( \varphi^* \varphi -1 )^2.
\end{gather}
So the stronger the coupling and potential wall, the larger the zero-point energy would be. Also, for the radial mean-field value $r = \bar{r}$, the energy levels are
\begin{gather}
E_n ( g \to \infty, \theta) = \frac{1}{2m} \left ( n - \frac{\theta \ \bar{r}^2}{2\pi} \right )^2.
\end{gather}
So, one would expect that for $\theta = \frac{2\pi n}{\bar{r}^2}$, there would be exceptional angles and exceptional lines.

In \cite{Guan:2025xoj}, the non-analytic structures correspond to level crossings between different topological sectors, which are indexed by winding number $\nu$ and are given explicitly by $\theta r^2 = (2n-1)\pi$. This leads to cusp-like and non-differentiable behavior in the vacuum energy. The full potential can then be constructed by taking the minimum over a set of sector-wise potentials $V_n(r^2, \theta)$, where each corresponds to a fixed winding number $n$, as $V(r^2, \theta) = \text{min}_n V_n(r^2, \theta)$. This structure is the source of non-analyticities, as different branches dominate in different regions of parameter space.

Their method, which involves computing sector potentials with fixed winding number and recombining, is mathematically similar to building a multi-sheeted analytic continuation or choosing branch representatives per sector and gluing them. This is the correct procedure if one wants to reveal branch points or EPs that lie between sectors. These cusps and non-analyticities as a function of $\theta$, and the transitions between sectors (branch switches), are reminiscent of the topology around EPs and also could reflect how energy levels behave near an EP. Again, the summation over topological sectors is related to the coalescence of eigenvalues and eigenvectors.

Furthermore, based on this work and others such as \cite{Gaiotto:2017yup}, the case of $\theta = \pi$ is often a critical value where the vacuum structure changes dramatically. For example, it can lead to degenerate vacua, spontaneous symmetry breaking (e.g., CP or time-reversal symmetry), phase transitions, time-reversal anomalies, vanishing mass gaps, or level degeneracies. The partition function in this model, and also the effective potential, have $\theta$-dependences. Specifically, this shows that the effective potential exhibits non-analytic structures originating from its semi-topological properties, similar to the effects of the end-wall in confining geometries as in \cite{Ghodrati:2023uef, Ghodrati:2021ozc}.

Also, the piecewise-defined minimal and multi-sector potential and the resulting multivaluedness are directly analogous to the Riemann-surface picture around EPs. This is related to the non-Hermitian transition operator $\text{Tr}(\rho^{T_B}_{AB})^n$ and time-like entanglement entropy, as in \cite{Gong:2025pnu}, where it has been shown that the reflected entropy for timelike intervals is twice the timelike entanglement wedge cross section. This supports our picture of EPs and the $\theta$-vacuum as the summation of topological sectors and the coalescence of eigenvalues/eigenvectors.

Moreover, \cite{Gong:2025pnu} stresses that some flow setups such as flow with a simple complex frequency or without topological resummation fail to capture the ground-state level crossing. In non-Hermitian parameter space, a true EP is where two levels coalesce rather than cross trivially. Capturing this requires the correct analytic continuation as if the regulator or truncation distorts the analytic continuation, the EP or branch structure may be missed or misrepresented.

However, there are also differences as the non-analyticity in the $\theta$-vacuum arises from the topological summation over sectors, whereas EPs arise from the coalescence of eigenmodes in non-Hermitian parameter space. Nevertheless, under suitable mappings (e.g., synthetic gauge fields or complexified parameters), photonic analogues of $\theta$-vacuum cusps could be constructed, mimicking topological transitions as EPs or EP lines. Thus, level crossings in the $\theta$-vacuum could manifest as photonic EP-like transitions.

\subsection{Numerical investigation of EPs in $\theta$-vacuum QCD model}\label{sec:numEP}

For the $U(1)$-symmetric quantum system, the partition function can be written as  \cite{Guan:2025xoj}
\begin{gather}
Z_\phi \lbrack J_\phi \rbrack = \int \mathcal{D} \hat{\varphi} e^{-S \lbrack \hat{\varphi} \rbrack+ \Delta S_\phi \lbrack \hat{\phi}\rbrack + \int_\tau J_\phi^a \hat{\phi}^a },
\end{gather}
where here the cutoff term $\Delta S_\phi \lbrack \hat{\phi} \rbrack$ is quadratic in the field that is coupled to the source,
\begin{gather}
\Delta S_k \lbrack \phi \rbrack = \frac{1}{2} \int_p \phi_a(-p) R_k^{ab}(p) \phi_b(p), 
\end{gather}
where $R_k(p)$ is the momentum-dependent mass term, and this affects the dispersion of the field $\phi$.

The flow equation for the effective action, which is the Legendre transform of $\log Z_\phi \lbrack J_\phi \rbrack$, can be written as
\begin{gather}
\partial_t \Gamma_k \lbrack \phi \rbrack = \frac{1}{2} \int_{q,p} \left \lbrack \frac{1}{\Gamma_k^{(2)} \lbrack \phi \rbrack + R_k} \right \rbrack_{ba}  (-p,p) \ \partial_t R_k^{ab} (p),
\end{gather}
and $\Gamma^{(2)}_{k, ab}(p,q)$ is the full two-point function which can be written as
\begin{gather}
\Gamma_{k,ab}^{(2)} (p,q) = \frac{\delta^2 \Gamma_k}{\delta \phi^a(p) \delta \phi^b(q)}.
\end{gather}

When the theory flows towards the UV, the bare action is
\begin{gather}
\Gamma_{\Lambda \to \infty} \lbrack \phi \rbrack = S\lbrack \varphi \lbrack \phi \rbrack \rbrack + \Delta \Gamma_{\text{UV}}\lbrack \phi \rbrack.
\end{gather}
which can be the result of a cutoff term, such as the gluon mass term in QCD.

An interesting phenomenon is the `topological freezing' on the lattice, which its relation to EPs and $T_{AB}$ could be studied. In the functional renormalization setup which, instead of summing over topological sectors, we have a $\theta$-term in the effective action, the flow equation develops a pole and the RG flow breaks down at intermediate scales, as the computation terminates around $\theta \sim 2\pi$, and another EP similar to the confining end-wall would form there.
The flow equation would be
\begin{gather}
\partial_t V_k = \frac{1}{\pi} \frac{2 \ m \ k^3(m k^2 + 2V'_k + 2r^2 V''_k) }{(m k^2 + 2 V'_k)^2 + 4r^2 V''_k}.
\end{gather}

In order to investigate this phenomenon, the integro-differential form of the Cartesian field $\varphi$ could be written as
\begin{gather}
e^{- \Gamma_k \lbrack \varphi \rbrack} = \text{lim}_{\beta \to \infty} \sum_\nu \int \mathcal{D} \hat{\varphi} e^{- S \left \lbrack \hat{\varphi} e^{\frac{2\pi i \nu}{\beta} \tau} \right \rbrack + \Delta S_k \lbrack \hat{\varphi}- \varphi\rbrack} \times e^{\int_\tau (\hat{\varphi} - \varphi) \frac{\delta \Gamma\lbrack \varphi \rbrack}{\delta \varphi}}, 
\end{gather}
where $\hat{\varphi}$ has non-trivial windings, in the limit of $\beta \to \infty$ where the regulator term would diverge.

This failure is named \textit{topological freezing}, where the RG regulator and the chosen setup suppress the fluctuations that change a topological sector, so the flow cannot tunnel between sectors and in this method ``freezes'' the topology during the flow. This method has been contrasted with complexifying frequencies and performing a topological resummation which avoids that particular breakdown.

\begin{figure}[ht!]
 \centering
  \includegraphics[width=6.8cm] {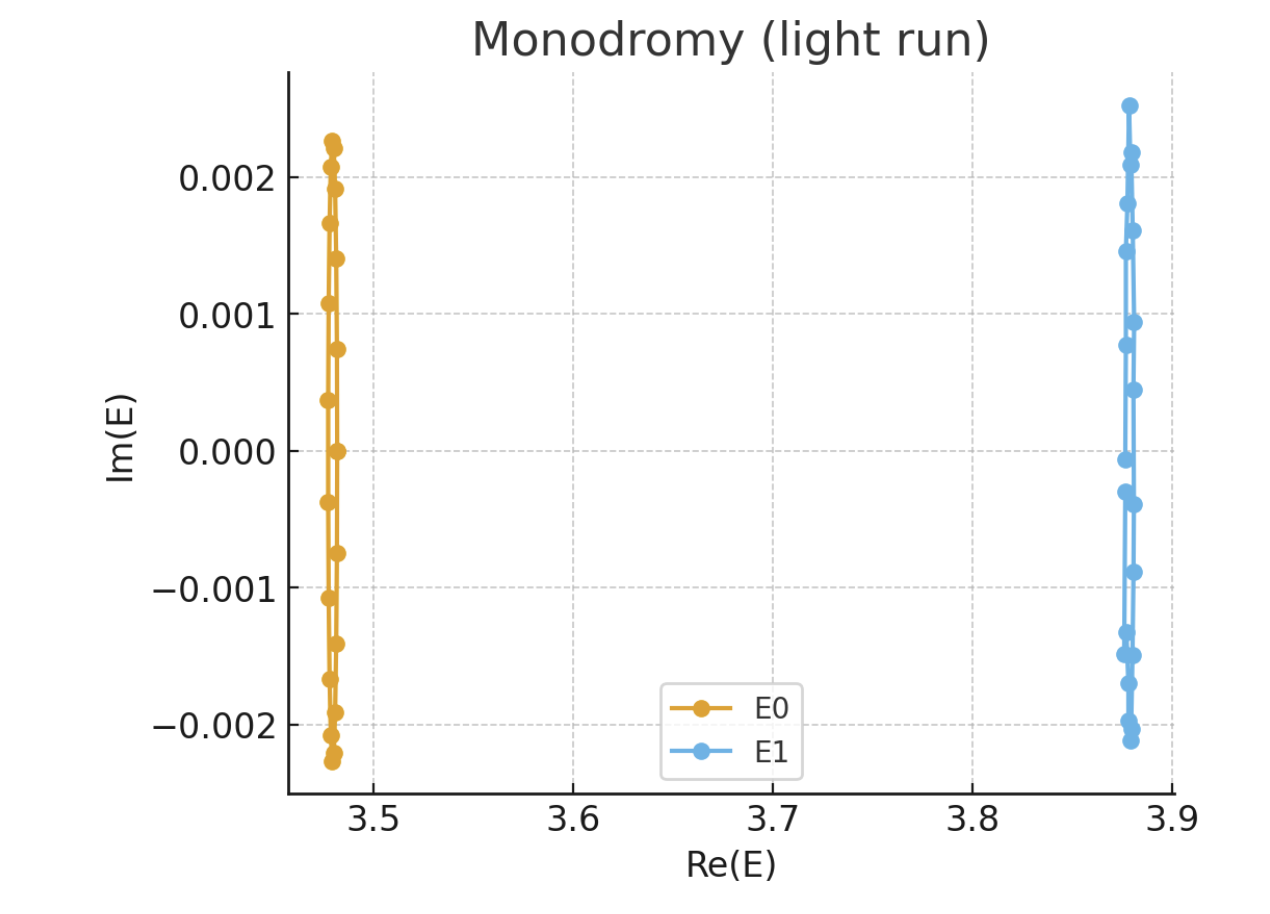} 
    \includegraphics[width=6.8cm] {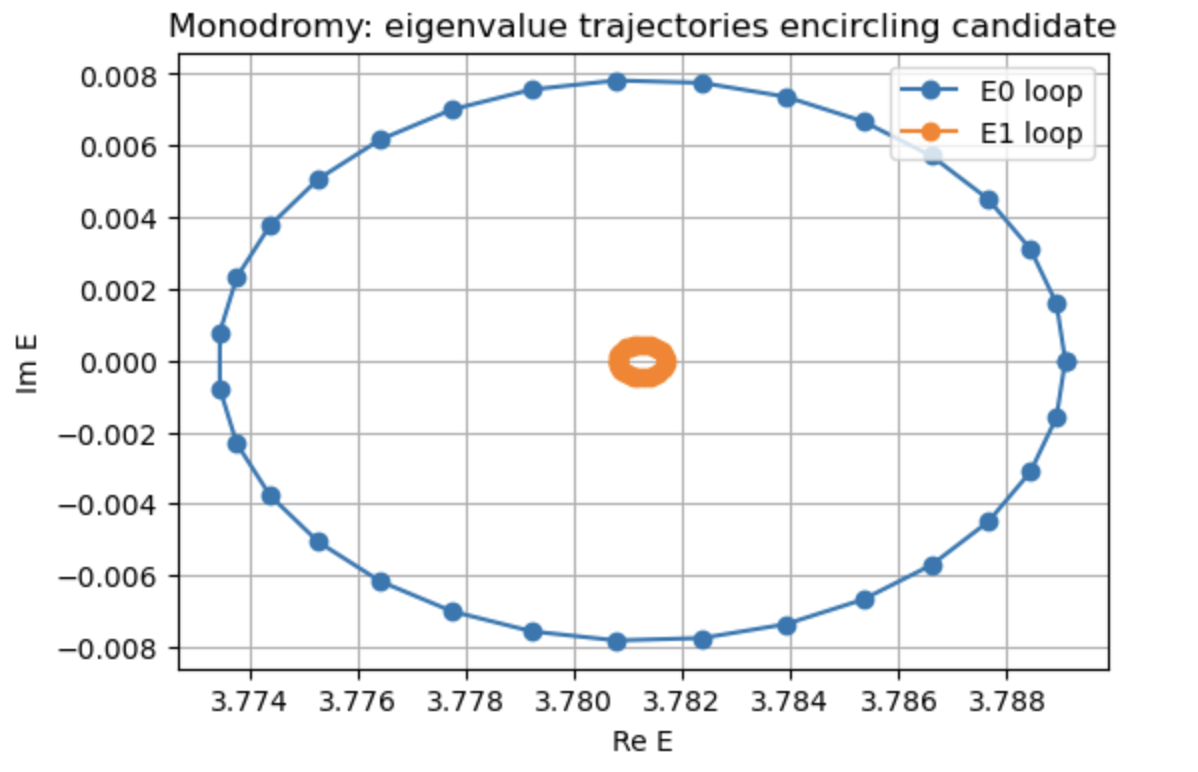}  
  \caption{The result of a numerical simulation of the Schrödinger eigenproblem that searches the complex $\theta$-plane for EPs for $g=30$. The crossing (no coalescence) is around $\theta_{\text{cand}} \approx 3.6416 - 3.27 \times 10^{-7} i$, and $E_0 \approx 3.4794107$ and $E_1 \approx 3.8785131 + 3.08 \times 10^{-4} i$.}
 \label{fig:NoEPdecoupled}
\end{figure}

In this model, if strictly one solves the Schrödinger equation, no real EP could be found as the crossing of $E_0$ and $E_1$ is decoupled, as one could see in figure \ref{fig:NoEPdecoupled}.

However, by adding a \textbf{small angular coupling} $\mathbf{\epsilon V(r) \cos \varphi}$ which mixes sectors, one can build the full block Hamiltonian and scan the complex $\theta$. Then, after a coarse scan near $\theta \approx \pi$, we could find a real second-order EP and check the behavior of monodromy plots for various radii which is the size of the circular path we draw around the suspected EP in the complex-parameter plane, as shown in figure \ref{fig:monodromy}.

With $r=0.03$, we find a real second-order EP around 
\begin{gather}
\boxed{\theta_{EP} \approx 4.21986 +1.4 \times 10^{-5} i}
\end{gather}
where $\theta(\Phi) = \theta_{EP}+ r e^{i \phi}$, $0 \le \phi < 2\pi$.

For bigger $r$, we actually enclose other singularities or regions where the eigenvalue surfaces are already well separated. So the results for $r=0.03$ would be enough for our case, and only after one loop do we have monodromy permutation and eigenvalues swap.

 \begin{figure}[ht!]
 \centering
  \includegraphics[width=5.8cm] {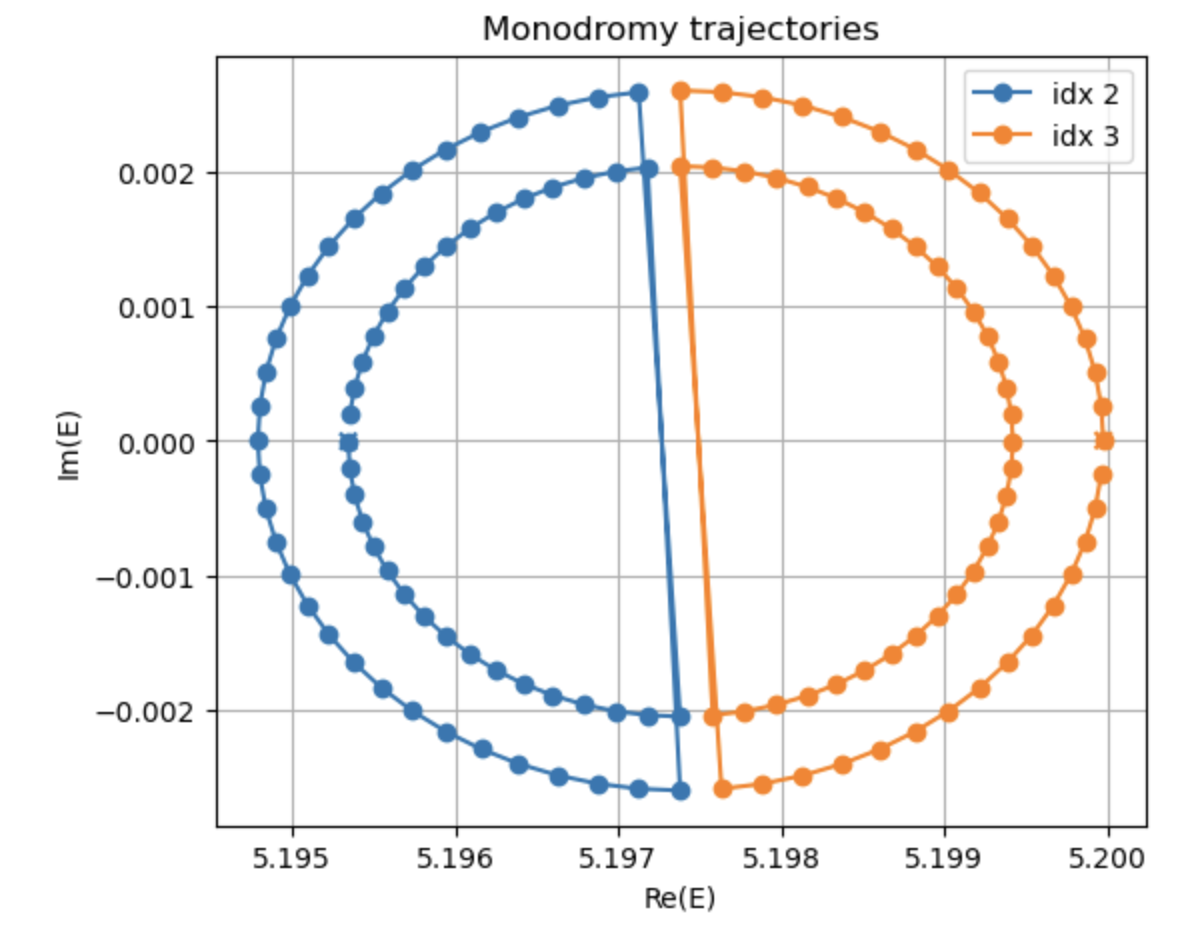} 
    \includegraphics[width=5.6cm] {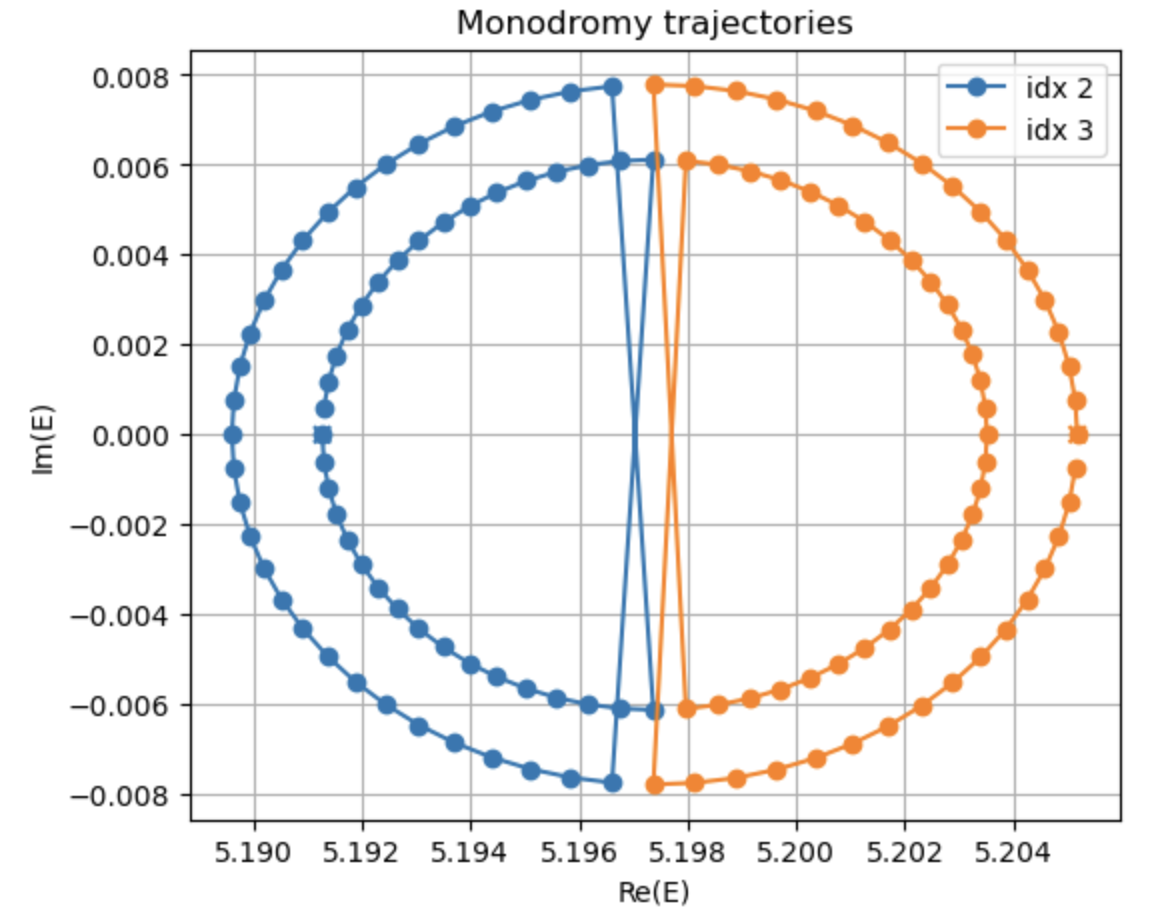}  
     \includegraphics[width=5.5cm] {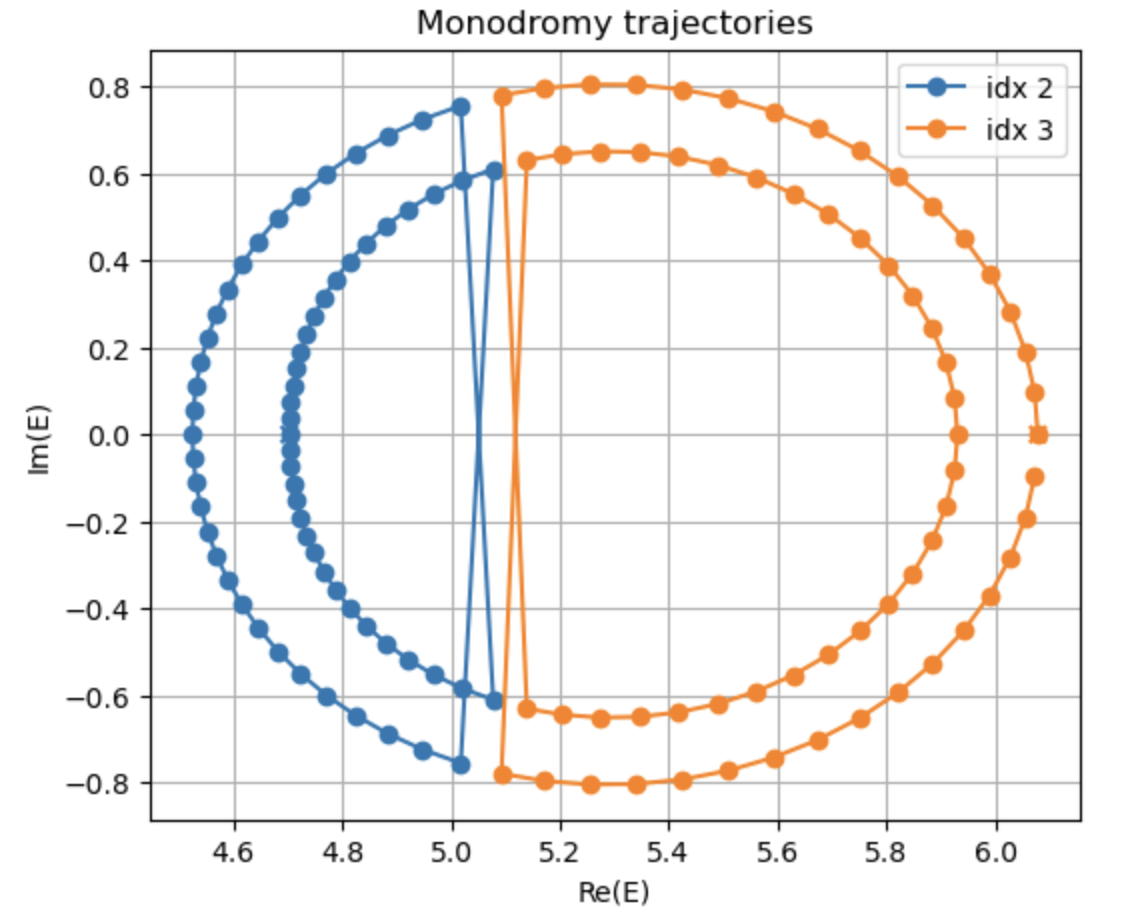}  \\
     \includegraphics[width=5.5cm] {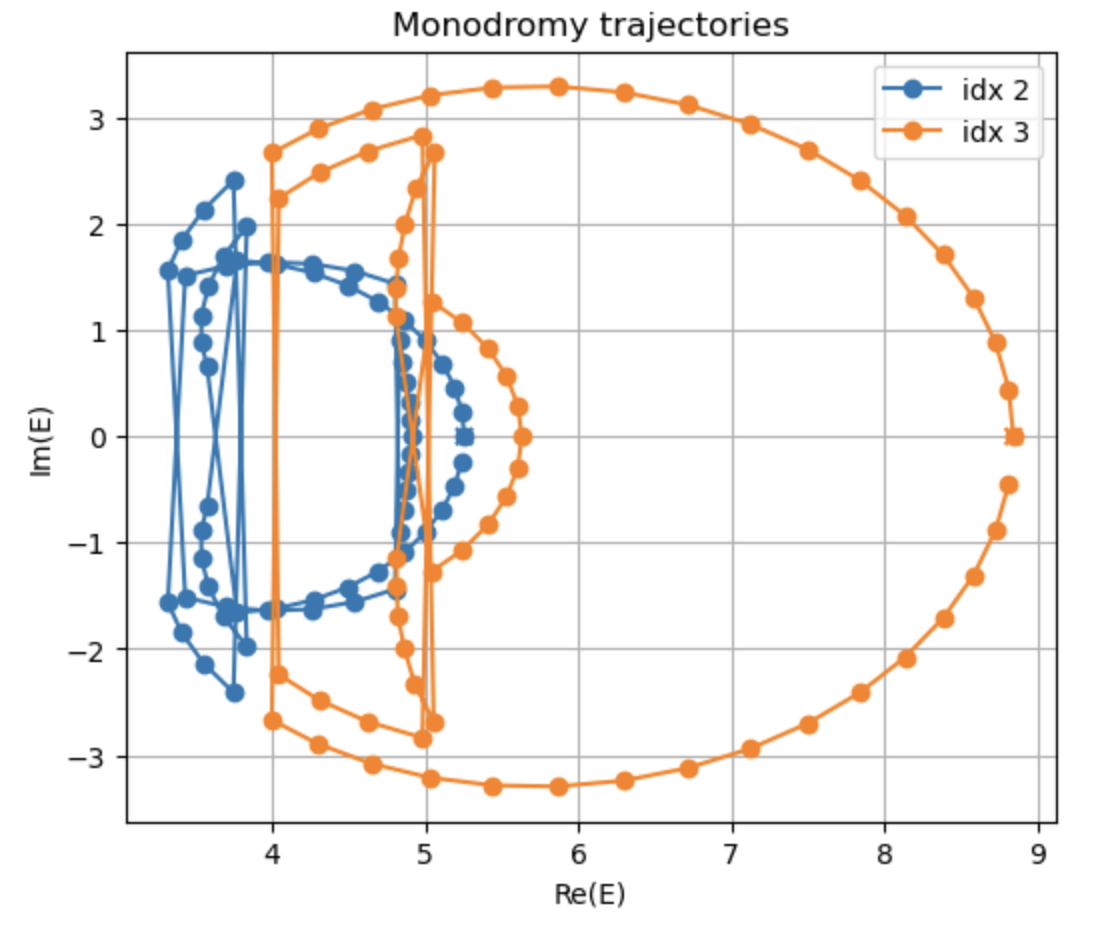}  
          \includegraphics[width=5.65cm] {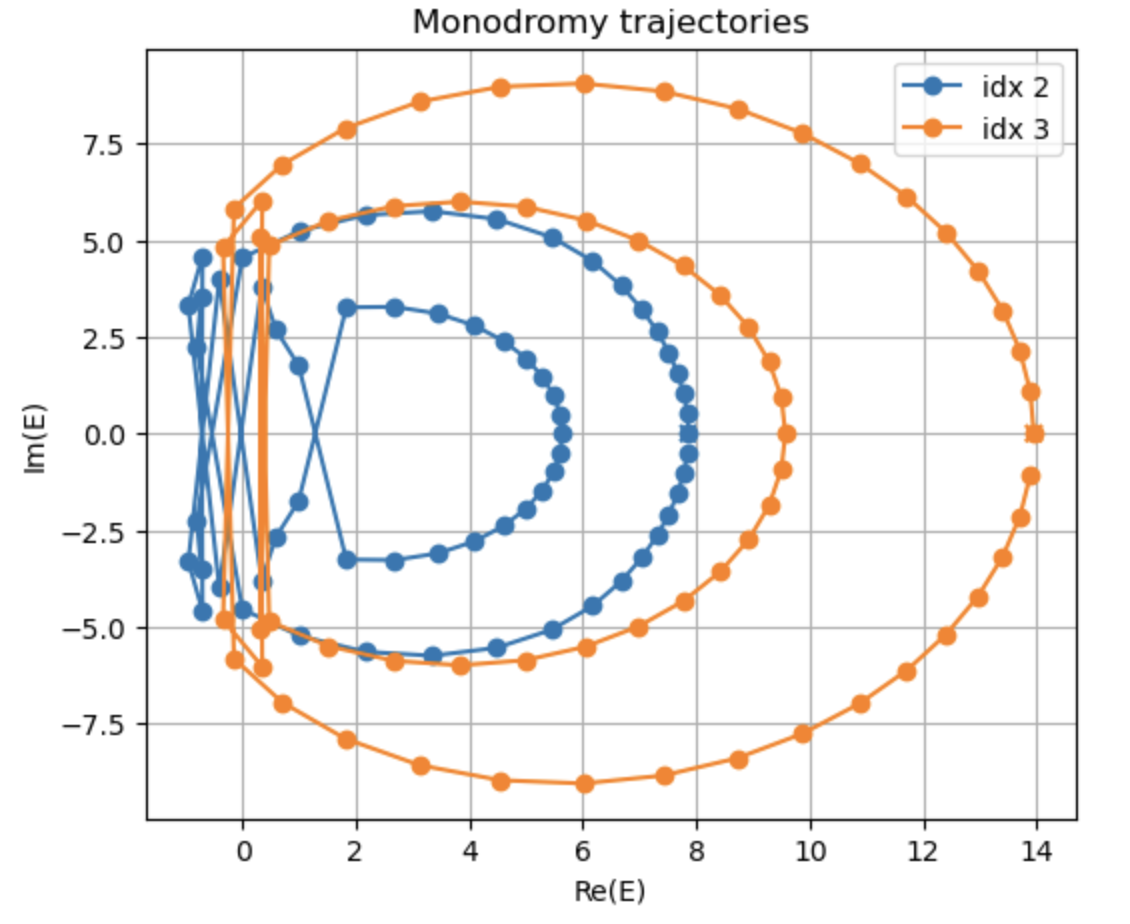}  
  \caption{Monodromy plots for various $r$. From top-left to bottom-right, $r=0.01$, $r=0.03$, $r=3$, $r=10$, and $r=20$. The results came from a numerical simulation of the Schrödinger eigenproblem of \cite{Guan:2025xoj}, with small coupling $\epsilon V(r) \cos \varphi$ and searching for exceptional points.}
 \label{fig:monodromy}
\end{figure}

The authors of \cite{Gong:2025pnu} also noted that the Litim regulator deforms the analytic structure in frequency space while a Callan-Symanzik (CS) regulator preserves it better. So regulator choice can also change whether the flow sees the analytic structure that would host EPs or not. In EP language, one could say that a badly chosen deformation or truncation can move branch cuts or eliminate the correct multi-sheet connection, hiding EPs.

\subsection{Controlled non-Hermitian deformation of $\theta$-vacuum toy model}\label{sec:controlled}
In the $\theta$-vacuum model, we know that at $\theta \approx \pi$, the effective theory has two nearly degenerate topological branches.

From the action \ref{eq:Sthetaaction}, the second term, which is the $\theta$-term becomes a topological term as
\begin{gather}
\frac{\theta}{4\pi} \int_0^{\beta} d\tau \ (\varphi^* \dot{\varphi}-\dot{\varphi}^* \varphi)= i \theta \ \nu,
\end{gather}
where $\nu$ is the winding number and $\nu \in \mathbb{Z}$.

Also, from the action, the extended Hamiltonian would be
\begin{gather}
H = \frac{1}{2m} \Pi_r^2 + \frac{1}{2 m r^2}  \left( \Pi_{\vartheta}- \frac{\theta+i \epsilon}{2\pi} r^2\right)^2 + \frac{g}{4}(r^2-1)^2,
\end{gather}
where the canonical momenta are
\begin{gather}
\Pi_r = m \dot{r}, \ \ \ \ \ \ \ \ \Pi_{\vartheta}= m r^2 \dot{\vartheta}+ \frac{\theta}{2\pi}r^2.
\end{gather}

Then, we can make $\theta$ complex as $\theta \to \theta+i \epsilon$, and then similar to works such as \cite{Zheng:2025apa, Liu:2026mox}, we can study Liouvillian, dissipative gap and the spectrum of the deformed operator as a function of $\epsilon$.

At low energy, $r$ is localized near $r=1$, so $\frac{1}{r^2} \to 1$, and the radial excitations decouple, and the angular energy levels would be
\begin{gather}
E_n(\epsilon) = \frac{1}{2m} \left ( n - \frac{\theta+ i \epsilon}{2\pi} \right).
\end{gather}
Near $\theta=\pi$, we have degeneracy, and can expand around it by defining $\delta \theta= \theta-\pi$.

Then, the effective Hamiltonian would be
\begin{gather}
H_{\text{eff}} = E_0 \mathbb{I} + \begin{pmatrix}
\alpha \ \delta \theta + i \gamma \epsilon & i \kappa \epsilon \\
i \kappa \epsilon & - \alpha \ \delta \theta - i \gamma \epsilon  
\end{pmatrix},
\end{gather}
where $\alpha \sim \frac{1}{2\pi m}$, and $\gamma$, $\kappa$ are real constants fixed by the angular sector.

In fact, if we consider the exact deformed Hamiltonian, we could write
\begin{gather}
\bra{n}H(\epsilon) \ket{n} = E_n^{(0)}+ i \epsilon \gamma_n + \mathcal{O}(\epsilon^2),
\end{gather}
which 
\begin{gather}
\gamma_n =  \frac{1}{2m \pi r^2} \left( n- \frac{\theta}{2\pi}\right).
\end{gather}
So $\gamma$ is the rate at which the energy of each angular sector moves into the complex plane. Also, $\kappa$ is a second order induced non-Hermitian mixing, coming from the projection onto finite subspace, integrating out higher angular modes or radial excitations, or allowing weak environmental coupling. So we can write
\begin{gather}
\kappa \sim \frac{\epsilon}{2m r^2 \pi} \sum_{\alpha} \frac{\bra{0} \Pi_{\vartheta} - \theta/2\pi \ket{\alpha} \bra{\alpha} \Pi_{\vartheta} - \theta/2\pi \ket{1}     }{E_0 - E_\alpha}.
\end{gather}
This parameter actually comes from virtual transitions through eliminated states and is zero if the system is closed. Note that in low energy and in subspace $r \to 1$.

The eigenvalues of the Hamiltonian are
\begin{gather}
\lambda_{\pm} = E_0 \pm \sqrt{\alpha^2 \delta \theta^2 - (\kappa^2 - \gamma^2)\epsilon^2+ 2 i\ \alpha  \gamma \delta \theta \ \epsilon},
\end{gather}
which for the EP, gives the condition of
\begin{gather}
\alpha^2 \delta \theta^2 = (\kappa^2-\gamma^2) \epsilon^2,
\end{gather}
where $\delta \theta$, and $\epsilon$ are real. At these points, the eigenvalues and eigenvectors coalesce, the Hamiltonian becomes non-diagonalizable and the spectrum exhibits a square-root branch point.

The Liouvillian, $\frac{d\rho}{dt}= \mathcal{L}(\rho)$, using the Lindblad formalism would be
\begin{gather}
\mathcal{L} = -i \lbrack H_\theta, \rho\rbrack + \sum_\mu \Big(L_\mu \rho L^\dagger_\mu - \frac{1}{2} \lbrace L^\dagger_\mu L_\mu, \rho \rbrace \Big).
\end{gather} 

Recently, in \cite{Liu:2024stj}, the operator size growth in Lindbladian SYK model has also been studied.

For our $\theta$-term model, as the angular degree of freedom is the main structure, a minimal, controlled choice for $L$ would be
\begin{gather}
L = \sqrt{\varepsilon} \ \left (\Pi_{\vartheta} - \frac{\theta}{2\pi} \right).
\end{gather}

The dissipative gap then would be
\begin{gather}
\Delta_{\text{diss}}= \frac{\varepsilon}{2} \left ( 1- \frac{\theta}{\pi} \right)^2,
\end{gather}
which closes near $\theta=\pi$.

If we define $A\equiv \Pi_{\vartheta}-\theta/2\pi$, the Liouvillian could be written as
\begin{gather}
\mathcal{L} \rho = -i \lbrack H_\theta, \rho \rbrack + \varepsilon \left \lbrack A \rho A - \frac{1}{2} \lbrace A^2, \rho \rbrace \right \rbrack,
\end{gather}
and in the angular momentum eigenbasis we have
\begin{gather}
A \ket{n} = a_n \ket{n}, \ \ \ \ a_n = n- \theta/2\pi.
\end{gather}

Then, for $n \ne m$, we have
\begin{gather}
\frac{d \rho_{nm} }{dt} = \left \lbrack -i (E_n - E_m) - \frac{\varepsilon}{2}(a_n - a_m)^2 \right \rbrack \rho_{nm},
\end{gather}
and therefore, the Liouvillian eigenvalues are
\begin{gather}
\lambda_{nm} = - \frac{\varepsilon}{2}(a_n-a_m)^2-i (E_n-E_m).
\end{gather}

The retarded Greens function then would be
\begin{gather}
G^R(\omega)= \frac{\omega + (\Delta- i \varepsilon \delta \gamma)\sigma_z + i \varepsilon \kappa \sigma_x}{\omega^2 - (\Delta - i \epsilon \delta \gamma)^2 + (\varepsilon \kappa)^2},
\end{gather}
where its poles are at
\begin{gather}
\omega_{\pm} = \pm \sqrt{(\Delta- i \varepsilon \delta \gamma)^2 -(\varepsilon \kappa)^2}.
\end{gather}

Based on the Liouvillian eigenvalues, one can decompose the Green's function as
\begin{gather}
\mathcal{G}(\omega) = \sum_{n \ne m} \frac{| n m ) ( n m |}{\omega + \frac{\varepsilon}{2}(a_n-a_m)^2 + i (E_n - E_m) },
\end{gather}
where $|n m)  \equiv \ket{n}\bra{m}$ are coherence modes which diagonalizes the Liouvillian superoperator, and $a_n \equiv n- \frac{\theta}{2\pi}$ is the $\theta$-shifted angular momentum eigenvalue.

From this Green's function, the closest pole to the origin would be
\begin{gather}
\Delta_{\text{diss}} = \min\limits_{n \ne m} \frac{\varepsilon}{2}(a_n-a_m)^2 = \frac{\varepsilon}{2}\left ( 1- \frac{\theta}{\pi} \right)^2,
\end{gather}
where it becomes long-lived at $\theta=\pi$, and signals dissipative criticality there.

\section{Conclusion}
In this note, based on the behavior of fidelity susceptibility, chiral symmetry breaking, and the number of spectra and coalescing levels, we proposed the holographic dual of optical exceptional points to be the end-wall in holographic confining geometries, and based on the order of EPs and their numbers in the system, we connected them with the hard-wall or soft-wall models. We also discussed the effects of both the EPs and end-wall on the chaos. Specifically, for different patterns of gain and loss we discussed the holographic spectra and showed that they match the behaviors coming from optical experiments. For symmetric pumping pattern in the ternary coupled microcavity model like the anti-$\mathcal{PT}$ sweep, we built the eigenvalue trajectories, radial profiles of the dominant lasing eigenmode, and also showed the behavior of growth rates of modes. We also built the phase diagram distinguishing unbroken $\mathcal{PT}$ versus broken $\mathcal{PT}$ phase.

Then, we specifically discussed a ``soft-wall'' holographic model of EPs and applied the Ferrell-Glover-Tinkham (FGT) sum rule for different patterns of pumping. In those systems, we swept the parameter space and studied the transfer fraction of spectral weight into the coherent lasing peak as a function of coupling and pumping gain/loss. Then, we extended the model into the inhomogeneous case of a lattice with three sites and studied the behavior of EPs and the FGT sum rule. We also showed how, by analyzing the behaviors of eigenvalues and eigenvectors, we could draw an analogy between the effects of the confining end-wall and EPs on the chaos in the system and showed they behaved similarly. There, we also studied the phase rigidity and Petermann factor numerically, using the holographic model we built based on AdS/QCD. By changing coupling $g$ or gain/loss $\gamma$ for each case, we found an exceptional point for each scenario.

Next, we studied the connections between timelike entanglement entropy and photonic exceptional points. Specifically, for this connection, we used the Kirkwood-Dirac-type distribution, which we used to describe both EPs and timelike entanglement entropy. We also sketched the real and imaginary parts of the KD distribution for different $\gamma$ (gain/loss), and then we went through the details of the connections between EP and TEE.

Finally, in our last section, we studied the connections between EPs and the $\theta$-vacuum of QCD. For both cases, we studied the winding number and connected those terms together. Then, in a $\theta$-vacuum model which has been studied by ``functional renormalisation'' (fRG), which has different topological sectors based on Chern-Simons number, we searched for exceptional points which we could not find. Then, we perturbed the system by a small angular coupling and again searched for EPs, where we found a real 2nd-order EP. For this model, we also presented the monodromy plots for various radii around the exceptional point.

This work can be extended and continued in many directions. For instance, these EPs could be found in various other models of QCD and condensed matter. Exceptional points could be applied in studying various other decay processes in QCD and quantum field theory. The numerical results from our holographic model could be compared to other experimental results. Different patterns of pumping could be constructed holographically and numerically, and the best setup for experiments and applications could be proposed. The connections between various recent measures of chaos and EPs could be built more rigorously. The timelike entanglement entropy, through EPs, could be applied to probe various quantum and optical systems, and the connections between neglectons and EPs could be constructed more firmly. We plan to address some of these problems in our future works.

 \section*{Acknowledgments}
I would like to thank Shahin Rouhani, Jun Nian, Wu-zhong Guo, Aireza Manavi, Mostafa Mohajeri, and the Quantonix startup group for useful and interesting discussions. ChatGPT has been used for developing Python codes and editing.

 \medskip

\bibliography{EPquantonix.bib}
\bibliographystyle{JHEP}
\end{document}